\documentclass[cernpreprint, atlasdraft=false, UKenglish,orcidlogo, texmf, orcidlogo]{atlasdoc}
\usepackage[subfigure,abbrevref=false,multirow=true]{atlaspackage}
\usepackage{makecell}
\usepackage{comment}
\PreprintIdNumber{CERN-EP-2024-102}
\usepackage{atlasbiblatex}

\usepackage{atlasphysics}
\usepackage{atlasbsm}

\usepackage{overpic}
\usepackage{afterpage}
\usepackage{float}
\graphicspath{{logos/}{figures/}}

\usepackage{ANA-EXOT-2018-62-PAPER-defs}

\AtlasTitle{Constraints on simplified dark matter models involving an $s$-channel mediator with the ATLAS detector in $pp$ collisions at $\sqrt{s} = 13$~\TeV}

\AtlasVersion{3.0}

\AtlasAbstract{
This paper reports a summary of searches for a fermionic dark matter candidate in the context
of theoretical models characterised by a mediator particle exchange in the $s$-channel.
The data sample considered consists of $pp$ collisions delivered by the Large Hadron Collider during its Run 2 at a centre-of-mass
energy of $\sqrt{s} = 13\,\TeV$ and recorded by the ATLAS detector, corresponding to up to \MaxLumi.
The interpretations of the results are based on simplified models where the new mediator particles
can be spin-0, with scalar or pseudo-scalar couplings to fermions, or spin-1, with vector or axial-vector couplings to fermions.
Exclusion limits are obtained from various searches characterised by final states with resonant production of Standard Model particles, or production of Standard Model particles in association with large missing transverse momentum.
}

\AtlasRefCode{EXOT-2018-62}

\AtlasJournalRef{Eur. Phys. J. C 84 (2024) 1102}
\AtlasDOI{10.1140/epjc/s10052-024-13215-5}

\hypersetup{pdftitle={ATLAS document},pdfauthor={The ATLAS Collaboration}}

\begin{document}

\maketitle

\tableofcontents

%
%

%
%
%

\section{Introduction}
\label{sec:intro}

The presence of dark matter (DM) in the Universe has been established for several decades, through astrophysical observations. Measurements of the cosmic microwave background radiation ~\cite{Hinshaw:2012aka,Akrami:2018vks} and the observation of the gravitational lensing effect ~\cite{Trimble1987,Bertone:2004pz,Feng:2010gw,Clowe_2006} confirm DM as one of the leading components of the matter density of the Universe.
Knowing little about its nature, an intriguing option is that DM is a particle, with peculiar features to conform with the cosmological observations.

Although gravitational interactions between dark matter and ordinary matter have been observed, there is no evidence for other types of interaction. The weakly interacting massive particle (WIMP) paradigm postulates that DM consists of massive particles that also have weak-scale interactions with ordinary matter that result in a relic density consistent with observation. In this paper, the ATLAS data are interpreted using theoretical models extending the Standard Model (SM) with additional parameters governing the WIMP interaction with SM and beyond the SM (BSM) particles.

Assuming the particle nature of DM, there are three main and complementary approaches for its experimental search: observation of DM particle co-annihilation into SM particles, such as photons, with peculiar kinematic features (\emph{indirect detection}), observation of DM particle scattering on SM particles, such as nuclei, through the recoil of the latter (\emph{direct detection}), and DM particle production in SM particle collisions, as produced at the Large Hadron Collider (LHC) at CERN. The complementarity of these three approaches can only be assessed by assuming a common underlying model describing the DM-SM particle interactions.

Several BSM extensions predict a suitable WIMP candidate. Such a particle is stable over cosmological scales and weakly interacts with SM particles. For example, the lightest non-SM particle predicted in (almost) complete $R$-parity-conserving supersymmetric models~\cite{Fayet:1976et,Fayet:1977yc,Farrar:1978xj} is itself a reasonable WIMP candidate. Setting constraints on the set of parameters characterising each of these DM-related BSM models would be challenging for each analysis carried out at a collider.
Simplified DM models propose minimal extensions of the SM that include a single DM particle candidate, $\chi$, and a single massive mediator connecting the DM particle with other SM particles \cite{Abdallah:2015ter,Abercrombie_2019}.

During Run 1 of the LHC, a minimalistic approach was considered to describe the DM-SM particle interactions~\cite{Goodman:2010ku,Goodman:2010yf}, based on Effective Field Theory (EFT) operators, upon which multiple searches set constraints.
This approach, though pragmatic, prevented a detailed comparison between searches carried out by the ATLAS Collaboration exploiting different experimental signatures, such as searches for visible decays of the mediator (e.g. dijet, dilepton) and semi-visible final states (e.g. $X+\MET$).
The simplified DM models were adopted in Run 2 to enable the comparison of the exclusion reach of these searches.

The family of aforementioned models depends on the spin-parity nature of the mediator, of the WIMP, and on the corresponding couplings between the newly introduced particles and the SM ones.
These models are often non-renormalisable and hence cannot be considered as a complete theory. Instead, they are a useful proxy for more extended theories, which can profit from the exclusion provided in the context of simplified models.
The sensitivity reach as a function of the few parameters of simplified models provides a good insight into unexplored regions and indicates where to focus future searches on this topic.
In \thisdoc the reinterpretation of the ATLAS DM-related searches is considered, focusing on the simplified model's signature via \schannel mediator production, with the subsequent decay of the mediator into either SM particles or DM candidate particles.
As reported by the LHC Dark Matter Working Group \cite{dmwg_recs, Abercrombie_2019}, the \schannel processes are more promising relative to other DM production mechanisms due to a larger cross-section and clear experimental signatures.

With respect to a previous ATLAS publication~\cite{EXOT-2017-32}, \thisdoc collects the latest results produced by the ATLAS Collaboration with interpretation in the context of \schannel DM simplified models.
Additional analyses, originally targeting different dark matter models, are reinterpreted in the context of \schannel DM simplified models.
Several of the input analyses considered in \thisdoc searched for evidence of a DM candidate but found no significant deviations relative to the SM prediction; therefore, constraints were set on the presence of BSM signals.
To better highlight the sensitivity and complementarity of the searches, constraints are provided for new assumptions of the simplified model's parameters, extending the information usually provided in benchmark scenarios \cite{Abdallah:2015ter, Abercrombie_2019}. A semi-analytical technique for dynamic rescaling of existing limits to different values of model parameters is exploited, and the reinterpretation of the results in terms of DM-SM particle scattering is presented.

In \cref{sec:detector} the ATLAS detector is briefly described, while in \cref{sec:models} the details of the theoretical framework of \schannel simplified models considered for the interpretation of DM searches are presented.
\cref{sec:analysis} includes a summary of the DM \schannel experimental signatures and the corresponding ATLAS analyses.
The semi-analytical method to reinterpret analysis results is described in \cref{sec:rescaling}, while the approach considered to compare the ATLAS results with those corresponding to DM direct detection (DD) experiments is reported in \cref{sec:dd_reinterpret}.
The results regarding spin-0 and spin-1 mediated models are summarised in \cref{sec:result} whereas concluding remarks are provided in \cref{sec:conclusion}.


%
%
%
%

%

%
\section{ATLAS detector}
\label{sec:detector}
\newcommand{\AtlasCoordFootnote}{%
ATLAS uses a right-handed coordinate system with its origin at the nominal interaction point (IP)
in the centre of the detector and the \(z\)-axis along the beam pipe.
The \(x\)-axis points from the IP to the centre of the LHC ring,
and the \(y\)-axis points upwards.
Polar coordinates \((r,\phi)\) are used in the transverse plane,
\(\phi\) being the azimuthal angle around the \(z\)-axis.
The pseudorapidity is defined in terms of the polar angle \(\theta\) as \(\eta = -\ln \tan(\theta/2)\).
Angular distance is measured in units of \(\Delta R \equiv \sqrt{(\Delta\eta)^{2} + (\Delta\phi)^{2}}\).}

\label{sec:atlas2b}

The ATLAS detector~\cite{PERF-2007-01} at the LHC covers nearly the entire solid angle around the collision point.\footnote{\AtlasCoordFootnote}
It consists of an inner tracking detector surrounded by a thin superconducting solenoid, electromagnetic and hadron calorimeters,
and a muon spectrometer incorporating three large superconducting air-core toroidal magnets.

The inner-detector system (ID) is immersed in a \qty{2}{\tesla} axial magnetic field and provides charged-particle tracking in the pseudorapidity range \(|\eta| < 2.5\).
The high-granularity silicon pixel detector covers the vertex region and typically provides four measurements per track,
with the first hit normally being in the insertable B-layer (IBL) installed before Run~2~\cite{ATLAS-TDR-19,PIX-2018-001}.
It is followed by the silicon microstrip tracker (SCT), which usually provides eight measurements per track.
These silicon detectors are complemented by the transition radiation tracker (TRT),
which enables radially extended track reconstruction up to \(|\eta| = 2.0\).
The TRT also provides electron identification information
based on the fraction of hits (typically 30 in total) above a higher energy-deposit threshold corresponding to transition radiation.

The calorimeter system covers the pseudorapidity range \(|\eta| < 4.9\).
Within the region \(|\eta|< 3.2\), electromagnetic calorimetry is provided by barrel and
endcap high-granularity lead/liquid-argon (LAr) calorimeters,
with an additional thin LAr presampler covering \(|\eta| < 1.8\)
to correct for energy loss in material upstream of the calorimeters.
Hadron calorimetry is provided by the steel/scintillator-tile calorimeter,
segmented into three barrel structures within \(|\eta| < 1.7\), and two copper/LAr hadron endcap calorimeters.
The solid angle coverage is completed with forward copper/LAr and tungsten/LAr calorimeter modules
optimised for electromagnetic and hadronic energy measurements respectively.

The muon spectrometer (MS) comprises separate trigger and
high-precision tracking chambers that measure the deflection of muons in a magnetic field generated by the superconducting air-core toroidal magnets.
The field integral of the toroids ranges between \num{2.0} and \qty{6.0}{\tesla\metre}
across most of the detector.
Three layers of precision chambers, each consisting of layers of monitored drift tubes, cover the region \(|\eta| < 2.7\) and are
complemented by cathode-strip chambers in the forward region where the background is highest.
The muon trigger system covers the range \(|\eta| < 2.4\) with resistive-plate chambers in the barrel and thin-gap chambers in the endcap regions.

The events of interest are selected by the first-level (L1) trigger system implemented in custom hardware,
followed by selections made by algorithms implemented in software in the high-level trigger (HLT)~\cite{TRIG-2016-01}.
The first-level trigger accepts events from the \qty{40}{\MHz} bunch crossings at a rate below \qty{100}{\kHz},
which the high-level trigger further reduces to record events to disk at about \qty{1}{\kHz}.

An extensive software suite~\cite{ATL-SOFT-PUB-2021-001} is used in data simulation, in the reconstruction
and analysis of real and simulated data, in detector operations, and in the trigger and data acquisition
systems of the experiment.


%
%
%
%

%

\section{Theoretical models}
\label{sec:models}
The results considered in \thisdoc are interpreted in the framework of simplified DM models \cite{Abdallah:2015ter,Abercrombie_2019}: extensions of the SM that add a WIMP-like particle suitable as a DM candidate and a new particle mediating the interaction of DM and SM particles. These DM simplified models, which overcome some of the shortcomings of the previous EFT-based DM models \cite{Bai:2010hh,Beltran:2010ww,Goodman:2010yf,Goodman:2010ku,Rajaraman:2011wf,Fox:2011pm}, can be classified according to the properties of the mediator between the DM and SM particles.
This gives rise to collider signatures with different kinematic characteristics and topologies.

Among the various types of simplified DM models, \thisdoc specifically considers those with a mediator sector composed of a single massive particle, either of spin-0 or spin-1, which is produced in the \schannel mode.
The DM candidate $\chi$ is always assumed to be a Dirac fermion. Different assumptions, such as $\chi$ being a Majorana fermion or a scalar particle, significantly change the set of allowed interactions, and hence the final state signatures and the total production cross-section. In addition, changes in the kinematic distributions of visible particles are expected when changing the assumption on the nature of the DM particle.

Assuming the aforementioned mediator is produced in \pp collisions at the LHC, it should couple both to SM particles and to yet undetected DM candidates.
Based on this principle, the resulting experimental signatures can be grouped into two categories:
\begin{itemize}
\item Semi-visible final states, where the mediator decays into undetected DM particles, and hence produces an imbalance in the observed momentum distribution in the ATLAS transverse plane (referred to as missing transverse momentum, with magnitude \MET). The study of these decays is possible by the reconstruction of at least one visible object, $X$, in the final state. This kind of signature, often referred to as $X+\MET$, is particularly sensitive to the mass of the DM candidate (\mDM) in connection with the mass of the mediator (\mMed). The latter can decay into a pair of the former only if $\mDM \leq \mMed/2$.
\item Visible final states, where the mediator decays into SM particles without direct production of the DM particles. These searches target evidence of the mediator itself, with mild dependency on the mass of the DM particle, which is mostly relevant only when the kinematic regime allows for direct decay of the mediator to DM particles.
\end{itemize}
Profiting from the interplay between these two search categories, it is possible to probe a large fraction of the (\mDM,~\mMed) plane, with each category covering different regions.

This paper assumes that the width of the mediator is consistently calculated as the smallest value possible from all other parameters, based on the minimal width assumption reported in Ref.~\cite{Abercrombie_2019}.
While the two categories of mediators are detailed in \cref{sec:spin_zeromodel,sec:spin_onemodel}, \cref{tab:models} summarises their main features.

\begin{table}[!htb]
\centering
\caption{Summary of the \schannel DM simplified model benchmarks considered in \thisdoc, along with the associated acronym, mediator symbol used throughout and model's parameter values. The last column lists the final-state signatures considered to set constraints on each model.}
\label{tab:models}
\begin{tabular}{p{3cm}cc|c|ccc|c}
\toprule
Mediator                          & Acronym & Symbol                        & $J^P$ & \multicolumn{3}{c|}{Couplings}       & Signatures \\
\midrule
Spin-0                            &         &                               &       & \multicolumn{2}{c}{\makecell{$g_q$}} & $g_\chi$     & \\
\midrule
\multirow{2}{*}{~ ~ Scalar}                        & \multirow{2}{*}{S}       & \multirow{2}{*}{$\phi$} & \multirow{2}{*}{$0^+$} & \multicolumn{2}{c}{\multirow{2}{*}{\makecell{1.0}}}   & \multirow{2}{*}{1.0}          & \multirow{4}{*}{\makecell{\Monojet, \ttmet,\\ \bbmet,\\~\monotop,~\fourtops}}\\
& & & & & &\\
\cline{1-7}
\multirow{2}{*}{~ ~ Pseudo-Scalar}                 & \multirow{2}{*}{PS}      & \multirow{2}{*}{$a$}    & \multirow{2}{*}{$0^-$} & \multicolumn{2}{c}{\multirow{2}{*}{\makecell{1.0}}}   & \multirow{2}{*}{1.0}          & \\
& & & & & &\\
\midrule
Spin-1                            &         &                               &       & $g_q$            & $g_l$        & $g_\chi$                               & \\
\midrule
~ ~ \multirow{4}{*}{Vector}       & V1      & \multirow{4}{*}{$Z^\prime_V$} & \multirow{4}{*}{$1^-$} & 0.25             & 0.0          & 1.0                                    & \multirow{8}{*}{\makecell{Jet/$\gamma$/\Wboson/\Zboson+\MET,\\ Dilepton resonances,\\ Dijet resonances}}\\
& V2      &                               &  & 0.1              & 0.01         & 1.0                                    & \\
& V3      &                               &  & 0.07             & 0.0          & 1.0                                    & \\ %
& V4      &                               &  & 0.15             & 0.03         & 1.0                                    & \\ %
\cline{1-7}
~ ~ \multirow{4}{*}{Axial-Vector} & A1      & \multirow{4}{*}{$Z^\prime_A$} & \multirow{4}{*}{$1^+$} & 0.25             & 0.0          & 1.0                                    & \\
& A2      &                               &  & 0.1              & 0.1          & 1.0                                    & \\
& A3      &                               &  & 0.07             & 0.0          & 1.0                                    & \\ %
& A4      &                               &  & 0.2              & 0.05         & 1.0                                    & \\ %
\bottomrule
\end{tabular}
\end{table}

\subsection{Spin-0 simplified models}
\label{sec:spin_zeromodel}

The first category of models considered is a set of simplified SM extensions that include a single spin-0 particle~\cite{Abercrombie_2019,Buckley_2015,Harris_2015,haisch2015simplified}, mediating the interaction between SM particles and the Dirac DM WIMP.
Assuming Minimal Flavour Violation~\cite{DAmbrosio:2002vsn}, this mediator has Higgs-like Yukawa couplings with SM particles and hence this class of models is a subset of the ultraviolet-complete theories predicting an extended Higgs sector
\footnote{The couplings of the mediator to W and Z bosons, as well as explicit dimension-4 $\phi$-h or a-h couplings, are set to zero in
this simplified model following ~\cite{Buckley_2015}. In addition, the coupling of the mediator to the dark sector are not taken to be proportional to the mass of
the DM candidate}.

These models assume either a scalar ($\phi$) or pseudo-scalar ($a$) mediator with colour-neutral couplings to SM particles with four free parameters: the mass of the mediator (\mphi or \ma), the DM mass (\mDM), the mediator coupling to DM (\gChi) and the flavour-universal term of the mediator coupling to SM particles\,(\gq).\footnote{A single universal parameter $\gq\equiv g_u = g_d = g_\ell$ is assumed for the sake of simplicity.}
The parameter \gq is combined with the corresponding SM-Yukawa coupling for each fermion to determine the mediator coupling to SM particles.

\begin{figure}[!tb]
\centering
\subfigure[]{\includegraphics[width=.23\textwidth]{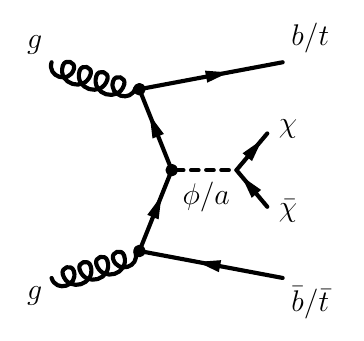}\label{fig:DMtt}}
\subfigure[]{\includegraphics[width=.23\textwidth]{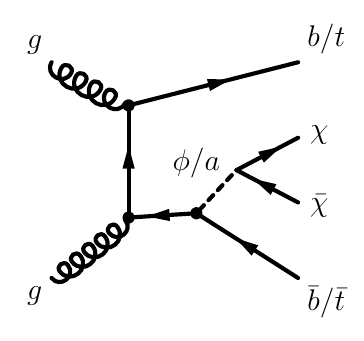}\label{fig:DMtt2}}
\subfigure[]{\includegraphics[width=.23\textwidth]{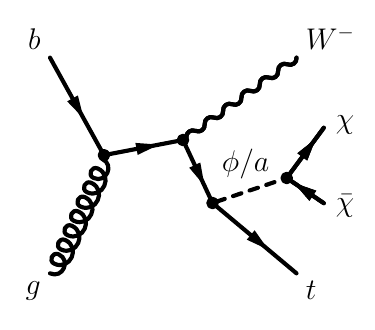}\label{fig:DMtW}}
\subfigure[]{\includegraphics[width=.23\textwidth]{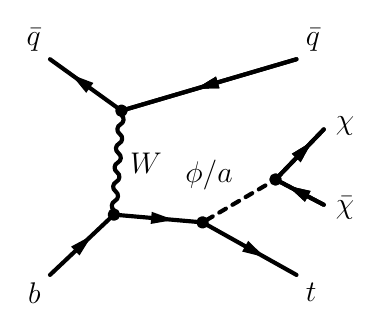}\label{fig:DMtj}}
\subfigure[]{\includegraphics[width=.23\textwidth]{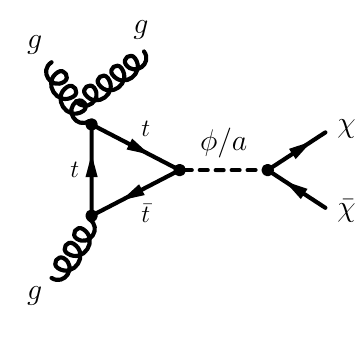}\label{fig:Monojet}}
\subfigure[]{\includegraphics[width=.23\textwidth]{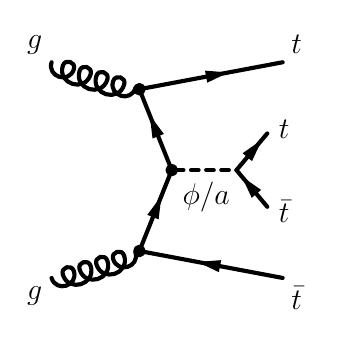}\label{fig:DM4top1}}
\subfigure[]{\includegraphics[width=.23\textwidth]{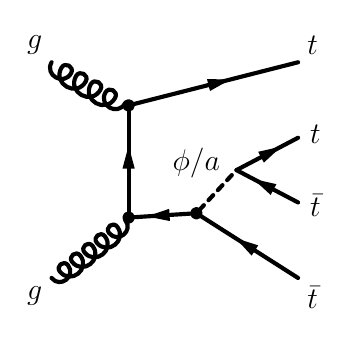}\label{fig:DM4top2}}
\caption{Representative Feynman diagrams for spin-0 mediator associated production with: \subref{fig:DMtt} and \subref{fig:DMtt2} a top/bottom quark pair (\ttdm, \bbdm),
\subref{fig:DMtW} a single top quark and a $W$ boson (\twdm), \subref{fig:DMtj} a single top quark and one (or more) jet(s) (\tjdm), \subref{fig:Monojet} monojet production (DM-monojet) and \subref{fig:DM4top1}, \subref{fig:DM4top2} 4 tops (\ttfourtop).}
\label{g:diagSpin0}
\end{figure}

\cref{g:diagSpin0} shows the most representative Feynman diagrams for the spin-0 \schannel simplified models. The mediator is primarily produced in association with heavy-flavour quarks (\cref{fig:DMtt,fig:DMtt2,fig:DM4top1,fig:DM4top2}) or through gluon-gluon fusion via a top-quark loop (\cref{fig:Monojet}).
The associated production of spin-0 mediators with a single top quark (\cref{fig:DMtW,fig:DMtj}) also has a sizeable, albeit non-dominant, cross-section~\cite{Pinna:2017tay,Pani:2017qyd,Haisch:2018bby}, in particular for higher-mass mediators.
In \thisdoc, results are presented for: associated production of a mediator particle $\phi/a$ (with $\phi/a\rightarrow\chi\chi$) with a pair of top/bottom quarks (\ttdm,~\bbdm), single top quark production (collectively referred to as \singletopdm), or with a jet (DM-monojet) as well as four-top-quark production (\ttfourtop).

Depending on the masses of the mediator, the SM, and DM particles, the mediator can decay into a pair of SM particles or a pair of DM particles.
This leads to different final states having complementary sensitivities used to constrain the model's parameter space.
Final states with multiple heavy quarks, associated or not with \MET, are useful to constrain these models. Due to the flavour non-universality of Yukawa-like couplings, the final states involving up-type or down-type quarks are studied separately when the corresponding searches are available.

\subsection{Spin-1 simplified models}
\label{sec:spin_onemodel}
These models assume the presence of an additional $U(1)$ gauge symmetry under which the DM particles are charged, with a resulting \ZprimeV (vector - V) or \ZprimeA (axial-vector - AV) boson mediator. Assuming the coupling of the mediator is independent of the fermion flavour, the model has five free parameters~\cite{Abercrombie_2019}: the masses of the mediator and of the DM particle (\mZprimeVA and \mDM, respectively), the flavour-universal coupling of the \Zprime boson to all quarks (\gq), the coupling to all leptons (\gl), and the coupling to DM (\gDM).

\begin{figure}[!tb]
\centering
\subfigure[]{\includegraphics[width=.25\textwidth]{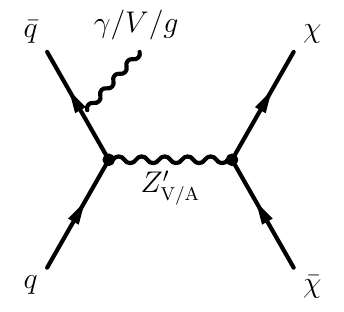}
\label{g:Diag-AVVneut1}
}
\subfigure[]{\includegraphics[width=.25\textwidth]{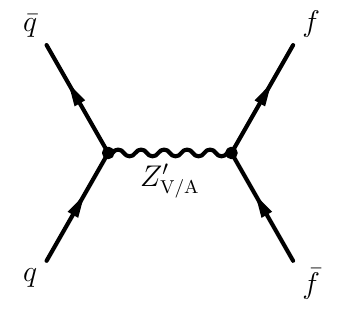}
\label{g:Diag-AVVneut2}
}
\caption{Representative Feynman diagrams of the dominant production and decay modes for the V/AV model. }
\label{g:Diag-AVVneut}
\end{figure}

\cref{g:Diag-AVVneut} shows the two most representative tree-level \schannel Feynman diagrams predicted by the aforementioned model in \pp collisions.
The \Zprime mediator can either decay into a pair of SM fermions or a pair of the DM particles resulting in either a fully visible or a semi-visible final state, respectively.
In the former case, the invariant mass reconstructed from the two final state fermions (quarks or leptons of any flavour family) are considered as a proxy for the mediator mass \mMed.
In the latter case, the final state DM particles are undetected, hence, a search for this kind of process considers an additional visible object produced in association with the mediator as initial-state radiation (ISR), as shown in \cref{g:Diag-AVVneut1}. Such a visible object can be a jet, a photon, or a \Wboson or \Zboson boson.
In \thisdoc, different coupling scenarios are considered for the interpretation of these models and to highlight the complementarity of each search. Of particular interest are benchmark models in which the mediator is leptophobic, with a null coupling to all SM leptons. These models are compared with other scenarios in which the coupling to leptons is non negligible.
As shown in \cref{tab:models}, two benchmark models with an axial-vector mediator (A1, A2) and two with a vector mediator (V1, V2) are considered, as suggested by the LHC DM Working Group \cite{dmwg_recs} and already considered in Ref.~\cite{EXOT-2017-32}.
Benchmark models (A1, V1) are selected based on Ref. \cite{Chala_2015} to balance the dijet and $X+\MET$ coverage, while benchmark models (A2, V2) are selected to highlight dilepton limits.

The results are extended to another new pair of benchmark models (A3, V3) chosen such that the semi-visible signatures ($X+\MET$) are sensitive in a region of parameters (\mDM, \mMed) where hadronic resonance searches are not.
The other new benchmark models (A4, V4) are considered to highlight the scenario in which the hadronic and leptonic final state searches have similar sensitivity in the (\mDM, \mMed) parameter space.


%
%
%
%

%

\section{Experimental signatures}
\label{sec:analysis}
Dark matter searches play a pivotal role in the overall ATLAS physics program.
This section provides an overview of searches that set constraints in the context of \schannel models, while more details can be found in the corresponding referenced papers.

\cref{tab:semivisible,tab:visible} summarise the DM searches for semi-visible and visible final states respectively, including an overview of the models constrained by each of these signatures.

Most of the analyses considered in \thisdoc exploited the proton-proton collision data at $\sqrt{s}=13$~\TeV collected during the entire Run 2 of the LHC, corresponding to an integrated luminosity of 139~\ifb.
A few recent analyses profited from a slightly extended data sample~\cite{DAPR-2021-01} including additional recovered data-taking periods, corresponding to \MaxLumi.
Other analyses exploited only a subset of the Run 2 data, as reported in the following section.
In the following, analyses used the data corresponding to 139~\ifb unless otherwise specified.

Electrons, muons, photons and jets were reconstructed by combining the signals from the different components of the ATLAS detector~\cite{PERF-2016-01,EGAM-2018-01,MUON-2018-03,PERF-2015-09,JETM-2018-05,JETM-2018-06,JETM-2018-02,ATLAS-CONF-2020-022,PERF-2014-03,ATL-PHYS-PUB-2019-026}.
Leptons ($\ell$) in the following refer to electrons or muons. In several analyses, events with identified leptons were rejected from the signal region. This is referred to here as a lepton veto. The analyses may have implemented different lepton and photon selection criteria for particle identification, and kinematic requirements (\pt, $\eta$).
Small-$R$ and large-$R$ jets were reconstructed from energy deposits in the calorimeters using the \antikt jet algorithm~\cite{Cacciari:2008gp,Fastjet} and using a radius parameter of $R=0.4$ and $R=1.0$, respectively.
Reclustered large-$R$ jets~\cite{Nachman:2014kla} were reconstructed from small-$R$ jets using a radius parameter of either $R=0.8$ or $R=1.2$.
Particle-flow (PFlow) jets~\cite{PERF-2015-09} were considered in some analyses, using neutral PFlow constituents and charged constituents associated with the primary vertex as input, and using the \antikt algorithm with $R=0.4$.
Multivariate algorithms were used to identify small-$R$ jets with $\pt > 20~\GeV$ containing $b$-hadrons (\bjets)~\cite{PERF-2012-04,ATL-PHYS-PUB-2016-012,FTAG-2018-01,FTAG-2019-07}, referred to as $b$-tagging. For large-$R$ jets, $b$-tagging was applied to their associated track-jets, which were built from tracks reconstructed in the inner detector using the \antikt jet algorithm with $R=0.2$ or with a variable cone size~\cite{Krohn_2009}.
The missing transverse momentum, \ptmissvec (with magnitude \MET), was calculated from the negative vector sum of transverse momenta (\pt) of electrons, muons, photons, jet candidates and an additional soft term~\cite{PERF-2016-07} that includes activity in the tracking system originating from the primary vertex but not associated with any reconstructed particle.
Some analyses also considered hadronically decaying $\tau$-leptons in the \ptmissvec reconstructions.
The object-based \MET significance~\cite{ATLAS-CONF-2018-038} was used in some analyses to compare the magnitude of the \MET relative to its resolution due to the constituent objects.

\subsection{Searches for semi-visible final states}
\label{ssec:semivisible_sig}

The presence of one or a pair of undetected DM particles from a collision would result in a significant energy imbalance in the transverse plane, reconstructed in the form of \MET. This invisible part of the final state was analysed together with the visible objects in searches for $X+\MET$, where $X$ denotes additional particles. A mediator can decay directly into DM particles only if its mass is above the kinematic threshold for DM pair production, $\mMed \gtrsim 2~\mDM$. In the $\mChi-\mMed$ plane, analyses aiming at the semi-visible final states were therefore sensitive below the $\mChi = \mMed/2$ threshold.

\begin{table}
\caption{Summary of searches for semi-visible final states used to constrain the different \schannel DM models defined in \cref{sec:models}. See the text for symbol definitions.
}
\label{tab:semivisible}
\centering
\resizebox{\textwidth}{!}{
\begin{tabular}{m{45mm}m{25mm}m{60mm}m{55mm}}
\toprule
Analysis                                             & Models targeted & Final state signature                                               & Key Characteristics                                    \\\midrule
\midrule
\bbmet~\cite{SUSY-2018-34}                                        & S/PS      & 2 $b$-jets, \met, 0 $\ell$                                          & Boosted decision tree and binned likelihood fit of $\cos\theta^*_{bb}$ \\[0.5ex]
\ttmet~\cite{SUSY-2018-12,SUSY-2023-22,SUSY-2018-08,SUSY-2019-12} & S/PS      & 0-1-2$\ell$, \met, $\geq 1$ \bjets                                  & Statistical combination of \ttmet final state analyses\\[0.5ex]
\twmet 0-1$\ell$~\cite{EXOT-2021-01}                              & S/PS      & 0-1$\ell$, \met, $\geq 1$~\bjets, \Wboson tagged jets               & Binned likelihood fit of \MET \\[0.5ex]
\twmet 2$\ell$~\cite{EXOT-2018-43}                                & S/PS      & 2$\ell$, $\geq 1$~\bjet, \MET                                       & Single bin likelihood fit \\[0.5ex]
\tjmet~\cite{EXOT-2018-43}                                        & S/PS      & 1$\ell$, 1-4~jet, 1-2~\bjet, \MET                                   & Binned likelihood fit of BDTs \\[0.5ex]
\Monojet~\cite{EXOT-2018-06}                                      & S/PS,V/AV & 1 high-\pt jet, \MET, 0 $\ell$                                      & Binned likelihood fit of \MET \\[0.5ex]
\monophoton~\cite{EXOT-2018-63}                                   & V/AV      & 1 high-\pt $\gamma$, \MET, 0 $\ell$                                 & Binned likelihood fit of \MET \\[0.5ex]
\monoZll~\cite{HIGG-2018-26}                                      & V/AV      & 2$\ell^+\ell^-$, \MET, 0 jets                                       & Binned likelihood fit of \MET \\[0.5ex]
\monoVhad~\cite{EXOT-2019-27}                                   & V/AV      & \MET, $\Wboson/\Zboson$ candidate (resolved and boosted topologies) & Binned likelihood fit of \MET \\[0.5ex] %
\bottomrule
\end{tabular}
}
\end{table}

\paragraph*{\bbmet}
\label{par:bbmet}
The ~\bbmet analysis ~\cite{SUSY-2018-34} exploited a selection with two \bjets in the final state, vetoing all events with any leptons satisfying the baseline selection criteria.
Events were further selected using a two-dimensional requirement based on \met and the \pT of the leading jet. %
Boosted decision trees (BDT) were trained to discriminate between the main background processes (top pair production, \Wjets , \Zjets) and two sets of kinematically similar signal models that were characterised by either low or high mediator mass.
The events were further divided into five bins using the $\cos\theta^*_{bb} = |\tanh{\Delta\eta_{bb}/2}|$ discriminating variable, where $\Delta\eta_{bb}$ is the pseudorapidity difference between the two leading b-jets.

\paragraph*{\ttmet} %
\label{par:ttmet}
The \ttmet final state has  been intensively investigated in different analyses~\cite{SUSY-2018-12,SUSY-2023-22,SUSY-2018-08,SUSY-2019-12}, taking advantage of the three possible decays of the \Wboson bosons originating from the top-quark-pair decay.
Events were considered separately based on the number of leptons produced by the decay of the \Wboson bosons:
zero leptons (both \Wboson bosons decayed hadronically), one lepton (one of the \Wboson bosons decayed leptonically and the other one hadronically), and two leptons (both \Wboson bosons decayed leptonically).
The \ttmet final state was also characterised by the presence of \(b\)-tagged jets and significant \met.
In \thisdoc, the result for the \ttmet final state is reported as the combination of the three leptonic final state analyses as in Ref.~\cite{SUSY-2019-12}. In addition a new and improved 1-lepton analysis targeting the \ttmet final state was published~\cite{SUSY-2023-22}. This updated analysis replaced the previous one reported in Ref.~\cite{SUSY-2019-12}.

The analysis targeting 0-lepton final state~\cite{SUSY-2018-12,SUSY-2019-12} used the $\met$ triggers to select events with large \MET and at least one highly energetic hadronically decaying top quark candidate. Events with lower-momentum jets that fail to meet the former trigger requirement were selected relying on a combination of $\met$ and $b$-tagged jet triggers.

The analysis targeting the 1-lepton final state~\cite{SUSY-2023-22}, based on a data sample corresponding to 140~\ifb, selected events with exactly one electron or muon using a combination of \MET and single lepton triggers. Two neural-network-based classifiers were used for the reconstruction of the hadronically decaying top quark and to discriminate signal and background processes.

Finally, the analysis targeting the 2-leptons channel~\cite{SUSY-2018-08} selected events with two opposite-sign leptons that were inconsistent with a \Zboson boson. Events were selected with dilepton triggers and at least one \bjet was also required in the selection.

\paragraph*{\monotop}
\label{par:monotop}
The monotop searches ~\cite{EXOT-2018-43,EXOT-2021-01} targeted events with one top quark and relatively large \met.
Three analyses with this signature are considered in \thisdoc.
The $\tjmet$ analysis~\cite{EXOT-2018-43} selected events with one electron or muon, collected using a single-lepton triggers, one to four jets with $\pT>30~\GeV$, one or two of which must be $b$-tagged and large \met. To further improve the sensitivity, a BDT was defined and a binned distribution of the BDT output was then used to extract the final results in the signal region.
The $\twmet$ 2-lepton analysis~\cite{EXOT-2018-43} selected events with two oppositely charged leptons (electron or muon) collected by dilepton triggers. Further selections using different kinematic variables were used to define a single bin signal region.
The $\twmet$ 0-leptons and 1-lepton analysis~\cite{EXOT-2021-01} selected events with zero or one charged lepton (electron or muon) collected using a \met trigger, at least one $b$-jet, and large \met. The selection required a large-$R$ jet consistent with the hadronic decay of a \Wboson boson.

\paragraph*{\Monojet}
\label{par:monojet}
The \monojet\ analysis~\cite{EXOT-2018-06} is characterised by the presence of an energetic jet and large \met.
Events were collected using an \met trigger and vetoed if any charged lepton or photon was reconstructed.
The dominant SM background for this search arises from the irreducible process \(\Zboson\rightarrow\nu\nu\) or \(\Wboson\rightarrow\ell\nu\) in association with jets, where the \Wboson boson decays into either hadronically decaying $\tau$-leptons or undetected electrons or muons. Additional contributions include top-quark-pair or single-top-quark production, diboson production, and non-collision and multijet backgrounds.
The estimate of the major SM processes in the analysis was based on a profile likelihood fit to the distribution of the \pT of the system recoiling against the jets reconstructed in the event, performed simultaneously in the signal region and in orthogonal control regions enriched with the targeted backgrounds.

\paragraph*{\monophoton}
\label{par:monophoton}
The \monophoton analysis~\cite{EXOT-2018-63} selected events with a $\ET> 150$~\GeV photon and no leptons in the final state. The leading photon was required not to overlap with the \ptmissvec by requiring $\Delta\phi(\gamma,\ptmissvec)>0.4$. Events were rejected if they contained more than one jet ($\pt > 30$~\GeV) or a jet fulfilling $\Delta\phi(\mathrm{jet},\ptmissvec)<0.4$. Four exclusive signal regions were defined with \MET ranges from 200~\GeV and above. To reduce the background from events with poorly reconstructed physics objects, \MET significance $>$ 8.5 was requested~\cite{ATLAS-CONF-2018-038}.
The $\Wboson\gamma$, $\Zboson\gamma$, and $\gamma$+jets backgrounds were normalised in their control regions using a simultaneous likelihood fit of all \MET regions, but with independent normalisation factors for each region. The backgrounds due to photons from the misidentification of electrons or jets in processes such as \Wjets, \Zjets, diboson, and multijet events were estimated by using data-driven techniques.

\paragraph*{\monoZll}
\label{par:monoZll}
The \monoZll analysis~\cite{HIGG-2018-26} selected events by requiring significant \MET and a pair of high-\pt leptons. Two opposite-sign, same-flavour leptons satisfying $\pt > 30$~\GeV and $\pt > 20~\GeV$ was required, with an invariant mass in the range $76~\GeV<\mll <106~\GeV$, as a proxy of the \Zboson boson mass. No additional leptons with $\pt > 7~\GeV$ nor \bjets with $\pt > 20~\GeV$ were allowed in the events. To target events consistent with a boosted \Zboson boson produced in the direction opposite to \ptmissvec, additional requirements were imposed on the azimuthal angle between the dilepton system and \ptmissvec and on the angular distance between the leptons. A single inclusive \MET signal region was defined with \MET$> 90~\GeV$ for each of the $ee$ and $\mu\mu$ channels. The dominant background, $ZZ$ production, was estimated from simulated events. The $WZ$ background was normalised to data in a three-lepton control region. The contributions from \Zjets and non-resonant-$\ell\ell$ backgrounds were estimated by using data-driven techniques. A statistical combination of the $ee$ and $\mu\mu$ channels was performed to obtain the final results.

\paragraph*{\monoVhad}
\label{par:monoVhad}
The \monoVhad analysis~\cite{EXOT-2019-27} selects events with $\MET>200~\GeV$ and a hadronically decaying \Wboson or \Zboson boson candidate from a data sample corresponding to 140~\ifb. The vector-boson candidate was defined with one large-$R$ jet with $\pt> 200~\GeV$ in a boosted topology ($\MET > 250~\GeV$) or with two small-$R$ jets with the leading (subleading) one having $\pt>45~\GeV$ ($\pt>20~\GeV$) in a resolved topology. In both cases, a lepton veto was applied. Additional requirements were applied to the invariant mass of the boson candidate and on the significance of the \MET with respect to the hadronic activity in the events~\cite{ATLAS-CONF-2018-038}. The boosted topology region was further split according to the high/low purity of the \Wboson/\Zboson tagger. Several control regions were defined according to lepton and \bjet multiplicity. The normalisation on the \ttbar and $\Wboson/\Zboson$+jets background processes were constrained using a simultaneous fit of the \MET distribution in all control and signal regions. The subdominant contribution from diboson production and other electroweak backgrounds were obtained from simulation. Multijet contributions were estimated with a data-driven technique.

\subsection{Searches for visible final states}
\label{ssec:visible_sig}

When the \schannel mediator decays into SM particles, analyses can target its visible objects: hadronic jets or leptons. By reconstructing such objects in the final state, it is possible to reconstruct the kinematic properties of the originating mediator candidate such as its mass \mMed. This suite of analyses is therefore characterised by the search for resonant pair of objects over a smooth background of non-resonant production of hadrons and leptons in \pp collisions.

\begin{table}
\caption{Summary of searches for visible final states used to constrain the different DM models defined in \cref{sec:models}. See the text for symbol definitions.
}
\label{tab:visible}
\centering
\resizebox{\textwidth}{!}{
\begin{tabular}{m{40mm}m{30mm}m{55mm}m{50mm}}
\toprule
Analysis                        		& Models targeted & Final state signature & Key Characteristics \\
\midrule
Dijet~\cite{EXOT-2019-03}                        & V/AV       & 2 jets, \mjj, $y^*$                                         & Sliding-window fit of \mjj \\[0.5ex]
Dijet angular~\cite{EXOT-2016-21}                & V/AV       & 2 jets, \mjj, $y^*$                                         & Binned likelihood fit of $\chi_{jj}$ \\[0.5ex]
Dijet ISR resolved~\cite{tmp_EXOT-2019-01}       & V/AV       & 2 jets, $\gamma$, \mjj, $y^*$                               & Sliding-window fit of \mjj \\[0.5ex]
Dijet ISR boosted~\cite{EXOT-2017-01}            & V/AV       & 1 small-$R$ jet, 1 large-$R$ jet, \mjj, $y^*$               & Data-driven extrapolation from control region via transfer factor \\[0.5ex]
Dijet TLA~\cite{EXOT-2016-20}                    & V/AV       & 2 trigger-level jets, \mjj, $y^*$                           & Sliding-window fit of \mjj \\[0.5ex]
Dijet + lepton~\cite{EXOT-2018-32}               & V/AV       & 2 jets, $\ell$, \mjj                                        & Fit of \mjj  \\[0.5ex]
Dilepton~\cite{EXOT-2018-08}                     & V/AV       & 2 e or 2 $\mu$                                              & $\Zboson/\gamma^*\to\ell\ell$ from fit of \mll \\[0.5ex]
\ttbar ~\cite{EXOT-2015-04,EXOT-2018-48}         & V/AV, S/PS & $\ell$+jets; 2 large-$R$ jets                               & Binned likelihood fit of $m_{\ttbar}$  \\[0.5ex]
\fourtops~\cite{EXOT-2019-26}                    & S/PS       & Same-sign $\ell^\pm\ell^\pm$ and $\ell^\pm\ell^\pm\ell^\mp$ & Binned likelihood fit of BDT \\[0.5ex]
\bottomrule
\end{tabular}
}
\end{table}

\paragraph*{Dijet and Di-b-jet}
\label{par:dijet}

The dijet search \cite{EXOT-2019-03} scrutinised events with at least two small-$R$ jets with the leading jet satisfying the requirement $\pt > 440~\GeV$ and the subleading jet $\pt > 60~\GeV$. The dijet selection required a rapidity difference between the two jets $|y^*|=|y_1-y_2|$ being $<0.6$ and the invariant mass of the dijet system to be $\mjj > 1.1~\TeV$. Events were categorised based on the presence of zero, one, two or more \bjets, to further optimise the sensitivity towards specific scenarios of the considered suite of models. The background estimate was obtained by fitting the falling \mjj distribution. A sliding-window fitting technique was used, where restricted regions of the spectrum were fitted with a functional form. The values from the full set of windows were then combined to create the background estimate for the full mass range. Model-independent limits on the visible cross-section for a Gaussian-shaped signal in the \mjj spectrum were extracted for different signal width hypotheses. This analysis observed the highest-\mjj event around 8.2$~\TeV$.

\paragraph*{Dijet angular}
\label{par:dijet_angular}

A dijet selection can also be exploited to search for deviations from the SM expectation in angular distributions, characteristic of wide resonances where the nominal dijet search would lose sensitivity. A dijet angular analysis~\cite{EXOT-2016-21} was performed on events with two jets following the \pt requirements of the dijet search, but relaxing the $|y^*|$ requirement to be below 1.7. The data collected by ATLAS during the first part of Run 2 was considered for this search, corresponding to 37~\ifb. Due to different kinematics in this loosened selection, the mass of the dijet pair was required to be \mjj $> 2.5~\TeV$. The analysis makes use of the variable $\chi_{jj}=e^{2|y^*|}\sim (1+\cos\theta^*)/(1-\cos\theta^*)$,\footnote{$\theta^*$ is defined as the polar angle relative to the direction of the initial partons in the dijet centre-of-mass frame.} constructed in such a way that, in the limit of massless parton scattering and when only the $t$-channel scattering contributes to the partonic cross-section, the angular distribution $dN/d\chi_{jj}$ is approximately independent of $\chi_{jj}$. Simulated events from multijet production were modelled at leading-order in QCD, and reweighted to next-to-leading-order predictions from NLOJET++~\cite{Nagy:2001fj,Nagy:2003tz} using mass- and angle-dependent correction factors. Additional electroweak mass- and angle-dependent correction factors were applied. The data were compared with a SM template in different $m_{jj}$ ranges and different $\chi_{jj}$ bins.

\paragraph*{Dijet TLA}
\label{par:dijet_tla}

The aforementioned dijet search was limited by the high-\pt requirement imposed on the leading jet due to the limited bandwidth available for the single-jet trigger algorithms. This limitation was overcome by recording only HLT jet information, rather than the full detector readout, which significantly reduced the average event size and bandwidth usage. This allowed a higher rate of events to be stored, including all events satisfying the single-jet L1 trigger, with a lower \pt threshold relative to the dijet search.
The dijet trigger-level analysis (TLA) \cite{EXOT-2016-20} considered events from data corresponding to 29.3~\ifb, required to have at least two trigger-level jets with $\pt>185~\GeV$. Two selection criteria were imposed: $|y^*|<0.6$ in the mass range $700~\GeV<\mjj<1.8~\TeV$ and $|y^*|<0.3$ for $450~\GeV<\mjj<700~\GeV$. The leading trigger-level jet was required to have $\pt>185~\GeV$ and $\pt>220~\GeV$ for the $|y^*|<0.3$ and $|y^*|<0.6$ selections, respectively, to ensure full efficiency of the L1 triggers. The search was then interpreted in terms of resonances with masses between 450 $\GeV$ and 1.8 $\TeV$. The background contribution was estimated with the same strategy as the previously described dijet search.

\paragraph*{Resolved dijet ISR}
\label{par:dijet_isr_res}

Another approach for resonance searches in the low-mass region is to select events with a pair of jets recoiling against a high-\pt object, such as a photon or an additional jet from ISR.
A recent analysis~\cite{tmp_EXOT-2019-01} considered the full Run 2 data sample, corresponding to \MaxLumi of data, to search for dijet resonances associated with either an ISR photon or an ISR jet. In the ISR photon case events were selected with a single-photon trigger algorithm, which was fully efficient for photons with $\pt > 150~\GeV$. In the ISR jet case events were selected with a single-jet trigger algorithm, which was fully efficient for events with a jet with $\pt > 475~\GeV$. Different event categories were defined based on whether the two jets from the resonance were tagged as \bjets or not. The dijet invariant mass spectrum was scrutinized to constrain the presence of BSM resonances in the range from 150~\GeV to 850~\GeV.

\paragraph*{Boosted dijet ISR}
\label{par:dijet_isr_boost}

In the case of a dijet+ISR selection, if the associated ISR photon or jet has large transverse momentum, the dijet resonance candidate was reconstructed as a large-$R$ jet~\cite{EXOT-2017-01} of radius 1.0 with mass $m$. To enhance the sensitivity to quark pair decays, jet substructure techniques were used to discriminate between a two-particle jet from a decay of a boosted resonance and a single-particle jet~\cite{Thaler:2010tr}. Events were required to have a large-$R$ jet, a resonance candidate, and at least one ISR object candidate. The azimuthal angular separation between the resonance candidate and the ISR object was required to satisfy $\Delta\phi > \pi/2$. A $\pt > 2m$ requirement ensures sufficient collimation of the resonance candidate. In the ISR jet (photon) channel, the large-$R$ jet satisfied $\pt > 450 (200)~\GeV$ and the ISR jet (photon) had $\pt > 450 (155)~\GeV$. A data-driven technique was used to estimate the expected background in the signal region via a transfer factor that extrapolates from a control region with inverted jet substructure requirements to the signal region. Data corresponding to an integrated luminosity of 36.1~\ifb were scrutinized to search for resonant BSM signals in the range from 100~\GeV to 220~\GeV.

\paragraph*{Dijet + lepton}\label{par:dijet_lep}

A search for hadronic resonances was carried out in events with an associated charged lepton~\cite{EXOT-2018-32}.
Events considered were required to have a single high-\pt charged lepton (either $e$ or $\mu$) that satisfied the single lepton trigger selection with $\pt > 60~\GeV$. At least two jets with $\pt > 20~\GeV$ were required to be reconstructed in the final state, and the two leading ones were considered as coming from the mediator decay, and hence were used to reconstruct its mass. A minimum requirement of $\mjj > 0.22~\TeV$ was applied to guarantee that the event rate of the background contribution due to uncorrelated non-resonant dijet production was monotonically decreasing and can therefore be modelled directly from data by performing a fit to the smoothly falling \mjj spectrum.

\paragraph*{Dilepton}\label{par:dilepton}

The dilepton analysis~\cite{EXOT-2018-08} selected events with at least two same-flavour leptons. The pair of electrons or muons with the highest \pt was chosen as the resonance candidate. Only the muon channel candidates were required to have opposite charge, due to higher charge misidentification for high-\pt electrons and the \pt misreconstruction associated with wrongly measured charge in muons. Background processes with two prompt leptons were modelled by fitting the dilepton mass distribution to data with a smooth functional form tested on simulated background templates. The analysis explored the dielectron and dimuon invariant mass spectra in the range $250~\GeV <\mll< 6~\TeV$.

\paragraph*{\ttbar~resonance}
\label{par:ttbar}

Two \ttbar resonance analyses were carried out by selecting events with two top-quark candidates, one in a final state including a lepton and hadronic jets~\cite{EXOT-2015-04}, and the other one considering a fully hadronic final state~\cite{EXOT-2018-48}.
The semileptonic search considered 36~\ifb of collected data, while the fully hadronic search considered the entire data sample collected during Run 2 of the LHC.

The semileptonic signature was characterised by events including a charged lepton and \MET consistent with a leptonic decay of a \Wboson boson, and a small-$R$ jet nearby. Events were classified as boosted or resolved depending on their hadronic activity. In the boosted selection, events contained one large-$R$ jet satisfying top-tagging requirements~\cite{JETM-2018-03}, while in the resolved selection events had at least four small-$R$ jets and failed to satisfy the boosted selection. The \ttbar invariant mass \mtt was reconstructed from the decay products of the two top-quark candidates in the events. The \bjet multiplicity was used for further event categorisation. The SM \ttbar production was estimated by using simulated samples and fixed-order theory calculations. The multijet and \Wjets background contributions were estimated by using data-driven techniques.

The fully hadronic signature was characterised by events with at least two large-$R$ jets having $\pt > 500~\GeV$ and $\pt > 350~\GeV$. Leptons were vetoed to maintain orthogonality to the semileptonic \ttbar search. The invariant mass of the two large-$R$ jets, \mJJ, was considered as a proxy of the mediator mass, and was scrutinised for events having $\mJJ > 1.4~\TeV$. Such jets were required to be top-tagged~\cite{JETM-2018-03} and were further categorised based on the number of small-$R$ jets tagged as \bjets. The dominant background originates from SM production of top-quark pairs and multijet events and was estimated from data by performing a fit to the smoothly falling \mJJ spectrum.

\paragraph*{4-top}
\label{par:fourtops}

The 4-top analysis ~\cite{EXOT-2019-26} targeted a final state with exactly two leptons with same-sign electric charges or at least three leptons. Multivariate analysis techniques were used to separate the signal from the SM backgrounds.
This was done through two sequential classifiers based on BDTs: the first one, called \textit{SM BDT}, separates SM $\fourtops$ events from other SM backgrounds. The second BDT, called \textit{BSM pBDT}, discriminates between BSM $\fourtops$ events and all background contributions. The \textit{BSM pBDT} was parameterised as a function of the mass of the heavy Higgs boson by introducing the mass as a labelled input in the training~\cite{Baldi:2016fzo}.
The reducible background contribution due to electron charge misidentified events was estimated by using a data-driven technique, while all other signal and background processes were modelled using simulated samples.
%
%


%
%
%
%

\section{Reinterpretation of experimental results to arbitrary couplings}
\label{sec:rescaling}

Spin-1 related results are usually expressed as limits on BSM signals by assigning a fixed value to most of the model parameters, such as the mediator couplings.
Traditionally, assessing the sensitivity to different model's parameter choices would require each of the DM-related analyses to generate multiple variations of the benchmark BSM signal for each parameter combination to be probed.
This is a time and resource-consuming process.
Instead, the semi-analytical technique described in Ref.~\cite{coupling_rescaling_tool} is exploited to rescale the published results and provide results corresponding to different parameter choices.

The semi-analytical rescaling procedure starts from an analysis constraint on a given DM model signal strength, assuming a given choice of the mediator coupling values, \mDM and \mMed, and evaluates a scale factor to derive the corresponding constraints on the signal strength for different mediator coupling values. This rescaling always assumes \mMed is unchanged, while \mDM can vary together with the mediator couplings. The procedure is different when considering a fully visible or a semi-visible final state, but both cases are handled in a coherent scenario.
The rescaling procedure was validated by comparing the rescaled limits to those set by the original analysis, covering the cases with the largest considered mediator widths and between axial-vector and vector Lorentz structures.
Checks of the mediator width in the considered phase space were carried out to verify that the underlying assumptions of the rescaling procedure are valid.

The resonant final state searches considered in \thisdoc typically set upper limits on the cross-section of generic Gaussian-shaped signals with different relative widths, as a function of the central value of such signals. The range of considered widths was constrained by the background estimate technique: the estimate obtained from a fit to data may be biased by wide resonance signals.
Limits on Gaussian-shaped signals should be translated in the context of \schannel model parameters.
To do so, simulated signals are smeared by the experimental di-fermion mass resolution and are considered for different choices of DM model parameters. The resonance width is evaluated and the corresponding limit on that signal is obtained comparing the limit on the Gaussian signal with a similar width and the theoretically predicted cross-section.

The procedure described in Ref.~\cite{coupling_rescaling_tool} is based on the semi-analytical rescaling technique that parametrises few relevant observables. This approach considers variations of the BSM signal cross-section, the intrinsic width of the mediator, and the branching fraction of its decay into a specific final state, while assuming the mediator mass unchanged. A change of the intrinsic width affects the analysis reach only when this becomes comparable with the experimentally reconstructed mediator mass resolution, hence affecting searches for leptonic resonances more than hadronic ones.

For dilepton resonance searches the shape of the DM signal is not easily reproduced by a Gaussian and SM-BSM interference effects produce a lower tail relative to the core of the signal peak~\cite{dmwg_recs}. This contribution was studied in detail and the largest effect observed is of the order of 5\% when considering $d\sigma/dm_{\ell\ell}$ in a mass window within a distance of five times the signal width ($\Gamma$) distance from the signal peak.
For this reason fiducial cuts of $m_{\ell\ell}>\mMed - 2\Gamma$ are applied when the DM signal and its corresponding fiducial cross-section in leptonic final states are considered.

The rescaling starting point considered for dijet and dilepton resonances are of two different kinds: 95\% confidence level (CL) upper limits on either the cross-section of a new Gaussian BSM signal ($\sigma_{BSM}$) as a function of \mMed (assuming a choice of the other coupling parameters), or on the mediator's coupling values to SM particles, $\gq$ or $\gl$ for hadronic and leptonic searches, respectively. In both cases, limits as a function of \mMed are rescaled to determine corresponding exclusion depths in the (\mDM, \mMed) plane, for arbitrary choices of DM model couplings. For resonant and visible final states, a good knowledge of the cross-section for each process is available, given the on-shell decays of the mediator to SM particles in the \mMed range considered.

When considering $X+\MET$ final states, the mediator decays into undetected DM particles, and hence the off-shell mediator decay becomes more relevant. Following the rescaling procedure described in Ref.~\cite{coupling_rescaling_tool}, the transition between on-shell and off-shell regimes is handled by integrating the full Breit--Wigner propagator term over the allowed phase space~\cite{Abercrombie_2019}.
The rescaling of $X+\MET$ searches usually starts from exclusion depths (95\% CL upper limit on the signal strength $\mu$) on a grid of (\mDM, \mMed) points. In this case, rescaling factors are evaluated for each (\mDM, \mMed) combination and interpolation is applied between grid points.

The rescaling of limits to different mediator structures considers the parton-level cross-sections as functions of the partonic centre-of-mass energy $\hat{s}$, together with the effect of the parton distribution functions. Ratios of such parton-level cross-sections can be used to rescale results from axial to axial-vector assumptions and vice versa.

\section{Comparison with direct detection experiments}
\label{sec:dd_reinterpret}

Limits on the DM model signal strength as a function of the model parameters can be translated into limits on the spin-dependent (spin-independent) $\chi-$proton and $\chi-$neutron ($\chi-$nucleon) scattering cross-sections as a function of \mDM, following the procedure described in Ref.~\cite{Boveia:2016mrp}. This enables a comparison of the results obtained in \thisdoc to the limits set by direct detection experiments.

For the spin-independent (SI) case, the scattering cross-section for the \schannel simplified model considered can be written as a function of the model parameters as:

\begin{equation}
\sigma_{SI} = \frac{f^2(\gq)\gDM^2\mu^2_{n\chi}}{\pi \mMed^4},
\end{equation}

where $\mu_{n\chi} = m_n \mDM/(m_n+\mDM)$ is the $\chi-$nucleon reduced mass. The $f(\gq)$ term is the mediator-nucleon coupling and depends on the mediator-quark coupling.
This is equal to $3\gq$ when considering a vector mediator while for a scalar mediator, assuming it couples to all quarks, is evaluated to be $f(\gq)=1.16\cdot10^{-3}\gq$~\cite{Boveia:2016mrp}.

For the spin-dependent (SD) case, the scattering cross-section can be written as:

\begin{equation}
\sigma_{SD} = \frac{3f^2(\gq)\gDM^2\mu^2_{n\chi}}{\pi \mMed^4},
\end{equation}

where in this case $f_{p,n}(\gq)$ differs between protons and neutrons but, assuming that the coupling \gq is equal for all quarks, it can be set to $0.32\gq$ when considering the axial-vector mediator scenario.

When considering pseudo-scalar mediators, the rate of direct detection experiments is suppressed due to the introduction of supplementary velocity-dependent terms in the cross-section. Consequently, direct detection experiments exhibit limited sensitivity in this scenario, rendering it impractical to show LHC results with standard constraints on scalar and spin-dependent cross-sections. Instead, it proves more meaningful to evaluate LHC bounds against those established by indirect detection experiments.

These limits are expressed in terms of the cross-section $\langle\sigma\upsilon_{\text{rel}}\rangle_{(q,g)}$~\cite{Boveia:2016mrp} for the annihilation into a $q\bar{q}$ final state:

\begin{equation}
\langle\sigma\upsilon_{\text{rel}}\rangle_q=\frac{3m^2_q}{2\pi\upsilon^2}\frac{\gq^2\gDM^2\mDM^2}{\left(\mMed^2-4\mDM^2\right)^2+\mMed^2\Gamma^2_{\mathrm{Med}}}\sqrt{1-\frac{\mq^2}{\mMed^2}},
\label{eq:sigmavq}
\end{equation}

or into a pair of gluons:

\begin{equation}
\langle\sigma\upsilon_{\text{rel}}\rangle_g=\frac{\alpha_s^2}{2\pi^3\upsilon^2}\frac{\gq^2\gDM^2}{\left(\mMed^2-4\mDM^2\right)^2+\mMed^2\Gamma^2_{\mathrm{Med}}}\cdot\left|\sum_q\mq^2f_{PS}\left(\frac{\mq^2}{\mDM^2}\right)\right|^2,
\label{eq:sigmavg}
\end{equation}

where $f_{PS}(\tau)=\arctan^2\left(\frac{1}{\sqrt{\tau-1}}\right)$ and $\alpha_s$ is the strong coupling constant. The total annihilation cross-section is given by the sum of the quark and gluon contributions calculated using Eqs.~(\ref{eq:sigmavq}) and (\ref{eq:sigmavg}).


%
%
%
%

\section{Results}
\label{sec:result}

%
%
%
%
%

\subsection{Spin-0}
\label{sec:result_spin0}
The most stringent limits on scalar and pseudo-scalar models considered are obtained from the \ttmet final state combination.
This combination exploits all top-quark-pair channels: fully hadronic, single lepton, and dilepton final states.
The results exclude at $95\%$ CL scalar (pseudo-scalar) mediators with unitary couplings $g_q=g_\chi=g=1$ up to mediator masses of 400~\GeV, assuming a specific $m_\chi$ of $1~\GeV$.
The strongest upper limit of the ratio of the signal production cross-section to the nominal cross-section ($\sigma_\text{Obs}/\sigma_\text{Th}(g=1)$), referred to as \textit{signal strength}, is obtained at the lowest mediator mass considered, as shown in~\cref{fig:scalarTTTW,fig:pseudoTTTW}.
The limits are derived considering both contributions from \ttdm and \singletopdm models.
The limits of the three \monotop analyses are similar, extending between $0.4$ and $1.1$ on the cross-section ratio for both the scalar and the pseudo-scalar mediator masses below $200~\GeV$.

For mediator masses greater than $350~\GeV$, the decay channel of the mediator into a pair of top quarks is open.
In the high mediator mass region, the limit produced by the \ttmet analysis is weaker while the limit from the \fourtops analysis becomes the dominant.
Models with mediators produced through loop-induced gluon fusion are constrained by the \monojet analysis.
Pseudo-scalar mediator models with unitary coupling are excluded up to $402~\GeV$ by the \monojet analysis.
For scalar mediators, the limits are weaker due to the lower \monojet cross-section.
Finally, limits from \bbmet final states also constrain the \bbdm simplified models.
The \bbmet search set upper limits on the signal strength between $5$ and $10$, specifically for mediator masses below $20~\GeV$, for both the scalar and the pseudo-scalar mediators. These findings provide a quantitative measure of the sensitivity of these models when up-type couplings are suppressed.

\begin{figure}[!tb]
\centering
\subfigure[]{\includegraphics[width=0.7\textwidth]{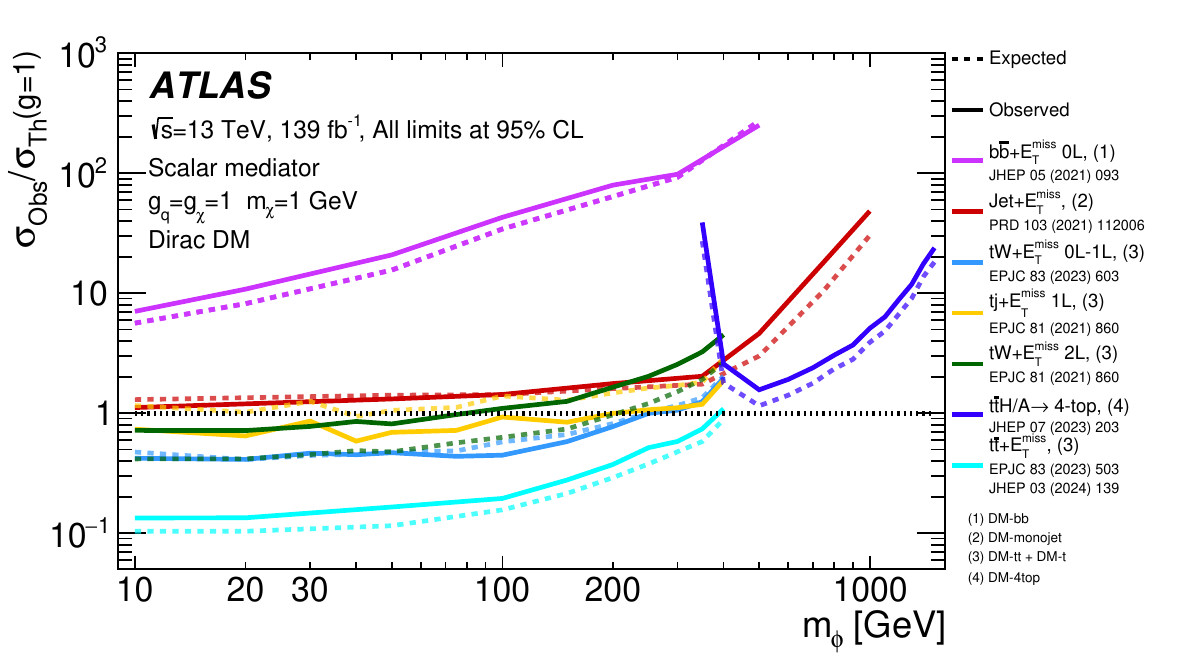}\label{fig:scalarTTTW}}
\subfigure[]{\includegraphics[width=0.7\textwidth]{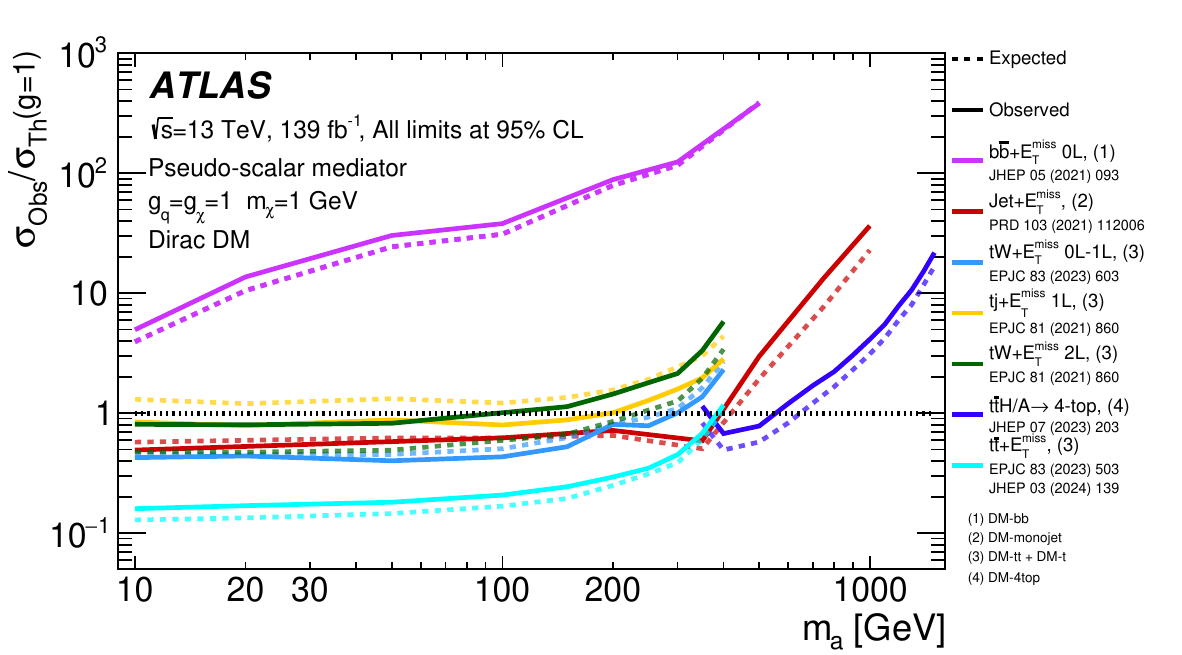}\label{fig:pseudoTTTW}}
\caption{Exclusion limits for colour-neutral \subref{fig:scalarTTTW} scalar and \subref{fig:pseudoTTTW} pseudo-scalar mediator dark matter models as a function of the mediator mass $m_\phi (m_a)$ for a dark matter mass $m_\chi$ of 1~\GeV. The limits are calculated at 95\% CL and are expressed in terms of the ratio of the excluded cross-section to the nominal cross-section for a coupling assumption of $g_q=g_\chi=g=1$. The solid (dashed) lines show the observed (expected) exclusion limits for different analyses. The \monojet constraint is strengthened around $m_a=350$~GeV due to an enhancement in the signal cross-section.
}
\end{figure}

\subsubsection{Comparison with direct detection}
The exclusion limit on the production cross-section of colour-neutral scalar mediator particles can be converted into a limit on the spin-independent DM--nucleon scattering cross-section using the procedure described in~\cref{sec:dd_reinterpret}.
The derivation of the limits is based on the assumption of constant acceptance as a function of \mchi in the $m_{\phi/a}<2\mDM$ phase-space~\cite{Abercrombie_2019}.
\cref{fig:XsecDMscalarTT} shows the resulting constraints\footnote{90\% CL limits are reported to be consistent with the convention used to report results from direct detection experiments.} in the plane defined by the dark-matter mass and the scattering cross-section, which are derived from the $\ttmet$ analysis considering only the contribution from the \ttdm model~\cite{SUSY-2023-22}.
The most stringent direct detection limits to date from DarkSide-50 QF MIGD~\cite{Agnes_2018}, PandaX-4T~\cite{Panda2023,Meng_2021}, and LZ~\cite{LZ:2022lsv} are overlaid for comparison. %
\cref{fig:XsecDMpseudoTT} shows the translation of the pseudo-scalar limit to the $\mchi$-$\langle\sigma v_\text{rel}\rangle$ plane considering the $\ttmet$ analysis.
The limit is calculated using the contribution derived from Eqs.~(\ref{eq:sigmavq}) and (\ref{eq:sigmavg}) assuming $m_a=384~\GeV$ and \mchi between $1~\GeV$ and $m_a/2$.
The limits from the gamma-ray telescopes Fermi-LAT~\cite{Ackermann:2015zua} are overlaid for comparison.

\begin{figure}[!tb]
\centering
\includegraphics[width=0.6\linewidth]{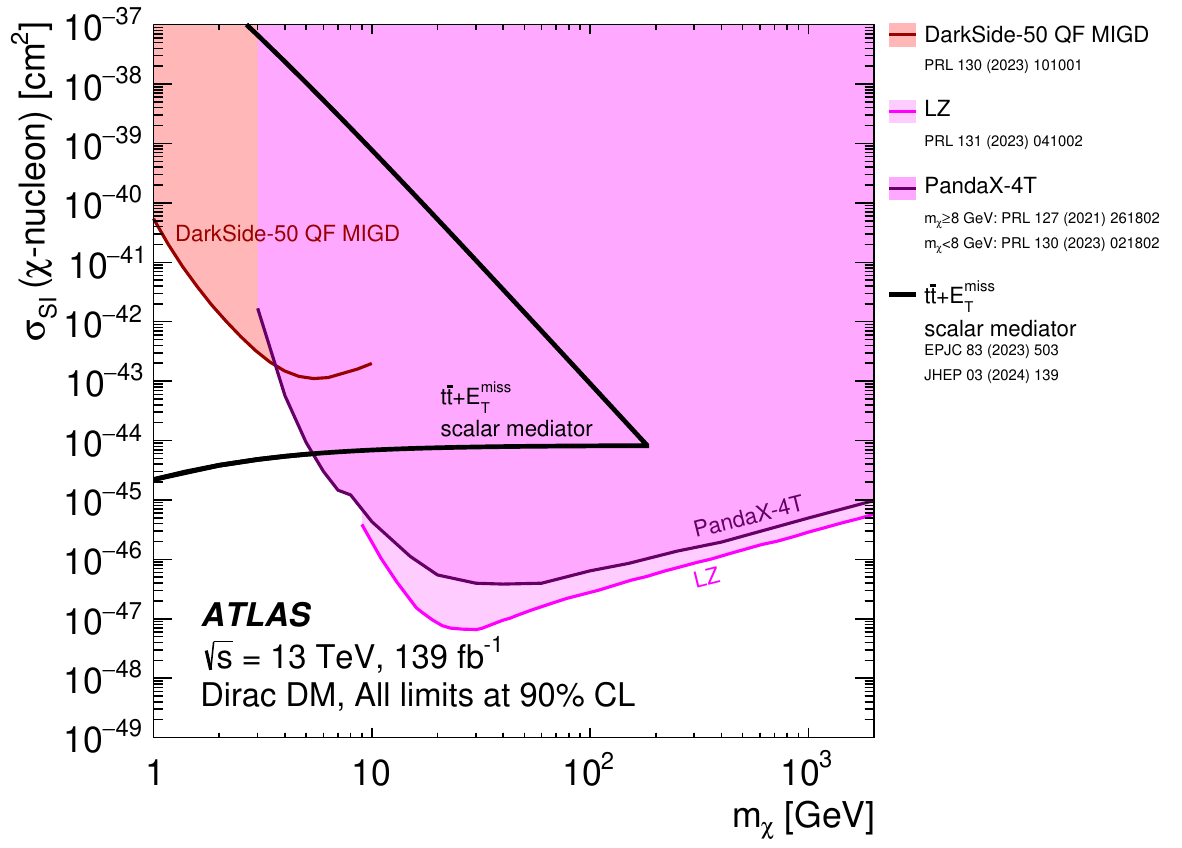}
\caption{
\label{fig:XsecDMscalarTT}
Comparison of the 90\% CL limits on the spin-independent DM--nucleon cross-section as a function of the DM mass between these results and the direct detection experiments, in the context of the colour-neutral simplified model with a scalar mediator.
The lower horizontal line of the DM--nucleon scattering cross-section for the \ttmet scalar mediator contour corresponds to the value of the cross-section for $m_\phi=366~\GeV$.
The results are compared with limits from direct detection experiments from DarkSide-50 QF MIGD~\cite{Agnes_2018}, PandaX-4T~\cite{Panda2023,Meng_2021}, and LZ~\cite{LZ:2022lsv}. %
}
\end{figure}

\begin{figure}[!tb]
\centering
\includegraphics[width=0.6\linewidth]{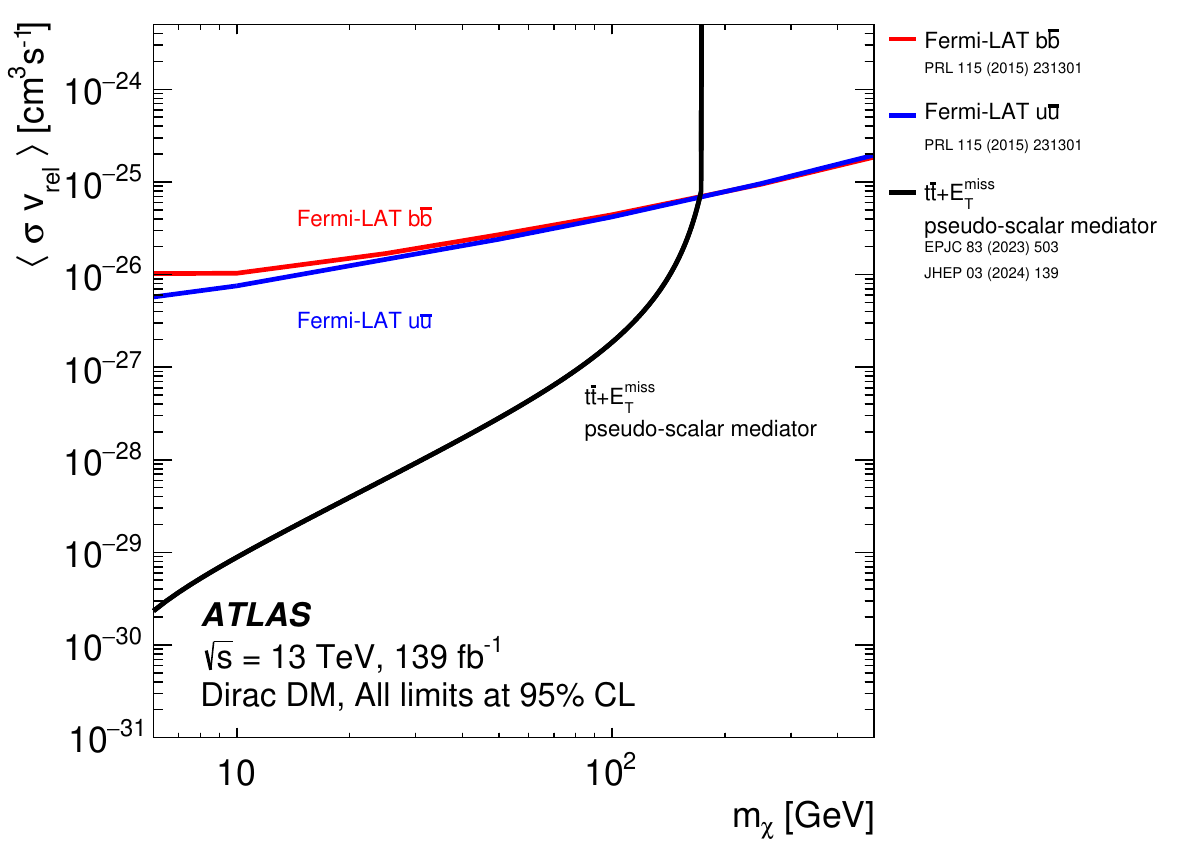}
\caption{
\label{fig:XsecDMpseudoTT}
Inferred 95\% CL limits on the WIMP annihilation rate as a function of the DM mass, for the pseudo-scalar mediator model. The annihilation rate is defined as the product of cross-section $\sigma$ and relative velocity $v_\text{rel}$, averaged over the DM velocity distribution ($\langle\sigma v_\text{rel}\rangle$). Both the \qqbar annihilation and the $gg$ fusion channels are considered in the calculation of the annihilation rate.
Results from gamma-ray telescopes~\cite{Ackermann:2015zua} are also shown.
The comparison is model-dependent and solely valid in the context of this model, assuming minimal mediator width and the coupling values $g_q=1$ and $g_\chi=1$.
}
\end{figure}

\FloatBarrier
%


\subsection{Spin-1}
\label{sec:result_spin1}

\subsubsection{Coupling limits versus mediator mass}

The hadronic and leptonic resonance searches were exploited to set limits on a spin-1 axial-vector mediator's coupling to quarks, \gq, and leptons, \gl.
The 95\% CL upper limits on the respective couplings are presented in \cref{fig:spin1-results-limitCouplings} as a function of the mediator mass $m_{Z_A^\prime}$.
The remaining parameters of the simplified model are fixed to specific values: in \cref{fig:spin1_gq} $m_\chi=10$~\TeV, $g_\ell=0$, and $g_\chi=1.0$ according to the A1 benchmark model, while in \cref{fig:spin1_gl} $m_\chi=10$~\TeV, $g_q=0.1$, and $g_\chi=1.0$ according to the A2 benchmark model.
The expected limits from each search are indicated by dotted lines while observed ones are indicated by solid lines.
In \cref{fig:spin1_gq} the limit from the TLA dijet analysis has two parts, employing different data sets with different selections in the rapidity difference $y^*$, as indicated. The limit from the dijet+ISR($\gamma$) analysis also has two parts, each using a different trigger strategy, and each further studied in inclusive and b-tagged channels. Two lines are also shown for the di-\bjet search, resulting from two separate analyses: one used \bjet triggers and sets the limit in the low mediator mass region, while the other used inclusive jet triggers and sets the limit in the high mediator mass region.
The coupling values above the solid lines are excluded for signals narrow enough to be detected using each search.
The TLA dijet search with $|y^*|< 0.6$ was sensitive up to mediator relative width $\Gamma/m_{Z^\prime}$= 7\%.
The TLA dijet search with $|y^*|<0.3$ and the dijet+ISR searches were sensitive up to 10\% of the mediator relative width.
The dijet and di-\bjet searches were sensitive up to 15\% of the mediator relative width.
The dijet angular search was sensitive up to 50\% of the mediator relative width while no limitation in sensitivity arises from large width resonances in the \ttbar analysis. Benchmark width lines are indicated in the plot; the 50\% mediator relative width line lies beyond the canvas borders.

\begin{figure}[!tb]
\begin{center}
\subfigure[]{\label{fig:spin1_gq}\includegraphics[width=0.48\textwidth]{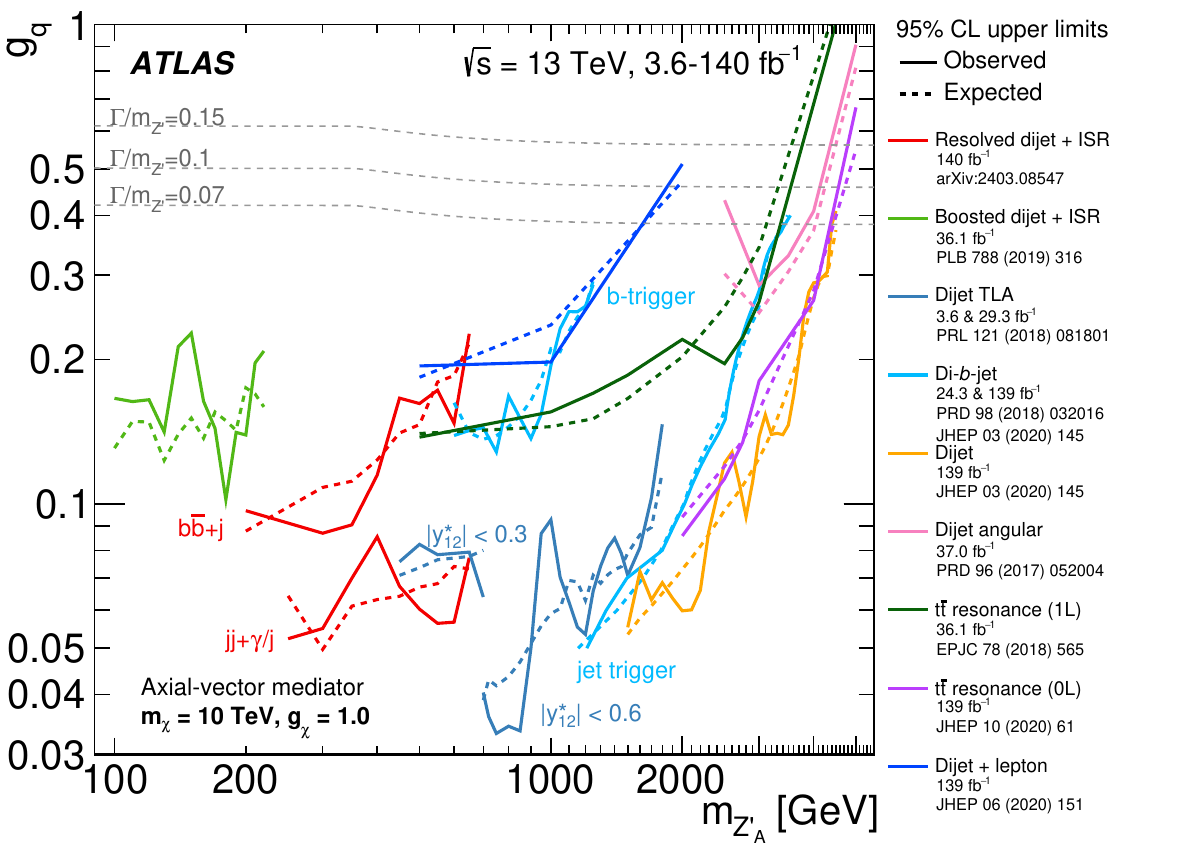}}
\subfigure[]{\label{fig:spin1_gl}\includegraphics[width=0.48\textwidth]{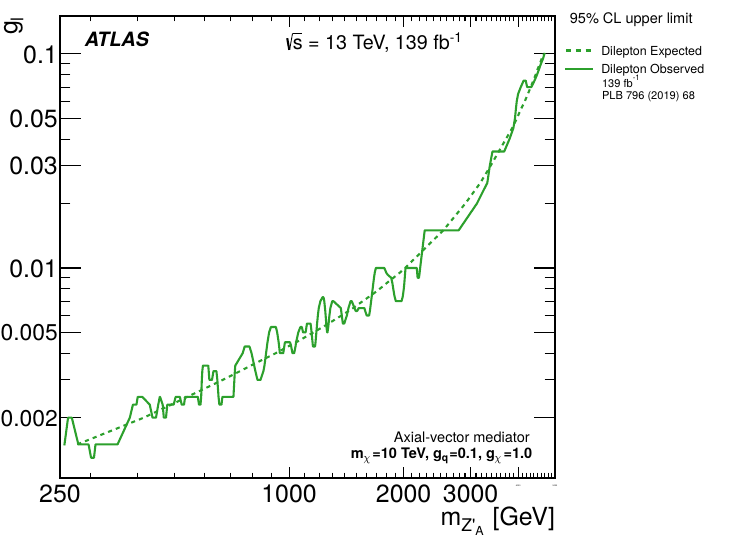}}
\end{center}
\caption{95\% CL upper limits on \subref{fig:spin1_gq} \gq and on \subref{fig:spin1_gl} \gl as a function of $m_{Z_A'}$. In these figures, \mDM is set to 10~\TeV to prevent the mediator from decaying into dark matter in the considered range of masses. The other parameters of the model are fixed to the values indicated in the figure, or equal to zero if not specified.}
\label{fig:spin1-results-limitCouplings}
\end{figure}

\subsubsection{Models constraints in the \mDM versus \mMed plane}
\label{sec:results_massmass}

Limits from both the resonant searches and the X+\MET searches are considered to set constraints on \schannel simplified models for a spin-1 mediator in the (\mDM, \mMed) plane. The sensitivity of each analysis depends on the specific choice of the mediator spin-parity and couplings. Using the rescaling technique described in~\cref{sec:rescaling}, the exclusion reach of the most sensitive searches is evaluated for different coupling hypotheses.
The exclusions from a set of analyses are grouped as the union of excluded points as indicated in the figure captions.
\cref{fig:spin1-results-coupling2} reports 95\% CL exclusion limits in the (\mDM, \mMed) plane for different choices of \gq for the hadronic signatures and \gl for the leptonic signatures.
From these figures it is possible to appreciate the dependence of the sensitivity of each analysis on the mediator coupling and how the excluded region changes accordingly.

\begin{figure}[!tb]
\centering
\subfigure[]{\includegraphics[width=0.49\textwidth]{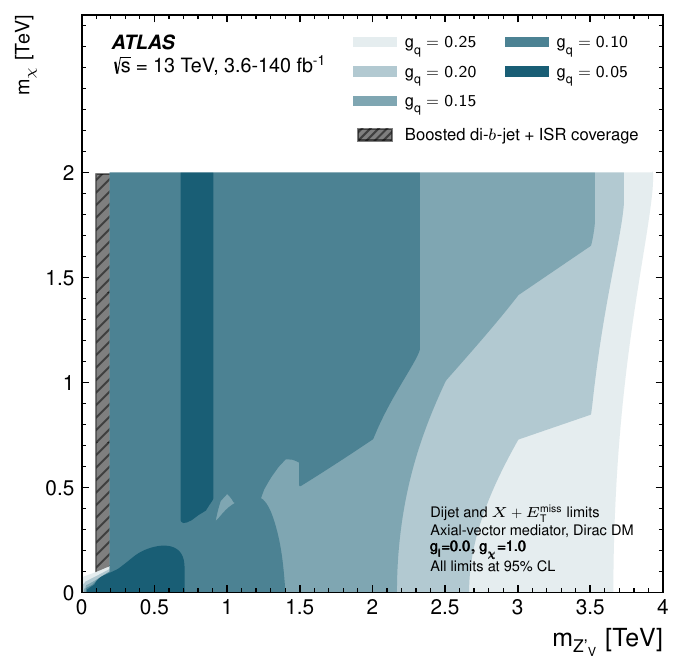}               \label{fig:dijet-multi-axial}}
\subfigure[]{\includegraphics[width=0.49\textwidth]{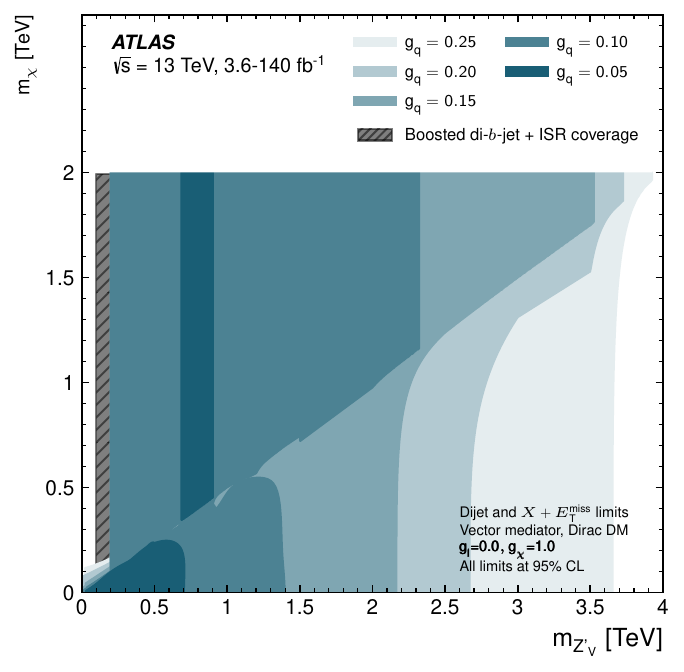}              \label{fig:dijet-multi-vector}}\\
\subfigure[]{\includegraphics[width=0.49\textwidth]{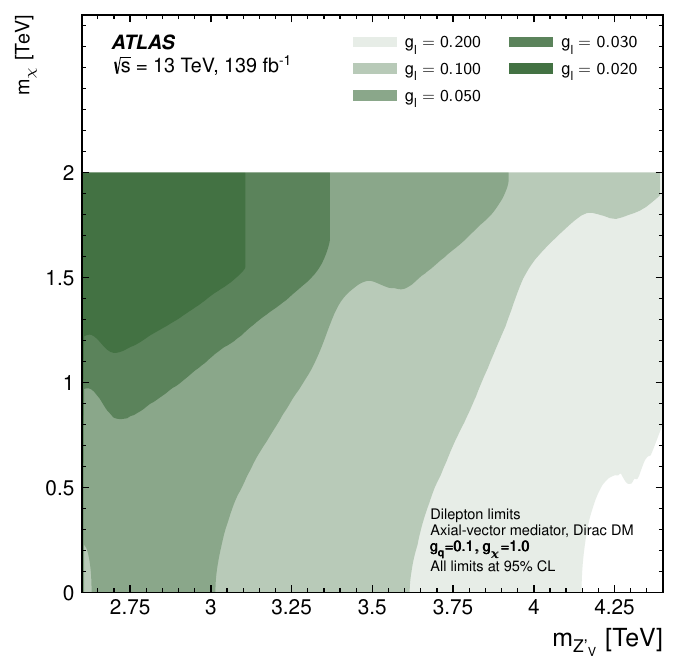}        \label{fig:dilepton_axial}}
\subfigure[]{\includegraphics[width=0.49\textwidth]{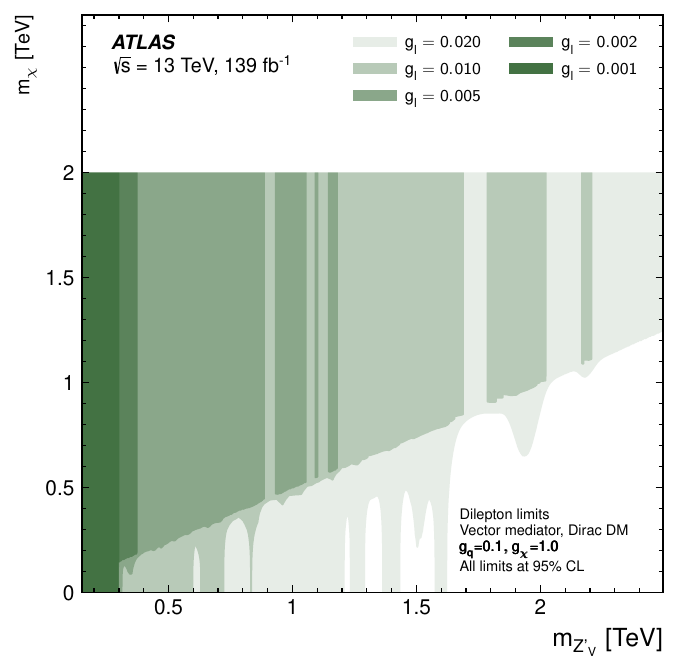}       \label{fig:dilepton_vector}}
\caption{Comparison of 95\% CL exclusion contours for signal hypotheses \subref{fig:dijet-multi-axial},\subref{fig:dijet-multi-vector} varying in $g_q$ for the dijet and X+\met analyses (dijet, dijet TLA, dijet ISR resolved, \monophoton, and \Monojet), and \subref{fig:dilepton_axial},\subref{fig:dilepton_vector} varying $g_\ell$ for the dilepton analysis, for Axial-Vector (left plots) and Vector (right plots) mediator hypotheses. The unions of multiple contours from the aforementioned analyses are displayed together.}
\label{fig:spin1-results-coupling2}
\end{figure}

As shown in \cref{tab:models}, eight benchmark combinations of model coupling choices are considered to highlight the complementarity of the different searches considered spanning the (\mDM, \mMed) plane.
\cref{fig:spin1-results-Overlay-most,fig:spin1-results-Overlay-mono} depict the different 95\% CL exclusion contours from the considered analyses for the four benchmarks considered for the axial-vector and vector mediators.
In these plots each shaded region shows the contribution to exclusion coming from a particular set of results: dijet resonances, \ttbar resonances, \bbbar resonances, dilepton resonances, and X+\met final states.\footnote{The rescaling procedure only considers the set of most sensitive signatures for which the needed inputs are available.}
The X+\met final state curves are interpolated by splines.
Dashed curves labelled ``thermal relic'' correspond to combinations of DM and mediator mass values that are consistent with a DM density of $\Omega h^2= 0.12$ and a standard thermal history, as computed in MadDM~\cite{dmwg_recs,maddm}.
Between the two curves, annihilation processes described by the simplified model deplete $\Omega h^2$ to values below 0.12. A dotted line indicates the kinematic threshold where the mediator can decay on-shell into DM. Regions that are in tension with perturbative unitarity considerations are shaded in the upper left corner.

Searches carried out in visible final states (e.g. dijet, dilepton) produced exclusions that are sensitive to the presence of a resonance of a given mass. These exclusions have a reduced dependency on the DM particle mass \mDM, resulting in almost vertical exclusion regions.
When the exclusions approach the end of the sensitivity of their corresponding search, the decay branching fraction of the mediator to DM particles, allowed when $\mDM < \mMed/2$, indirectly affects the branching fraction to SM particles.
This creates a significant reduction of the aforementioned sensitivity for low \mDM values.
Searches considering semi-visible final states ($X+\MET$) produced exclusions within the $\mDM < \mMed/2$ parameter space, with a modest extension into a slightly off-shell region.
These exclusions extend towards high \mMed values with minimal dependence on \mDM.

\begin{figure}[!tb]
\centering
\subfigure[]{\includegraphics[width=0.49\textwidth]{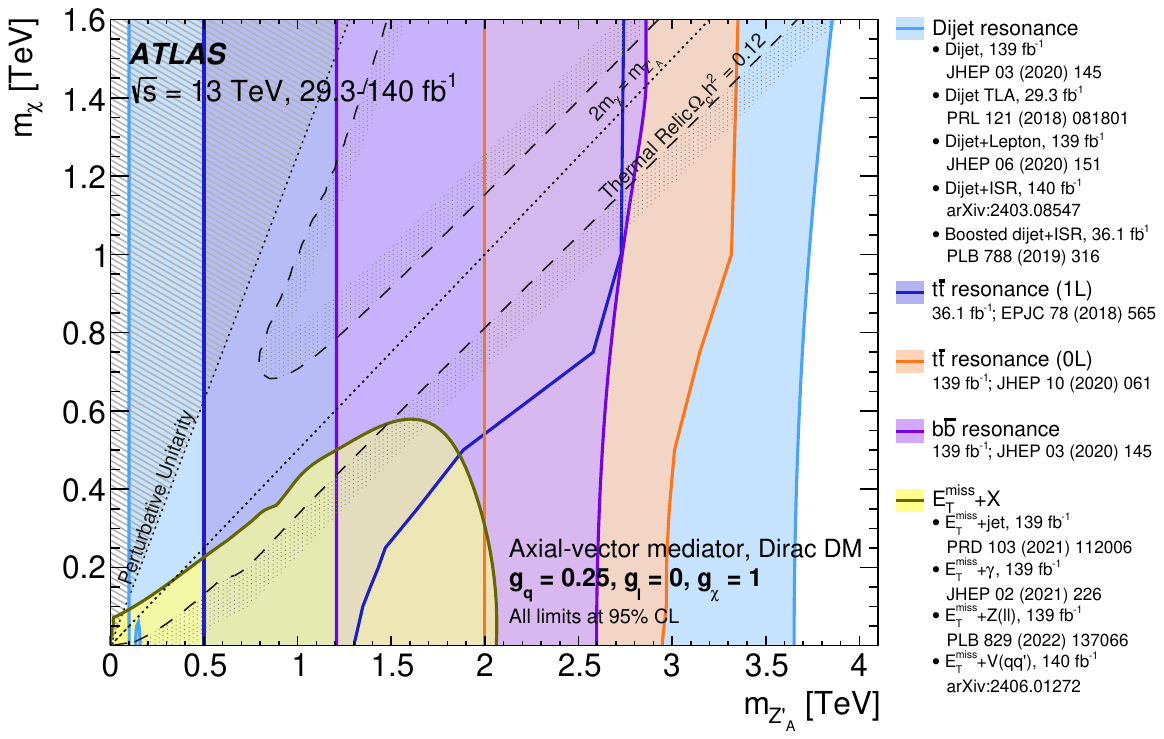}\label{fig:DMCombination_A1}}
\subfigure[]{\includegraphics[width=0.49\textwidth]{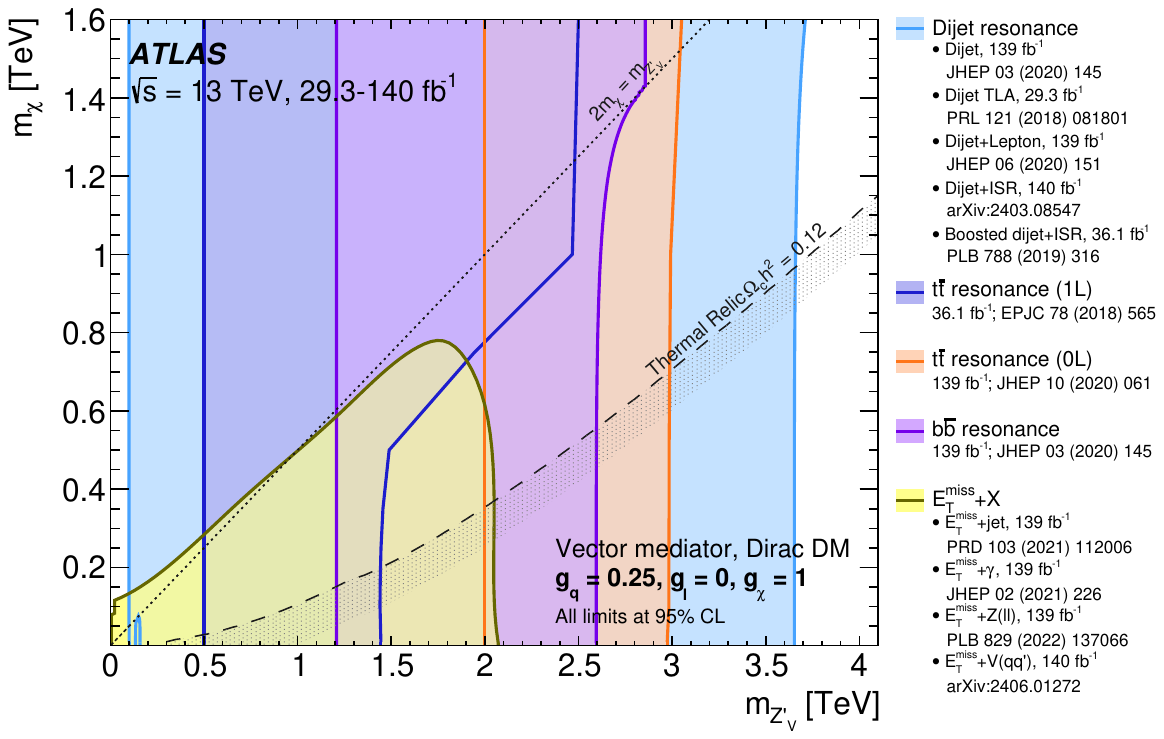}\label{fig:DMCombination_V1}} \\
\subfigure[]{\includegraphics[width=0.49\textwidth]{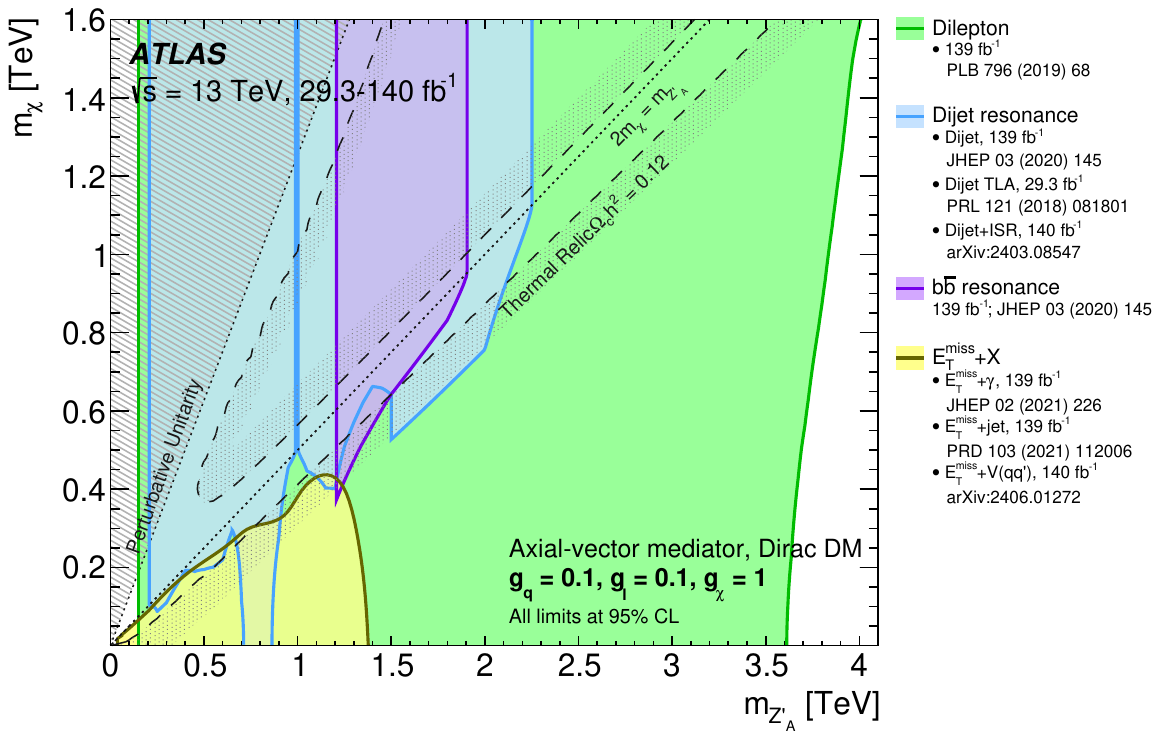}\label{fig:DMCombination_A2}}
\subfigure[]{\includegraphics[width=0.49\textwidth]{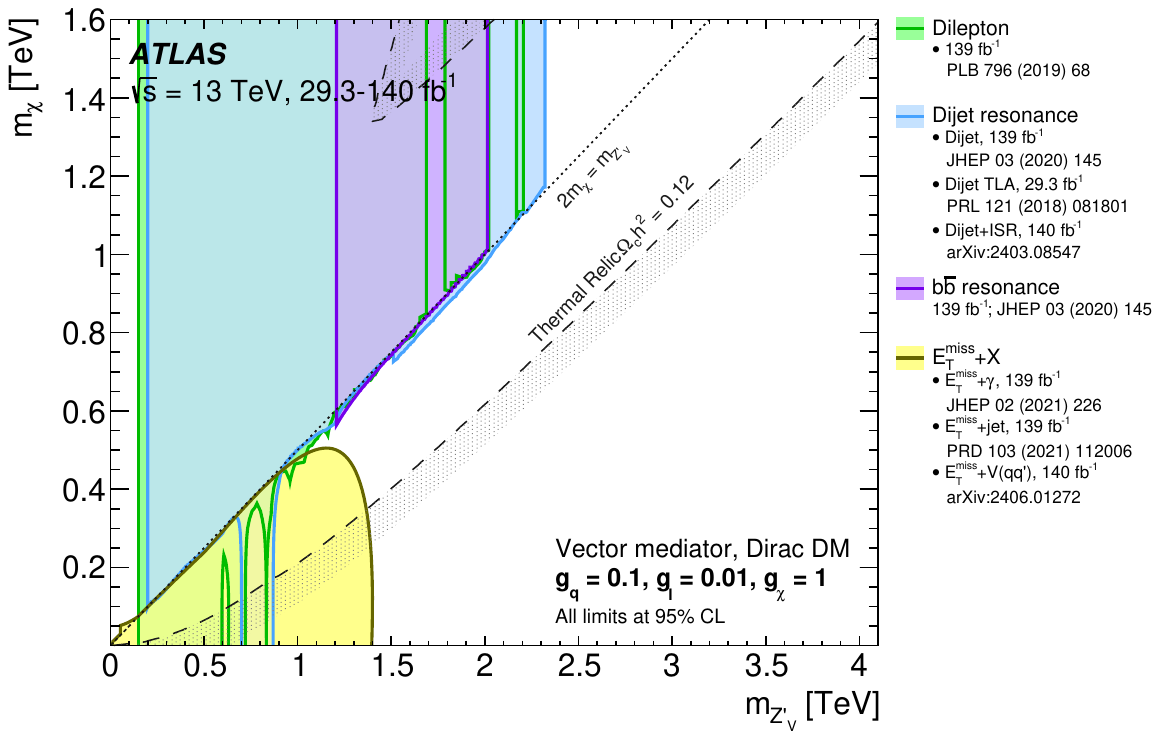}\label{fig:DMCombination_V2}} \\
\subfigure[]{\includegraphics[width=0.49\textwidth]{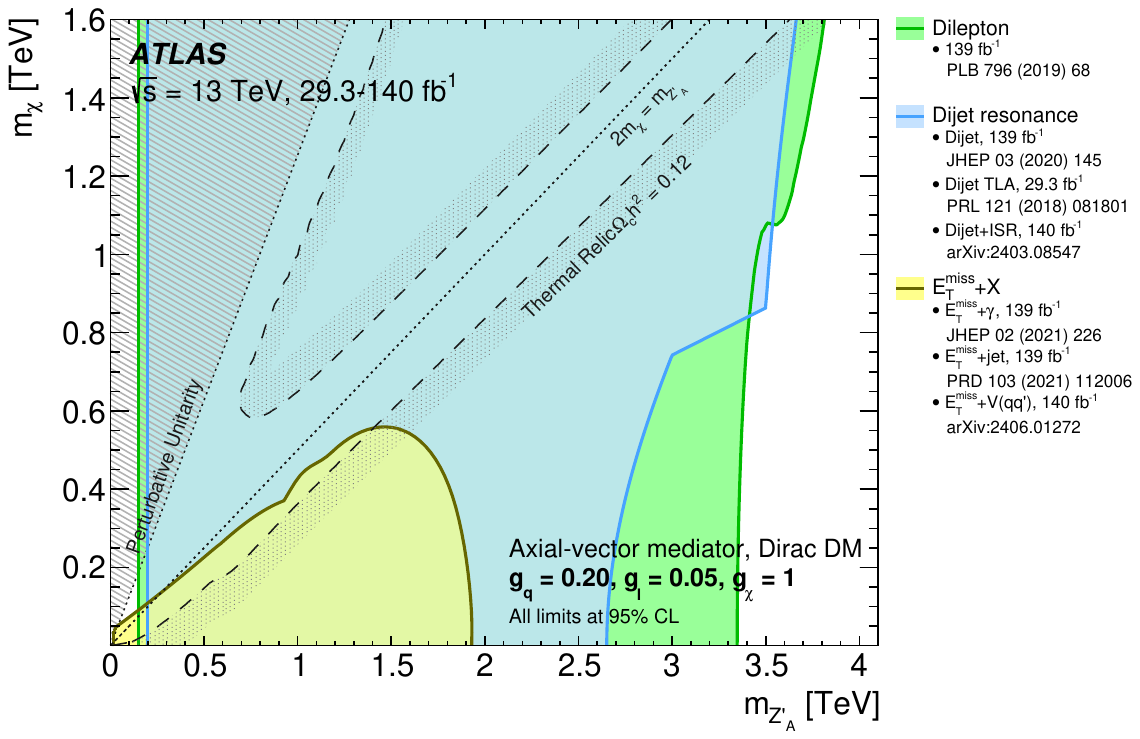}\label{fig:DMCombination_A4}}
\subfigure[]{\includegraphics[width=0.49\textwidth]{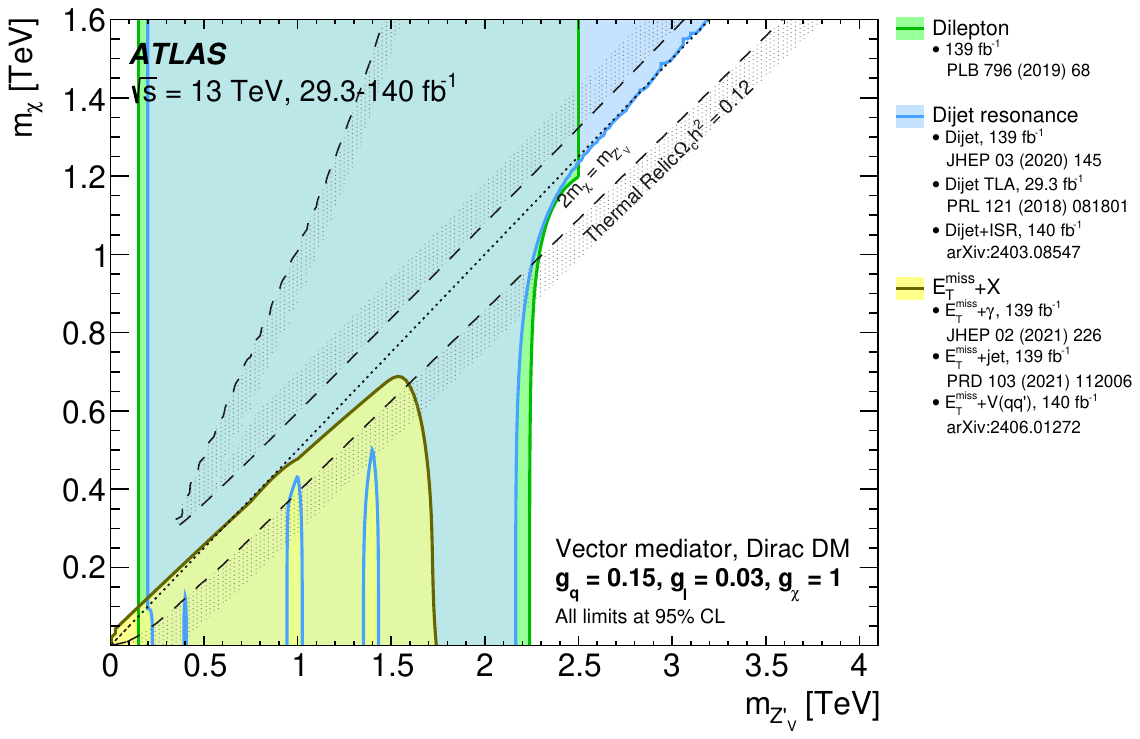}\label{fig:DMCombination_V4}}
\caption{Exclusions in the (\mDM, \mMed) plane for axial-vector benchmark models \subref{fig:DMCombination_A1} A1, \subref{fig:DMCombination_A2} A2, and \subref{fig:DMCombination_A4} A4, and vector benchmark models \subref{fig:DMCombination_V1} V1, \subref{fig:DMCombination_V2} V2, and \subref{fig:DMCombination_V4} V4. The unions of multiple contours are plotted together as indicated in the figure legends. The dashed curves indicate points consistent with a thermal relic DM density of $\Omega h^2=0.12$, with the over-dense side shaded.}
\label{fig:spin1-results-Overlay-most}
\end{figure}

\begin{figure}[!tb]
\centering
\subfigure[]{\includegraphics[width=0.49\textwidth]{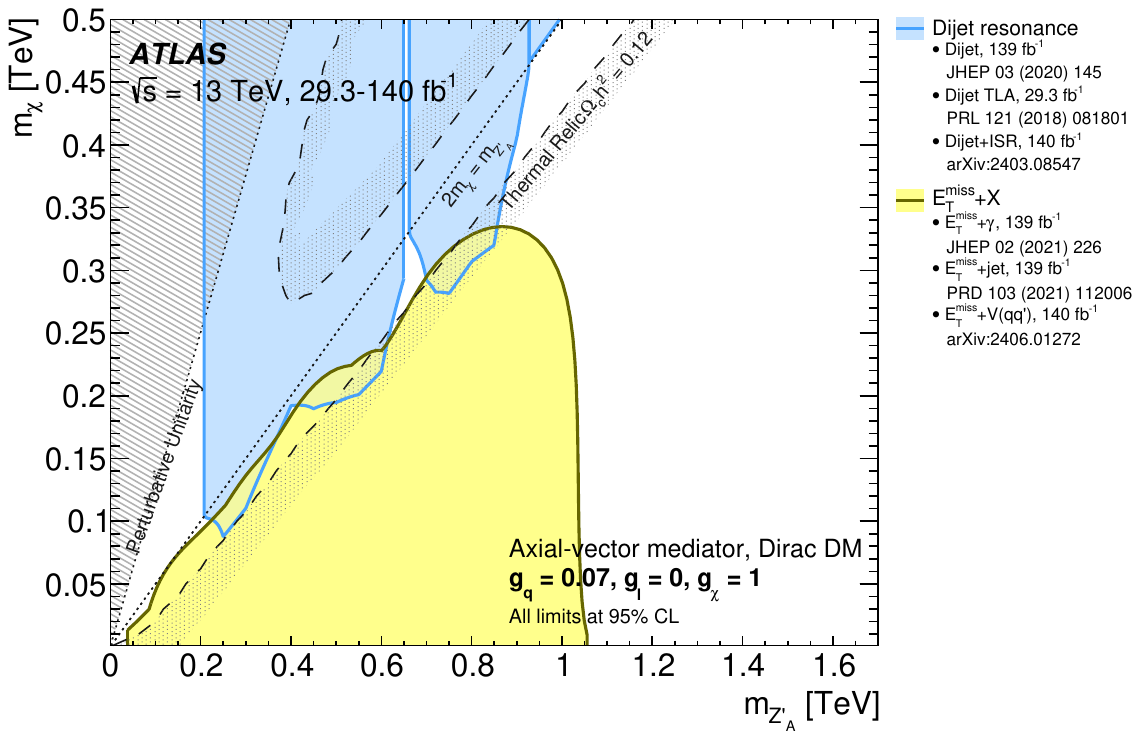}\label{fig:DMCombination_A3}}
\subfigure[]{\includegraphics[width=0.49\textwidth]{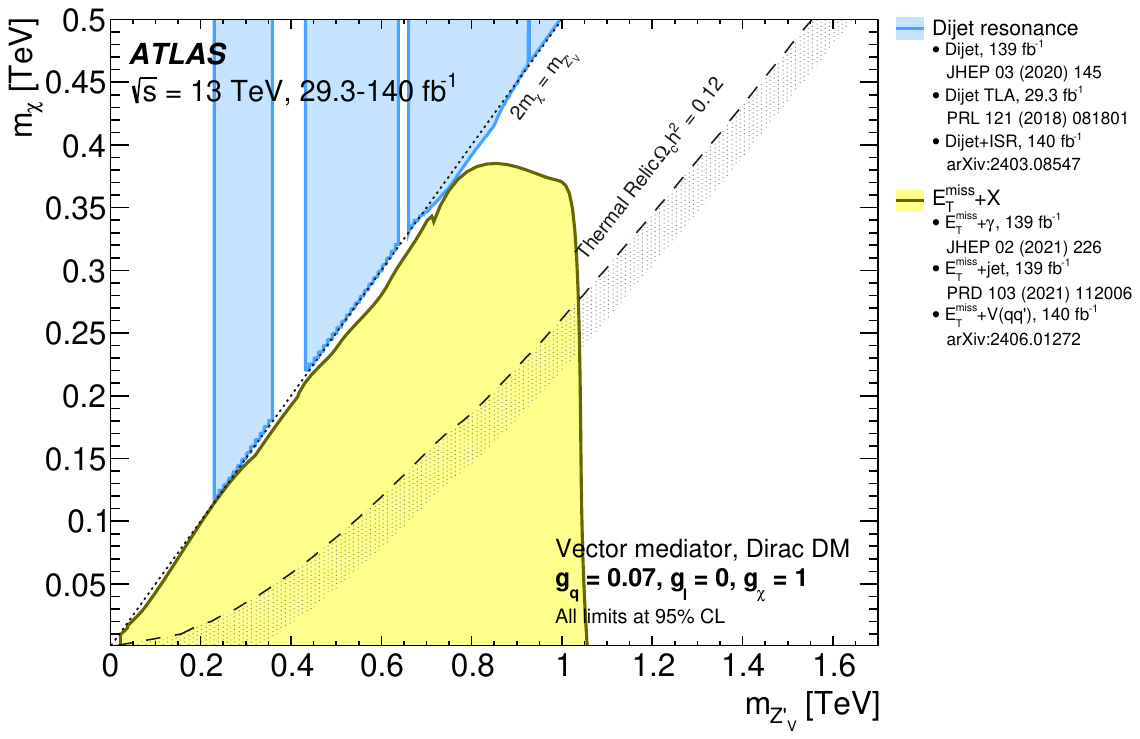}\label{fig:DMCombination_V3}}
\caption{Exclusions in the (\mDM, \mMed) plane for benchmark models \subref{fig:DMCombination_A3} A3 and \subref{fig:DMCombination_V3} V3. The unions of multiple contours are plotted together as indicated in the figure legends. The dashed curves indicate points consistent with a thermal relic DM density of $\Omega h^2=0.12$, with the over-dense side shaded. The interplay between the coverage of visible and semi-visible final state searches is visible here.}
\label{fig:spin1-results-Overlay-mono}
\end{figure}

\subsubsection{Limits on the mediator mass in the coupling-coupling plane}
\label{sec:results_couplingcoupling}

The rescaling procedure described in \cref{sec:rescaling} is used to produce 95\% CL lower limits on the mediator mass, \mMed, in the (\gq, \gl) coupling plane.
Each figure corresponds to the sensitivity of a specific analysis, assuming either an axial-vector or vector mediator. The dijet and dilepton resonant searches are reported as the most sensitive ones to hadronic and leptonic couplings.
The mass of the DM candidate is fixed to a value of 100~\GeV, allowing the mediator to decay into a light DM particle.
For each (\gq, \gl) combination in the plane, the rescaling procedure is used to evaluate the exclusion depth as a function of the mediator mass. The lowest non-excluded mediator mass is extracted for the given choice of (\gq, \gl).
\cref{fig:spin1-results-couplingPlane} reports the limits obtained from the dijet and dilepton searches under the axial-vector or vector assumption on the mediator spin-parity.

\begin{figure}[!tb]
\centering
\subfigure[]{\includegraphics[width=0.49\textwidth]{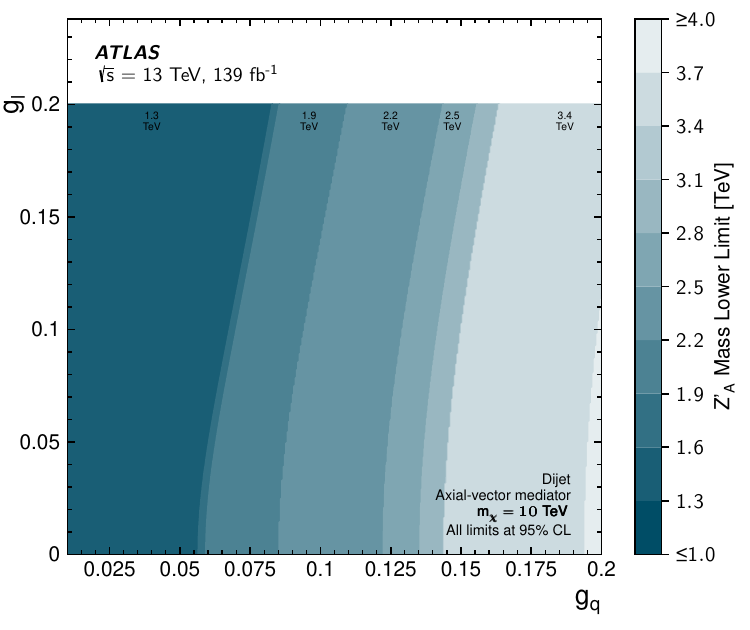}\label{fig:dijet_axial_mx100_gqgl}}
\subfigure[]{\includegraphics[width=0.49\textwidth]{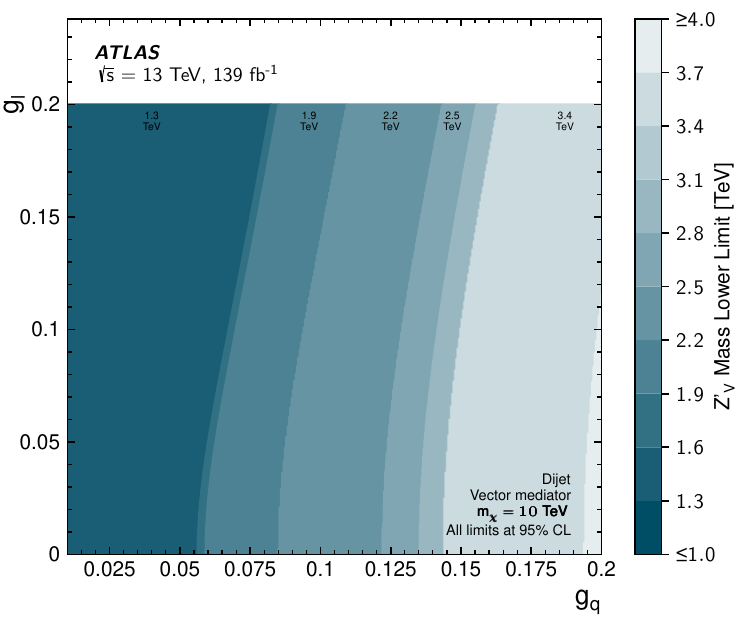}\label{fig:dijet_vector_mx100_gqgl}}\\
\subfigure[]{\includegraphics[width=0.49\textwidth]{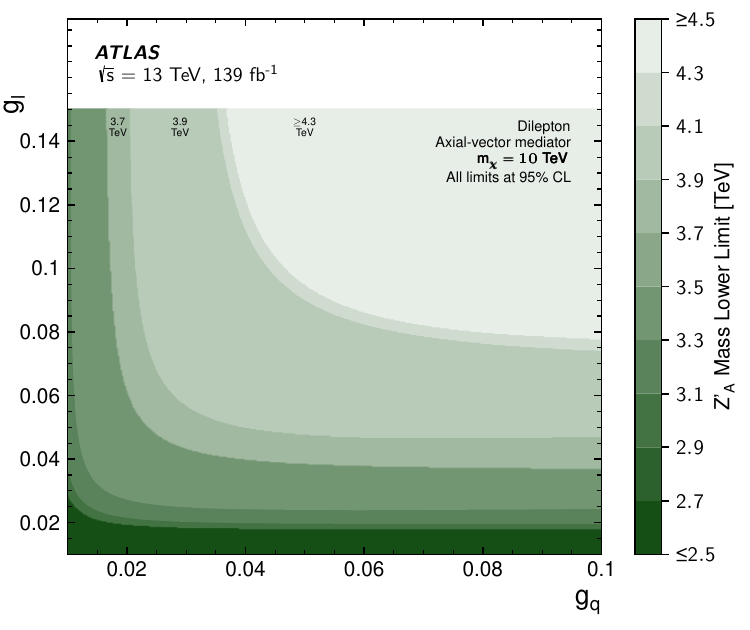}\label{fig:dilepton_axial_mx100_gqgl}}
\subfigure[]{\includegraphics[width=0.49\textwidth]{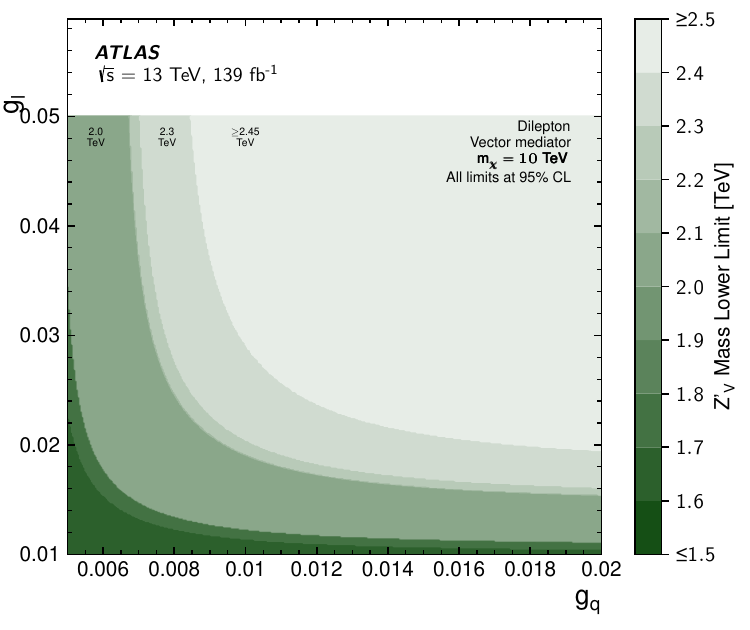}\label{fig:dilepton_vector_mx100_gqgl}}
\caption{95\% CL lower limits on the mediator mass \mMed as a function of \gq and \gl. The mass of the DM is fixed to 1~$\TeV$. The figures report the results for \subref{fig:dijet_axial_mx100_gqgl},\subref{fig:dijet_vector_mx100_gqgl} the dijet search and \subref{fig:dilepton_axial_mx100_gqgl},\subref{fig:dilepton_vector_mx100_gqgl} the dilepton searches under the \subref{fig:dijet_axial_mx100_gqgl}-\subref{fig:dilepton_axial_mx100_gqgl} axial-vector or \subref{fig:dijet_vector_mx100_gqgl}-\subref{fig:dilepton_vector_mx100_gqgl} vector assumption on the mediator spin-parity.
}
\label{fig:spin1-results-couplingPlane}
\end{figure}

\subsubsection{Comparison with direct detection}

Selected results from \cref{fig:spin1-results-Overlay-most,fig:spin1-results-Overlay-mono,fig:spin1-results-Overlay-mono} are translated into constraints on the DM scattering cross-section into nucleons using the procedure described in \cref{sec:dd_reinterpret}. This permits a comparison with direct detection experiments under spin-dependent and spin-independent scattering cross-section assumptions.
\Cref{fig:spin1-results-dd} shows the constraints from selected analyses as a function of the DM mass \mDM for four of the benchmarks described in \cref{tab:models}.
The regions excluded by the ATLAS searches can be extended to lower values of \mDM but it has to be noted that below $\mDM\simeq 1~\GeV$ some of the WIMP assumptions are not fully valid. The results achievable for massless DM is not significantly different relative to the $\mDM=1~\GeV$ case since the DM is relativistic in both cases.

\begin{figure}[!tb]
\centering
\subfigure[]{\includegraphics[width=0.45\textwidth]{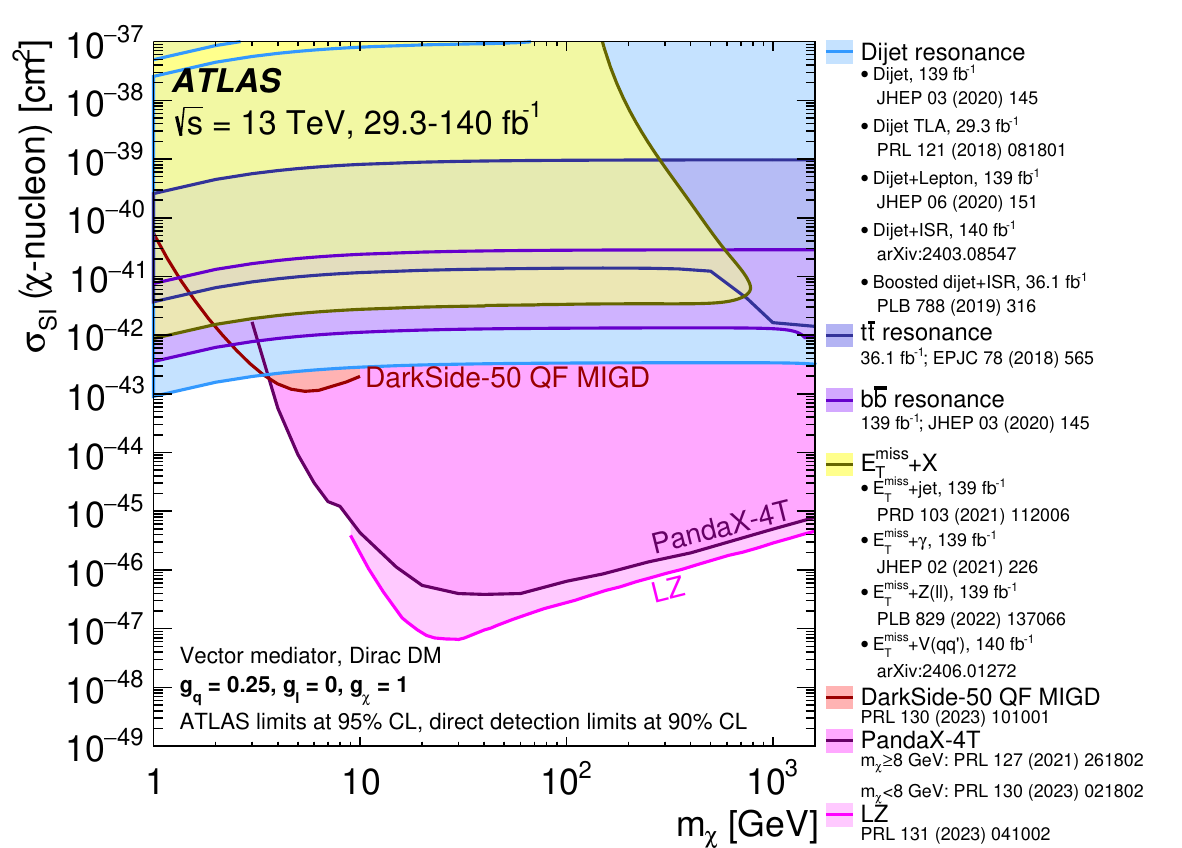}\label{fig:DMCombinationDD_Internal_V1SInucleon}}
\subfigure[]{\includegraphics[width=0.45\textwidth]{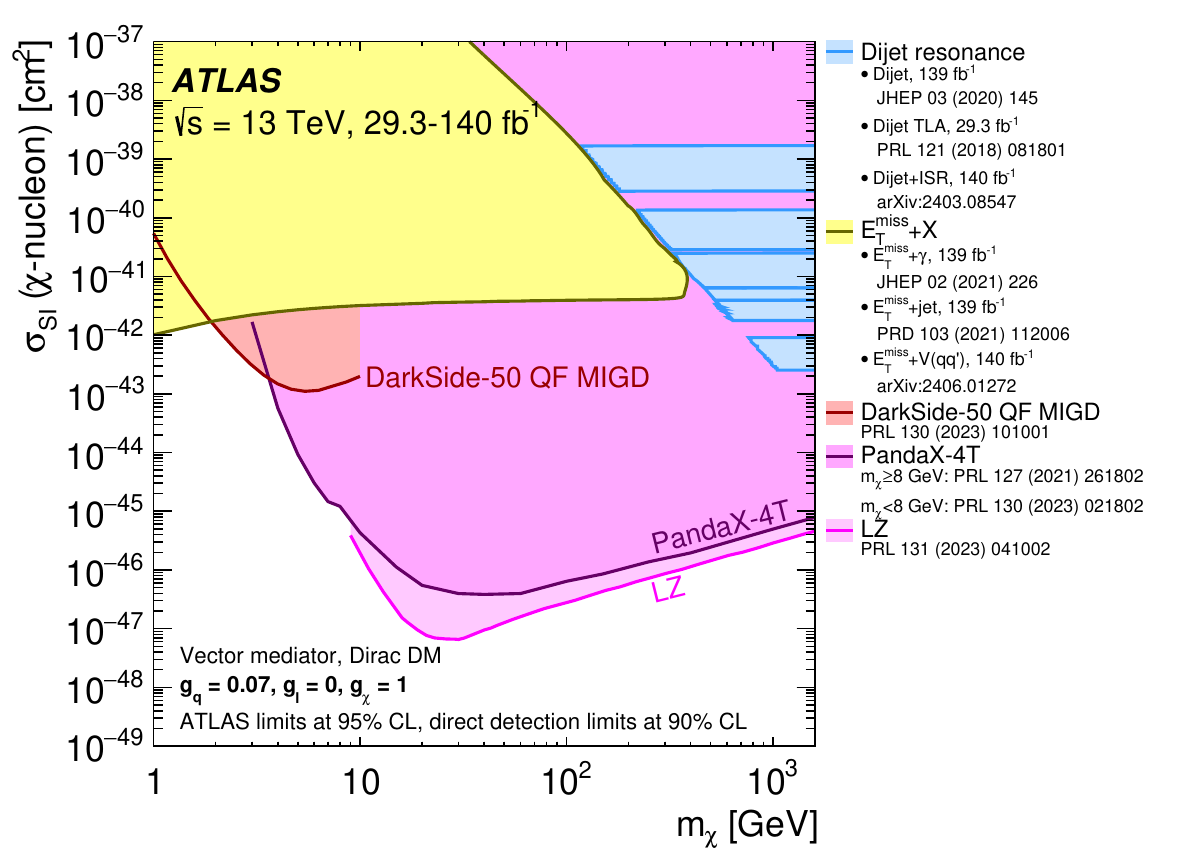}\label{fig:DMCombinationDD_Internal_V3SInucleon}}
\subfigure[]{\includegraphics[width=0.45\textwidth]{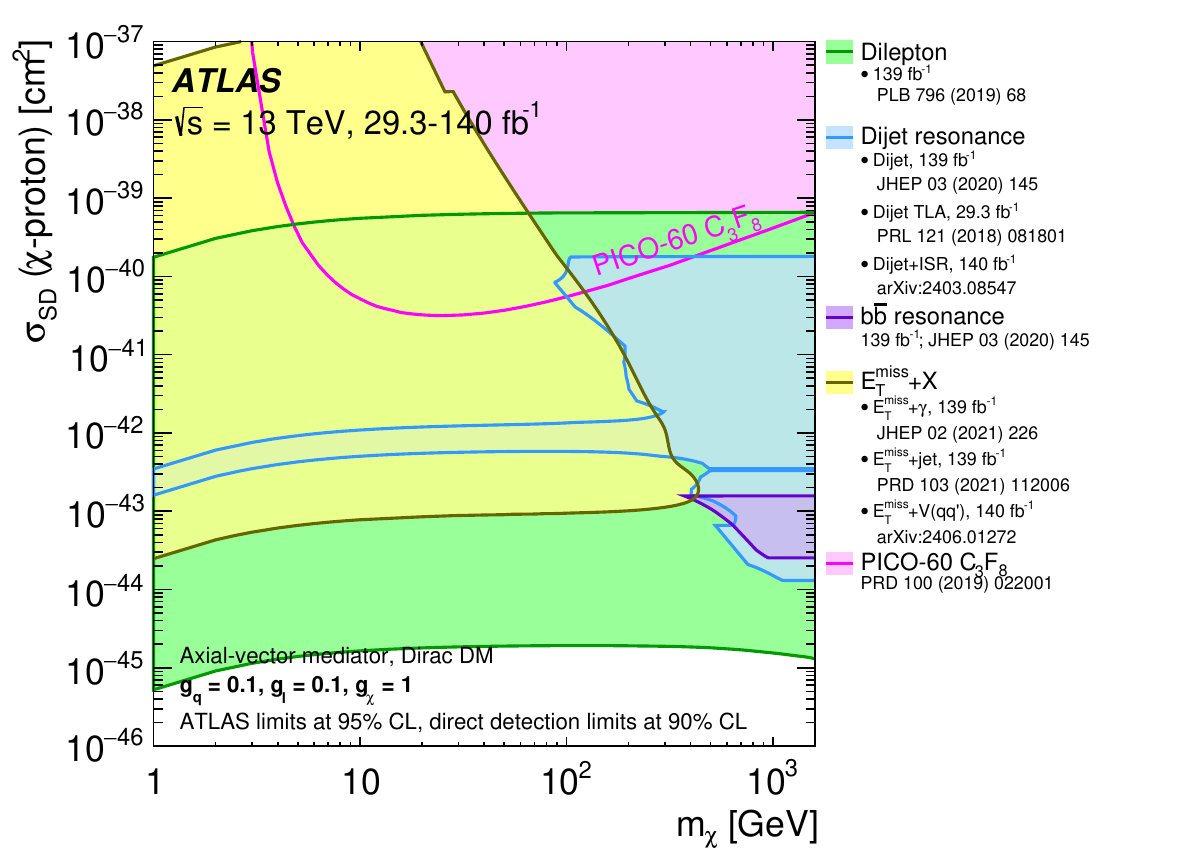}\label{fig:DMCombinationDD_Internal_A2SDproton}}
\subfigure[]{\includegraphics[width=0.45\textwidth]{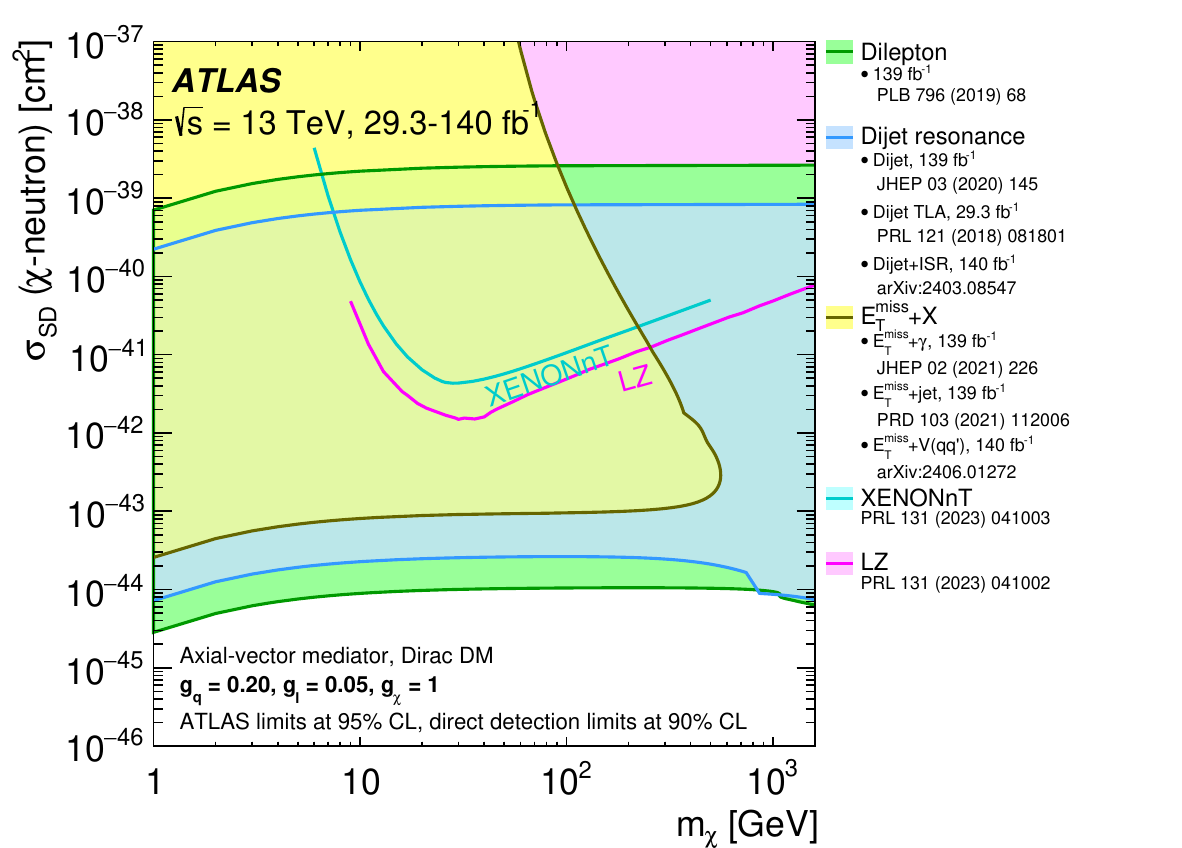}\label{fig:DMCombinationDD_Internal_A4SDneutron}}
\caption{Comparison of limits with direct detection experiments in four scenarios: \subref{fig:DMCombinationDD_Internal_V1SInucleon} V1 and \subref{fig:DMCombinationDD_Internal_V3SInucleon} V3 with spin-independent $\chi$-nucleon cross-sections, \subref{fig:DMCombinationDD_Internal_A2SDproton} A2 with spin-dependent $\chi$-proton cross-sections, and \subref{fig:DMCombinationDD_Internal_A4SDneutron} A4 with spin-dependent $\chi$-neutron cross-sections. The results are compared with limits from direct detection experiments from DarkSide-50 QF MIGD~\cite{Agnes_2018}, PandaX-4T~\cite{Meng_2021,Panda2023}, XENONnT~\cite{PhysRevLett.131.041003}, LZ~\cite{LZ:2022lsv}, LUX~\cite{PhysRevLett.118.251302}, and PICO-60 C$_3$F$_8$~\cite{PhysRevD.100.022001}.}
\label{fig:spin1-results-dd}
\end{figure}



%

%
%
%
\FloatBarrier
%
%
%

\section{Conclusion}
\label{sec:conclusion}

The ATLAS Collaboration has searched for evidence of dark matter, carrying out several analyses over a wide variety of final-state signatures.
These searches are based on proton-proton collision data at a centre-of-mass energy $\sqrt{s}=13$~TeV provided by the LHC, and correspond to an integrated luminosity of up to \MaxLumi.
The simplified dark matter models proposed by the LHC Dark Matter Working Group provide a framework to interpret the results of these searches together, highlighting their complementarity. In \thisdoc spin-0 and spin-1 mediators produced through the \schannel with four parity types are considered: scalar, pseudo-scalar, vector, and axial-vector.
The 95\% CL limits of several resonant and $X+\MET$ analyses are rescaled for the first time using the procedure described in \cref{sec:rescaling}, to further extend the interpretation of these results in the model's parameters space.
The constraints from the different analyses cover much of the accessible parameter space in DM mass, mediator mass, and couplings, hinting at where to focus for the next searches.

For models with spin-0 mediators, exclusion limits for the simplified model of dark matter production including a colour neutral scalar (pseudo-scalar) mediator are compared for \ttmet,~\monotop,~\monojet, and~\fourtops final states. These limits span the mediator mass range from 10~\GeV to 1500~\GeV.
The results are also interpreted as constrains on spin-independent DM--nucleon cross-section and $\langle\sigma v_\text{rel}\rangle$ for a comparison with the results from several direct detection experiments. ATLAS searches extend beyond the exclusion reach of direct detection experiments in the range of small DM mass.

For models with a spin-1 mediator, limits are derived separately for couplings to hadronic and leptonic particles for axial-vector mediator masses up to 5~\TeV.
Exclusion contours in the (\mDM, \mMed) plane are set for the eight spin-1 benchmark models with $\mDM<1.6$~\TeV and $\mZprime<4$~\TeV. These benchmark scenarios extend those considered in the previous publications, highlighting the complementarity and maximum reach of the different signatures when varying model parameter values.
For the first time, lower limits on the mediator mass are provided in the (\gl, \gq) plane for the most sensitive corresponding resonance searches, reaching exclusion up to 5~\TeV.
The constraints from different analyses are compared in several scenarios to the limits set by direct detection experiments. ATLAS limits are particularly stringent in the case of spin-dependent scattering cross-sections.

The results in \thisdoc provide a comprehensive summary of the most recent limits on DM and \schannel mediators by the ATLAS Collaboration.


\clearpage
\appendix

\part*{Appendix}
\addcontentsline{toc}{part}{Appendix}
\section{Signal model generation details}
\label{app:SignalModelDetails}
The model implementations, settings and parameter scans used in this paper follow the prescriptions of the DM Forum/LHC DM Working Group~\cite{Abercrombie_2019,Boveia:2016mrp,dmwg_recs,2HDMWGproxi} and all generation settings used for signal models in \thisdoc are summarised in \cref{tab:MCvectorsignal,tab:MCscalarsignal}.

\begin{table}[h!]
\caption{Details of the generator configuration and Universal FeynRules Output (UFO) model used for the spin-1 mediator simplified models.}
\label{tab:MCvectorsignal}
\scalebox{0.75}{
\begin{tabular}{m{38mm}m{28mm}m{74mm}m{15mm}m{45mm}}
\toprule
Model and Final State & UFO & Generator and Parton Shower & Cross-section & Additional details \\
\toprule
$Z'(\chi\bar{\chi})+j$       & DMV \cite{EXOT-2018-06,urldmv} &\POWHEGBOX[v2]~\cite{Alioli:2010xd} +  \PYTHIA[8.205]~\cite{Sjostrand:2007gs} & NLO & Particle-level rescaling of leptophobic \ZprimeA scenario of Ref.~\cite{EXOT-2018-06}  \\\hline
$Z'(\chi\bar{\chi})+\gamma$  & DMSimp \cite{Backovic:2015soa,urldmsimp} & \MGNLO[2.4.3] (NLO)~\cite{Alwall:2014hca} \newline +  \PYTHIA[8.212] & NLO & Leptophobic \ZprimeA scenario simulated, other scenarios obtained by cross-section rescaling \\\hline
$Z'(qq)$ or $Z'(qq)$+ISR & DMSimp          & \MGNLO[2.2.3] (NLO) \newline+  \PYTHIA 8.210                       & NLO &Leptophobic \ZprimeA scenario simulated, other scenarios obtained by Gaussian resonance limits and cross-section rescaling \\\hline
$Z'(\bbbar)$ & DMSimp          & \MGNLO[2.2.3] (NLO) \newline +  \PYTHIA[8.210]                       & NLO &Leptophobic \ZprimeA scenario simulated, other scenarios obtained by Gaussian resonance limits and cross-section rescaling \\\hline
$Z'(\ell\ell)$ & DMSimp          & \MGNLO[2.2.2] (NLO) \newline + \PYTHIA[8.212]   & NLO & Gaussian resonance limits and cross-section rescaling \\\hline
$Z'(t\bar t)$            & DMSimp  & \MGNLO[2.4.3] (LO) \newline +  \PYTHIA[8.186] & LO & Particle-level rescaling of the topcolour-assisted technicolour samples~\cite{EXOT-2015-04}\\\hline
\bottomrule
\end{tabular}
}
\end{table}

\begin{table}
\caption{Details of the generator configuration and Universal FeynRules Output (UFO) model used for the spin-0 mediator models.}
\label{tab:MCscalarsignal}
\scalebox{0.75}{
\begin{tabular}{m{38mm}m{31mm}m{72mm}m{15mm}m{45mm}}
\toprule
Model and Final State & UFO & Generator and Parton Shower & Cross-section & Additional details \\
\toprule
DM-monojet          & DMS\_tloop~\cite{Buckley:2014fba,urlscalmj} & \POWHEGBOX[v2] +  \PYTHIA[8.205] & NLO & Ref.~\cite{EXOT-2018-06}  \\\hline
\ttdm, \twdm, \tjdm, \ttfourtop    & DMScalarMed\_loop \cite{Buckley:2014fba,urldmhf} & \MGNLO[2.3.3] (LO) \newline + \PYTHIA[8.186] & NLO~\cite{Backovic:2015soa} & Up to one additional parton. Ref.~\cite{SUSY-2019-12}  \\\hline
\bbdm                 & DMScalarMed\_loop & \MGNLO[2.3.3] (LO) \newline + \PYTHIA[8.186] & NLO~\cite{Afik:2018rxl} & Up to one additional parton. Ref.~\cite{SUSY-2019-12}  \\\hline
\bottomrule
\end{tabular}
}
\end{table}


%
%
%
\FloatBarrier
\section*{Acknowledgements}

%
%

%
%

We thank CERN for the very successful operation of the LHC and its injectors, as well as the support staff at
CERN and at our institutions worldwide without whom ATLAS could not be operated efficiently.

The crucial computing support from all WLCG partners is acknowledged gratefully, in particular from CERN, the ATLAS Tier-1 facilities at TRIUMF/SFU (Canada), NDGF (Denmark, Norway, Sweden), CC-IN2P3 (France), KIT/GridKA (Germany), INFN-CNAF (Italy), NL-T1 (Netherlands), PIC (Spain), RAL (UK) and BNL (USA), the Tier-2 facilities worldwide and large non-WLCG resource providers. Major contributors of computing resources are listed in Ref.~\cite{ATL-SOFT-PUB-2023-001}.

We gratefully acknowledge the support of ANPCyT, Argentina; YerPhI, Armenia; ARC, Australia; BMWFW and FWF, Austria; ANAS, Azerbaijan; CNPq and FAPESP, Brazil; NSERC, NRC and CFI, Canada; CERN; ANID, Chile; CAS, MOST and NSFC, China; Minciencias, Colombia; MEYS CR, Czech Republic; DNRF and DNSRC, Denmark; IN2P3-CNRS and CEA-DRF/IRFU, France; SRNSFG, Georgia; BMBF, HGF and MPG, Germany; GSRI, Greece; RGC and Hong Kong SAR, China; ISF and Benoziyo Center, Israel; INFN, Italy; MEXT and JSPS, Japan; CNRST, Morocco; NWO, Netherlands; RCN, Norway; MNiSW, Poland; FCT, Portugal; MNE/IFA, Romania; MESTD, Serbia; MSSR, Slovakia; ARIS and MVZI, Slovenia; DSI/NRF, South Africa; MICIU/AEI, Spain; SRC and Wallenberg Foundation, Sweden; SERI, SNSF and Cantons of Bern and Geneva, Switzerland; NSTC, Taipei; TENMAK, T\"urkiye; STFC/UKRI, United Kingdom; DOE and NSF, United States of America.

Individual groups and members have received support from BCKDF, CANARIE, CRC and DRAC, Canada; CERN-CZ, FORTE and PRIMUS, Czech Republic; COST, ERC, ERDF, Horizon 2020, ICSC-NextGenerationEU and Marie Sk{\l}odowska-Curie Actions, European Union; Investissements d'Avenir Labex, Investissements d'Avenir Idex and ANR, France; DFG and AvH Foundation, Germany; Herakleitos, Thales and Aristeia programmes co-financed by EU-ESF and the Greek NSRF, Greece; BSF-NSF and MINERVA, Israel; Norwegian Financial Mechanism 2014-2021, Norway; NCN and NAWA, Poland; La Caixa Banking Foundation, CERCA Programme Generalitat de Catalunya and PROMETEO and GenT Programmes Generalitat Valenciana, Spain; G\"{o}ran Gustafssons Stiftelse, Sweden; The Royal Society and Leverhulme Trust, United Kingdom.

In addition, individual members wish to acknowledge support from CERN: European Organization for Nuclear Research (CERN PJAS); Chile: Agencia Nacional de Investigaci\'on y Desarrollo (FONDECYT 1190886, FONDECYT 1230812, FONDECYT 1230987); China: Chinese Ministry of Science and Technology (MOST-2023YFA1605700), National Natural Science Foundation of China (NSFC - 12175119, NSFC 12275265, NSFC-12075060); Czech Republic: Czech Science Foundation (GACR - 24-11373S), Ministry of Education Youth and Sports (FORTE CZ.02.01.01/00/22\_008/0004632), PRIMUS Research Programme (PRIMUS/21/SCI/017); EU: H2020 European Research Council (ERC - 101002463); European Union: European Research Council (ERC - 948254, ERC 101089007), Horizon 2020 Framework Programme (MUCCA - CHIST-ERA-19-XAI-00), European Union, Future Artificial Intelligence Research (FAIR-NextGenerationEU PE00000013), Italian Center for High Performance Computing, Big Data and Quantum Computing (ICSC, NextGenerationEU); France: Agence Nationale de la Recherche (ANR-20-CE31-0013, ANR-21-CE31-0013, ANR-21-CE31-0022, ANR-22-EDIR-0002), Investissements d'Avenir Labex (ANR-11-LABX-0012); Germany: Baden-Württemberg Stiftung (BW Stiftung-Postdoc Eliteprogramme), Deutsche Forschungsgemeinschaft (DFG - 469666862, DFG - CR 312/5-2); Italy: Istituto Nazionale di Fisica Nucleare (ICSC, NextGenerationEU), Ministero dell'Università e della Ricerca (PRIN - 20223N7F8K - PNRR M4.C2.1.1); Japan: Japan Society for the Promotion of Science (JSPS KAKENHI JP21H05085, JSPS KAKENHI JP22H01227, JSPS KAKENHI JP22H04944, JSPS KAKENHI JP22KK0227); Netherlands: Netherlands Organisation for Scientific Research (NWO Veni 2020 - VI.Veni.202.179); Norway: Research Council of Norway (RCN-314472); Poland: Ministry of Science and Higher Education (IDUB AGH, POB8, D4 no 9722), Polish National Agency for Academic Exchange (PPN/PPO/2020/1/00002/U/00001), Polish National Science Centre (NCN 2021/42/E/ST2/00350, NCN OPUS nr 2022/47/B/ST2/03059, NCN UMO-2019/34/E/ST2/00393, UMO-2020/37/B/ST2/01043, UMO-2021/40/C/ST2/00187, UMO-2022/47/O/ST2/00148, UMO-2023/49/B/ST2/04085); Slovenia: Slovenian Research Agency (ARIS grant J1-3010); Spain: Generalitat Valenciana (Artemisa, FEDER, IDIFEDER/2018/048), Ministry of Science and Innovation (MCIN \& NextGenEU PCI2022-135018-2, MICIN \& FEDER PID2021-125273NB, RYC2019-028510-I, RYC2020-030254-I, RYC2021-031273-I, RYC2022-038164-I), PROMETEO and GenT Programmes Generalitat Valenciana (CIDEGENT/2019/023, CIDEGENT/2019/027); Sweden: Swedish Research Council (Swedish Research Council 2023-04654, VR 2018-00482, VR 2022-03845, VR 2022-04683, VR 2023-03403, VR grant 2021-03651), Knut and Alice Wallenberg Foundation (KAW 2018.0157, KAW 2018.0458, KAW 2019.0447, KAW 2022.0358); Switzerland: Swiss National Science Foundation (SNSF - PCEFP2\_194658); United Kingdom: Leverhulme Trust (Leverhulme Trust RPG-2020-004), Royal Society (NIF-R1-231091); United States of America: U.S. Department of Energy (ECA DE-AC02-76SF00515), Neubauer Family Foundation.

%
%


%

%

%
%
%
%
%
%
%

%

%

%
%

%
%
%
\clearpage
\printbibliography

@article{Backovic:2015soa,
      author         = "Backović, Mihailo and Krämer, Michael and Maltoni,
                        Fabio and Martini, Antony and Mawatari, Kentarou and
                        Pellen, Mathieu",
      title          = "{Higher-order QCD predictions for dark matter production
                        at the LHC in simplified models with s-channel mediators}",
      journal        = "Eur. Phys. J. C",
      volume         = "75",
      year           = "2015",
      number         = "10",
      pages          = "482",
      doi            = "10.1140/epjc/s10052-015-3700-6",
      eprint         = "1508.05327",
      archivePrefix  = "arXiv",
      primaryClass   = "hep-ph",
      reportNumber   = "MCNET-15-24, CP3-15-25, TTK-15-19",
}

@article{dmwg_recs,
    author = "Albert, A. and Backovi\'c, M. and Boveia, A. and Buchmueller, O. and Busoni, G. and De Roeck, A. and Doglioni, C. and DuPree, T. and Fairbairn, M. and Genest, M.~H. and others",
    title = "{Recommendations of the LHC Dark Matter Working Group: Comparing LHC searches for dark matter mediators in visible and invisible decay channels and calculations of the thermal relic density}",
    eprint = "1703.05703",
    archivePrefix = "arXiv",
    primaryClass = "hep-ex",
    reportNumber = "CERN-LPCC-2017-01",
    doi = "10.1016/j.dark.2019.100377",
    journal = "Phys. Dark Univ.",
    volume = "26",
    pages = "100377",
    year = "2019"
}

@article{maddm,
    author = "Backovi\'c, Mihailo and Martini, Antony and Kong, Kyoungchul and Mattelaer, Olivier and Mohlabeng, Gopolang",
    title = "{MadDM: New dark matter tool in the LHC era}",
    eprint = "1509.03683",
    archivePrefix = "arXiv",
    primaryClass = "hep-ph",
    doi = "10.1063/1.4953318",
    journal = "AIP Conf. Proc.",
    volume = "1743",
    number = "1",
    pages = "060001",
    year = "2016"
}

@article{Buckley_2015,
  title         = {Scalar simplified models for dark matter},
  volume        = {91},
  doi           = {10.1103/physrevd.91.015017},
  journal       = {Phys. Rev. D},
  author        = {Buckley, Matthew R. and Feld, David and Gon\c{c}alves, Dorival},
  year          = {2015},
  eprint        = {1410.6497},
  archiveprefix = {arXiv},
  primaryclass  = {hep-ph},
}

@article{haisch2015simplified,
  title         = {Simplified dark matter top-quark interactions at the LHC},
  author        = {Ulrich Haisch and Emanuele Re},
  doi           = {10.1007/JHEP06(2015)078},
  year          = {2015},
  journal       = {JHEP},
  volume        = {06},
  pages         = {078},
  eprint        = {1503.00691},
  archiveprefix = {arXiv},
  primaryclass  = {hep-ph},
}

@article{Abercrombie_2019,
  title         = {Dark Matter benchmark models for early LHC Run-2 Searches: Report of the ATLAS/CMS Dark Matter Forum},
  doi           = {10.1016/j.dark.2019.100371},
  journal       = {Phys. Dark Univ.},
  volume        = {27},
  author        = {Abercrombie, Daniel and others},
  year          = {2020},
  pages         = {100371},
  eprint        = {1507.00966},
  archiveprefix = {arXiv},
  primaryclass  = {hep-ex},
}

@article{Trimble1987,
author = {Trimble, Virginia},
title = {Existence and Nature of Dark Matter in the Universe},
journal = {Ann. Rev. Astron. Astrophys.},
volume = {25},
number = {1},
pages = {425-472},
year = {1987},
doi = {10.1146/annurev.aa.25.090187.002233},
}

@article{Hinshaw:2012aka,
      author         = "Hinshaw, G. and others",
      title          = "{Nine-Year Wilkinson Microwave Anisotropy Probe (WMAP)
                        Observations: Cosmological Parameter Results}",
      collaboration  = "WMAP",
      journal        = "Astrophys. J. Suppl.",
      volume         = "208",
      year           = "2013",
      pages          = "19",
      doi            = "10.1088/0067-0049/208/2/19",
      eprint         = "1212.5226",
      archivePrefix  = "arXiv",
      primaryClass   = "astro-ph.CO",
      SLACcitation   = "%%CITATION = ARXIV:1212.5226;%%"
}

@article{Akrami:2018vks,
    author = "{Planck Collaboration}",
    collaboration = "Planck",
    title = "{Planck 2018 results - I. Overview and the cosmological legacy of Planck}",
    eprint = "1807.06205",
    archivePrefix = "arXiv",
    primaryClass = "astro-ph.CO",
    doi = "10.1051/0004-6361/201833880",
    journal = "A\&A",
    volume = "641",
    pages = "A1",
    year = "2020",
}

@article{Boveia:2016mrp,
    author = "Boveia, Antonio and others",
    editor = "Buchmueller, Oliver and Doglioni, Caterina and Hahn, Kristian and Haisch, Ulrich and Kahlhoefer, Felix and Mangano, Michelangelo and McCabe, Christopher and Tait, Tim M. P.",
    title = "{Recommendations on presenting LHC searches for missing transverse energy signals using simplified $s$-channel models of dark matter}",
    eprint = "1603.04156",
    archivePrefix = "arXiv",
    primaryClass = "hep-ex",
    reportNumber = "CERN-LPCC-2016-001",
    doi = "10.1016/j.dark.2019.100365",
    journal = "Phys. Dark Univ.",
    volume = "27",
    pages = "100365",
    year = "2020",
}

@article{2HDMWGproxi,
    author = "Abe, Tomohiro and others",
    collaboration = "LHC Dark Matter Working Group",
    title = "{LHC Dark Matter Working Group: Next-generation spin-0 dark matter models}",
    eprint = "1810.09420",
    archivePrefix = "arXiv",
    primaryClass = "hep-ex",
    reportNumber = "CERN-LPCC-2018-02",
    doi = "10.1016/j.dark.2019.100351",
    journal = "Phys. Dark Univ.",
    volume = "27",
    pages = "100351",
    year = "2020",
}

@article{tmp_EXOT-2019-01,
      title={Search for low-mass resonances decaying into two jets and produced in association with a photon or a jet at $\sqrt{s}=13$ TeV with the ATLAS detector}, 
      author="{ATLAS Collaboration}",
      %journal = "Submitted to: Phys.Rev.D",
      year={2024},
      eprint={2403.08547},
      archivePrefix={arXiv},
      primaryClass={hep-ex},
      reportNumber="CERN-EP-2024-044",
}

@article{SUSY-2023-22,
      title="{Search for new phenomena with top-quark pairs and large missing transverse momentum using 140 $\mathrm{fb}^{-1}$ of $pp$ collision data at $\sqrt{s}=13$ TeV with the ATLAS detector}",
      author="{ATLAS Collaboration}",
      eprint = "2401.13430",
      archivePrefix = "arXiv",
      primaryClass = "hep-ex",
      reportNumber = "CERN-EP-2024-003",
      doi = "10.1007/JHEP03(2024)139",
      journal = "JHEP",
      volume = "03",
      pages = "139",
      year = "2024"
}

@article{Haisch:2018bby,
      author         = "Haisch, Ulrich and Polesello, Giacomo",
      title          = "{Searching for production of dark matter in association
                        with top quarks at the LHC}",
      journal        = "JHEP",
      volume         = "02",
      year           = "2019",
      pages          = "029",
      doi            = "10.1007/JHEP02(2019)029",
      eprint         = "1812.00694",
      archivePrefix  = "arXiv",
      primaryClass   = "hep-ph",
      SLACcitation   = "%%CITATION = ARXIV:1812.00694;%%",
}

@Booklet{Afik:2018rxl,
      author         = "Afik, Yoav and Maltoni, Fabio and Mawatari, Kentarou and
                        Pani, Priscilla and Polesello, Giacomo and Rozen, Yoram
                        and Zaro, Marco",
      title          = "{DM+$b\bar b$ simulations with DMSimp: an update}",
      year           = "2018",
      eprint         = "1811.08002",
      archivePrefix  = "arXiv",
      primaryClass   = "hep-ex",
      SLACcitation   = "%%CITATION = ARXIV:1811.08002;%%",
}

@article{Nagy:2001fj,
      author         = "Nagy, Zoltan",
      title          = "{Three-Jet Cross Sections in Hadron-Hadron Collisions at Next-To-Leading Order}",
      journal        = "Phys. Rev. Lett.",
      volume         = "88",
      year           = "2002",
      pages          = "122003",
      doi            = "10.1103/PhysRevLett.88.122003",
      eprint         = "hep-ph/0110315",
      archivePrefix  = "arXiv",
      primaryClass   = "hep-ph",
      reportNumber   = "IPPP-01-48, DCPT-01-96",
      SLACcitation   = "%%CITATION = HEP-PH/0110315;%%",
}

@article{Nagy:2003tz,
      author         = "Nagy, Zoltan",
      title          = "{Next-to-leading order calculation of three-jet
                        observables in hadron-hadron collision}",
      journal        = "Phys. Rev. D",
      volume         = "68",
      year           = "2003",
      pages          = "094002",
      doi            = "10.1103/PhysRevD.68.094002",
      eprint         = "hep-ph/0307268",
      archivePrefix  = "arXiv",
      primaryClass   = "hep-ph",
      SLACcitation   = "%%CITATION = HEP-PH/0307268;%%",
}

@Article{Fox:2011pm,
      author         = "Fox, Patrick J. and Harnik, Roni and Kopp, Joachim and
                        Tsai, Yuhsin",
      title          = "{Missing energy signatures of dark matter at the LHC}",
      journal        = "Phys. Rev. D",
      volume         = "85",
      year           = "2012",
      pages          = "056011",
      doi            = "10.1103/PhysRevD.85.056011",
      eprint         = "1109.4398",
      archivePrefix  = "arXiv",
      primaryClass   = "hep-ph",
      reportNumber   = "FERMILAB-PUB-11-478-T",
      SLACcitation   = "%%CITATION = ARXIV:1109.4398;%%"
}

@Article{Rajaraman:2011wf,
      author         = "Rajaraman, Arvind and Shepherd, William and Tait, Tim M.
                        P. and Wijangco, Alexander M.",
      title          = "{LHC bounds on interactions of dark matter}",
      journal        = "Phys. Rev. D",
      volume         = "84",
      year           = "2011",
      pages          = "095013",
      doi            = "10.1103/PhysRevD.84.095013",
      eprint         = "1108.1196",
      archivePrefix  = "arXiv",
      primaryClass   = "hep-ph",
      reportNumber   = "UCI-HEP-TR-2011-15",
      SLACcitation   = "%%CITATION = ARXIV:1108.1196;%%"
}

@Article{Goodman:2010ku,
      author         = "Goodman, Jessica and Ibe, Masahiro and Rajaraman, Arvind
                        and Shepherd, William and Tait, Tim M. P. and Yu, Hai-Bo",
      title          = "{Constraints on dark matter from colliders}",
      journal        = "Phys. Rev. D",
      volume         = "82",
      year           = "2010",
      pages          = "116010",
      doi            = "10.1103/PhysRevD.82.116010",
      eprint         = "1008.1783",
      archivePrefix  = "arXiv",
      primaryClass   = "hep-ph",
      reportNumber   = "UCI-HEP-TR-2010-15",
      SLACcitation   = "%%CITATION = ARXIV:1008.1783;%%"
}

@Article{Goodman:2010yf,
      author         = "Goodman, Jessica and Ibe, Masahiro and Rajaraman, Arvind
                        and Shepherd, William and Tait, Tim M. P. and Yu, Hai-Bo",
      title          = "{Constraints on light Majorana dark matter from colliders}",
      journal        = "Phys. Lett. B",
      volume         = "695",
      year           = "2011",
      pages          = "185-188",
      doi            = "10.1016/j.physletb.2010.11.009",
      eprint         = "1005.1286",
      archivePrefix  = "arXiv",
      primaryClass   = "hep-ph",
      reportNumber   = "UCI-HEP-TR-2010-09",
      SLACcitation   = "%%CITATION = ARXIV:1005.1286;%%"
}

@Article{Beltran:2010ww,
      author         = "Beltr\'an, Maria and Hooper, Dan and Kolb, Edward W. and
                        Krusberg, Zosia A. C. and Tait, Tim M. P.",
      title          = "{Maverick dark matter at colliders}",
      journal        = "JHEP",
      volume         = "09",
      year           = "2010",
      pages          = "037",
      doi            = "10.1007/JHEP09(2010)037",
      eprint         = "1002.4137",
      archivePrefix  = "arXiv",
      primaryClass   = "hep-ph",
      reportNumber   = "FERMILAB-PUB-10-092-A",
      SLACcitation   = "%%CITATION = ARXIV:1002.4137;%%"
}

@Article{Pinna:2017tay,
      author         = "Pinna, Deborah and Zucchetta, Alberto and Buckley,
                        Matthew R. and Canelli, Florencia",
      title          = "{Single top quarks and dark matter}",
      journal        = "Phys. Rev. D",
      volume         = "96",
      year           = "2017",
      number         = "3",
      pages          = "035031",
      doi            = "10.1103/PhysRevD.96.035031",
      eprint         = "1701.05195",
      archivePrefix  = "arXiv",
      primaryClass   = "hep-ph",
      SLACcitation   = "%%CITATION = ARXIV:1701.05195;%%"
}

@article{LZ:2022lsv,
  author        = {{LUX-ZEPLIN Collaboration}},
  collaboration = {LZ},
  title         = {{First Dark Matter Search Results from the LUX-ZEPLIN (LZ) Experiment}},
  eprint        = {2207.03764},
  archiveprefix = {arXiv},
  primaryclass  = {hep-ex},
  doi           = {10.1103/PhysRevLett.131.041002},
  journal       = {Phys. Rev. Lett.},
  volume        = {131},
  number        = {4},
  pages         = {041002},
  year          = {2023}
}

@Booklet{urldmhf,
author = "{DM Forum repository}",
title = "{DMScalarMed\_loop UFO model}",
url = "https://svnweb.cern.ch/cern/wsvn/LHCDMF/trunk/models/HF_S%2BPS/",
}

@Booklet{urlscalmj,
author = "{DM Forum repository}",
title = "{DMS\_tloop UFO model}",
url = "https://svnweb.cern.ch/cern/wsvn/LHCDMF/trunk/models/Monojet_DMS_tloop/",
}

@Booklet{urldmsimp,
author = "{DMSimp Repository}",
title = "{ DMSimp UFO model}",
url = "http://feynrules.irmp.ucl.ac.be/wiki/DMsimp",
}

@Booklet{urldmv,
author = "{DM Forum repository}",
title = "{ DMV UFO model}",
url = "https://svnweb.cern.ch/cern/wsvn/LHCDMF/trunk/models/Monojet_DMV/?#ae98247b340ee12c1e7b0139c2062d807",
}

@Article{Pani:2017qyd,
      author         = "Pani, Priscilla and Polesello, Giacomo",
      title          = "{Dark matter production in association with a single
                        top-quark at the LHC in a two-Higgs-doublet model with a
                        pseudoscalar mediator}",
      journal        = "Phys. Dark Univ.",
      volume         = "21",
      year           = "2018",
      pages          = "8-15",
      doi            = "10.1016/j.dark.2018.04.006",
      eprint         = "1712.03874",
      archivePrefix  = "arXiv",
      primaryClass   = "hep-ph",
      SLACcitation   = "%%CITATION = ARXIV:1712.03874;%%"
}

@Article{DAmbrosio:2002vsn,
      author         = "D'Ambrosio, G. and Giudice, G. F. and Isidori, G. and
                        Strumia, A.",
      title={Minimal flavour violation: an effective field theory approach},
      journal        = "Nucl. Phys. B",
      volume         = "645",
      year           = "2002",
      pages          = "155-187",
      doi            = "10.1016/S0550-3213(02)00836-2",
      eprint         = "hep-ph/0207036",
      archivePrefix  = "arXiv",
      primaryClass   = "hep-ph",
      reportNumber   = "CERN-TH-2002-147, IFUP-TH-2002-17",
      SLACcitation   = "%%CITATION = HEP-PH/0207036;%%"
}

@Article{Buckley:2014fba,
      author         = "Buckley, Matthew R. and Feld, David and Gon\c{c}alves,
                        Dorival",
      title          = "{Scalar simplified models for dark matter}",
      journal        = "Phys. Rev. D",
      volume         = "91",
      year           = "2015",
      pages          = "015017",
      doi            = "10.1103/PhysRevD.91.015017",
      eprint         = "1410.6497",
      archivePrefix  = "arXiv",
      primaryClass   = "hep-ph",
      reportNumber   = "IPPP-14-92, DCPT-14-184",
      SLACcitation   = "%%CITATION = ARXIV:1410.6497;%%"
}

@Article{Bai:2010hh,
      author         = "Bai, Yang and Fox, Patrick J. and Harnik, Roni",
      title          = "{The Tevatron at the frontier of dark matter direct detection}",
      journal        = "JHEP",
      volume         = "12",
      year           = "2010",
      pages          = "048",
      doi            = "10.1007/JHEP12(2010)048",
      eprint         = "1005.3797",
      archivePrefix  = "arXiv",
      primaryClass   = "hep-ph",
      reportNumber   = "FERMILAB-PUB-10-143-T",
      SLACcitation   = "%%CITATION = ARXIV:1005.3797;%%"
}

@Article{Abdallah:2015ter,
      author         = "Abdallah, Jalal and others",
      title          = "{Simplified models for dark matter searches at the LHC}",
      journal        = "Phys. Dark Univ.",
      volume         = "9-10",
      year           = "2015",
      pages          = "8-23",
      doi            = "10.1016/j.dark.2015.08.001",
      eprint         = "1506.03116",
      archivePrefix  = "arXiv",
      primaryClass   = "hep-ph",
      reportNumber   = "FERMILAB-PUB-15-283-CD, CERN-PH-TH-2015-139",
      SLACcitation   = "%%CITATION = ARXIV:1506.03116;%%"
}

@Article{Feng:2010gw,
      author         = "Feng, Jonathan L.",
      title          = "{Dark Matter Candidates from Particle Physics and Methods
                        of Detection}",
      journal        = "Ann. Rev. Astron. Astrophys.",
      volume         = "48",
      year           = "2010",
      pages          = "495-545",
      doi            = "10.1146/annurev-astro-082708-101659",
      eprint         = "1003.0904",
      archivePrefix  = "arXiv",
      primaryClass   = "astro-ph.CO",
      reportNumber   = "UCI-TR-2009-13",
      SLACcitation   = "%%CITATION = ARXIV:1003.0904;%%"
}

@Article{Bertone:2004pz,
      author         = "Bertone, Gianfranco and Hooper, Dan and Silk, Joseph",
      title          = "{Particle dark matter: Evidence, candidates and
                        constraints}",
      journal        = "Phys. Rept.",
      volume         = "405",
      year           = "2005",
      pages          = "279-390",
      doi            = "10.1016/j.physrep.2004.08.031",
      eprint         = "hep-ph/0404175",
      archivePrefix  = "arXiv",
      primaryClass   = "hep-ph",
      reportNumber   = "FERMILAB-PUB-04-047-A",
      SLACcitation   = "%%CITATION = HEP-PH/0404175;%%"
}

@article{PhysRevD.100.022001,
  title = {Dark matter search results from the complete exposure of the PICO-60 ${\mathrm{C}}_{3}{\mathrm{F}}_{8}$ bubble chamber},
  author         = {{PICO Collaboration}},
  collaboration = {PICO Collaboration},
  journal = {Phys. Rev. D},
  volume = {100},
  issue = {2},
  pages = {022001},
  numpages = {9},
  year = {2019},
  month = {Jul},
  eprint = "1902.04031",
  archivePrefix = "arXiv",
  publisher = {American Physical Society},
  doi = {10.1103/PhysRevD.100.022001},
  % url = {https://link.aps.org/doi/10.1103/PhysRevD.100.022001}
}

@article{PhysRevLett.118.251302,
  author         = {{LUX Collaboration}},
  title = {Limits on Spin-Dependent WIMP-Nucleon Cross Section Obtained from the Complete LUX Exposure},
  collaboration = {LUX Collaboration},
  journal = {Phys. Rev. Lett.},
  volume = {118},
  issue = {25},
  pages = {251302},
  numpages = {7},
  year = {2017},
  month = {Jun},
  publisher = {American Physical Society},
  doi = {10.1103/PhysRevLett.118.251302},
  eprint         = "1705.03380",
  archivePrefix  = "arXiv",
  % url = {https://link.aps.org/doi/10.1103/PhysRevLett.118.251302}
}

@article{PhysRevLett.131.041003,
  title = {First Dark Matter Search with Nuclear Recoils from the XENONnT Experiment},
  author = {{XENON Collaboration}},
  collaboration = {XENON Collaboration},
  journal = {Phys. Rev. Lett.},
  volume = {131},
  issue = {4},
  pages = {041003},
  numpages = {9},
  year = {2023},
  month = {Jul},
  publisher = {American Physical Society},
  doi = {10.1103/PhysRevLett.131.041003},
  eprint         = "2303.14729",
  archivePrefix  = "arXiv",
  % url = {https://link.aps.org/doi/10.1103/PhysRevLett.131.041003}
}

@article{Krohn_2009,
	author = "Krohn, David and Thaler, Jesse and Wang, Lian-Tao",
	title = "{Jets with Variable R}",
	eprint = "0903.0392",
	archivePrefix = "arXiv",
	primaryClass = "hep-ph",
	doi = "10.1088/1126-6708/2009/06/059",
	journal = "JHEP",
	volume = "06",
	pages = "059",
	year = "2009"
}

@Article{Thaler:2010tr,
      author         = "Thaler, Jesse and Van Tilburg, Ken",
      title          = "{Identifying boosted objects with N-subjettiness}",
      journal        = "JHEP",
      volume         = "03",
      year           = "2011",
      pages          = "015",
      doi            = "10.1007/JHEP03(2011)015",
      eprint         = "1011.2268",
      archivePrefix  = "arXiv",
      primaryClass   = "hep-ph",
      reportNumber   = "MIT-CTP-4191",
      SLACcitation   = "%%CITATION = ARXIV:1011.2268;%%"
}

@article{Baldi:2016fzo,
      author         = "Baldi, Pierre and Cranmer, Kyle and Faucett, Taylor and
                        Sadowski, Peter and Whiteson, Daniel",
      title          = "{Parameterized neural networks for high-energy physics}",
      journal        = "Eur. Phys. J. C",
      volume         = "76",
      year           = "2016",
      number         = "5",
      pages          = "235",
      doi            = "10.1140/epjc/s10052-016-4099-4",
      eprint         = "1601.07913",
      archivePrefix  = "arXiv",
      primaryClass   = "hep-ex",
      SLACcitation   = "%%CITATION = ARXIV:1601.07913;%%"
}

@misc{coupling_rescaling_tool,
  title         = {Displaying dark matter constraints from colliders with varying simplified model parameters},
  author        = {Andreas Albert and Antonio Boveia and Oleg Brandt and Eric Corrigan and Zeynep Demiragli and Caterina Doglioni and Etienne Dreyer and Boyu Gao and Josh Greaves and Ulrich Haisch and Philip Harris and Greg Landsberg and Alexander Moreno and Katherine Pachal and Priscilla Pani and Federica Piazza and Tim M. P. Tait and David Yu and Felix Yu and Lian-Tao Wang},
  year          = {2022},
  eprint        = {2203.12035},
  archiveprefix = {arXiv},
  primaryclass  = {hep-ph}
}

@article{Harris_2015,
  title     = {Constraining dark sectors at colliders: Beyond the effective theory approach},
  author    = {Harris, Philip and Khoze, Valentin V. and Spannowsky, Michael and Williams, Ciaran},
  journal   = {Phys. Rev. D},
  volume    = {91},
  issue     = {5},
  pages     = {055009},
  numpages  = {19},
  year      = {2015},
  month     = {Mar},
  publisher = {American Physical Society},
  doi       = {10.1103/PhysRevD.91.055009},
  % url       = {https://link.aps.org/doi/10.1103/PhysRevD.91.055009},
  eprint    = {1411.0535},
  archiveprefix = {arXiv},
  primaryclass  = {hep-ex}
}

@article{Agnes_2018,
  title         = {Search for Dark-Matter--Nucleon Interactions via Migdal Effect with DarkSide-50},
  author        = {{DarkSide Collaboration}},
  journal       = {Phys. Rev. Lett.},
  volume        = {130},
  issue         = {10},
  pages         = {101001},
  numpages      = {7},
  year          = {2023},
  month         = {Mar},
  publisher     = {American Physical Society},
  doi           = {10.1103/PhysRevLett.130.101001},
  eprint        = {1802.06994},
  archiveprefix = {arXiv},
  xprimaryclass = {astro-ph.HE}
}

@article{Ackermann:2015zua,
  author        = {{Fermi-LAT Collaboration}},
  collaboration = {Fermi-LAT},
  title         = {{Searching for Dark Matter Annihilation from Milky Way Dwarf Spheroidal Galaxies with Six Years of Fermi Large Area Telescope Data}},
  eprint        = {1503.02641},
  archiveprefix = {arXiv},
  primaryclass  = {astro-ph.HE},
  reportnumber  = {FERMILAB-PUB-15-081-AE},
  doi           = {10.1103/PhysRevLett.115.231301},
  journal       = {Phys. Rev. Lett.},
  volume        = {115},
  number        = {23},
  pages         = {231301},
  year          = {2015}
}

@article{Panda2023,
  title = {Search for Solar $^{8}\mathrm{B}$ Neutrinos in the PandaX-4T Experiment Using Neutrino-Nucleus Coherent Scattering},
  author        = {{PandaX-4T Collaboration}},
  collaboration = {PandaX Collaboration},
  journal = {Phys. Rev. Lett.},
  volume = {130},
  issue = {2},
  pages = {021802},
  numpages = {8},
  year = {2023},
  month = {Jan},
  publisher = {American Physical Society},
  doi = {10.1103/PhysRevLett.130.021802},
  url = {https://link.aps.org/doi/10.1103/PhysRevLett.130.021802}
}

@article{Meng_2021,
  doi           = {10.1103/physrevlett.127.261802},
  eprint        = {2107.13438},
  archiveprefix = {arXiv},
  xprimaryclass = {hep-ex},
  year          = 2021,
  month         = {dec},
  publisher     = {American Physical Society ({APS})},
  volume        = {127},
  number        = {26},
  pages         = {261802},
  author        = {{PandaX-4T Collaboration}},
  collaboration = {PANDAX},
  title         = {Dark Matter Search Results from the {PandaX}-4T Commissioning Run},
  journal       = {Phys. Rev. Lett.}
}

@article{Chala_2015,
	author = "Chala, Mikael and Kahlhoefer, Felix and McCullough, Matthew and Nardini, Germano and Schmidt-Hoberg, Kai",
	title = "{Constraining Dark Sectors with Monojets and Dijets}",
	eprint = "1503.05916",
	archivePrefix = "arXiv",
	primaryClass = "hep-ph",
	reportNumber = "DESY-15-036, CERN-PH-TH-2015-057",
	doi = "10.1007/JHEP07(2015)089",
	journal = "JHEP",
	volume = "07",
	pages = "089",
	year = "2015"
}

@article{Nachman:2014kla,
  author        = {Nachman, Benjamin and Nef, Pascal and Schwartzman, Ariel and Swiatlowski, Maximilian and Wanotayaroj, Chaowaroj},
  title         = {{Jets from Jets: Re-clustering as a tool for large radius jet reconstruction and grooming at the LHC}},
  eprint        = {1407.2922},
  archiveprefix = {arXiv},
  primaryclass  = {hep-ph},
  doi           = {10.1007/JHEP02(2015)075},
  journal       = {JHEP},
  volume        = {02},
  pages         = {075},
  year          = {2015}
}

@article{Clowe_2006,
  doi       = {10.1086/508162},
  year      = {2006},
  month     = {aug},
  publisher = {},
  volume    = {648},
  number    = {2},
  pages     = {L109},
  author    = {Douglas Clowe and Maruša Bradač and Anthony H. Gonzalez and Maxim Markevitch and Scott W. Randall and Christine Jones and Dennis Zaritsky},
  title     = {A Direct Empirical Proof of the Existence of Dark Matter},
  journal   = {The Astrophysical Journal},
}

@misc{EXOT-2019-27,
      title={Search for new particles in events with a hadronically decaying W or Z boson and large missing transverse momentum at $\sqrt{s}$ = 13 TeV using the ATLAS detector}, 
      author={ATLAS Collaboration},
      year={2024},
      eprint={2406.01272},
      archivePrefix={arXiv},
      primaryClass={hep-ex}
}

@article{Fayet:1976et,
      author         = "Fayet, Pierre",
      title          = "{Supersymmetry and weak, electromagnetic and strong
                        interactions}",
      journal        = "Phys. Lett. B",
      volume         = "64",
      year           = "1976",
      pages          = "159",
      doi            = "10.1016/0370-2693(76)90319-1",
      reportNumber   = "PTENS 76/14",
      SLACcitation   = "%%CITATION = PHLTA,B64,159;%%"
}

@article{Fayet:1977yc,
      author         = "Fayet, Pierre",
      title          = "{Spontaneously broken supersymmetric theories of weak,
                        electromagnetic and strong interactions}",
      journal        = "Phys. Lett. B",
      volume         = "69",
      year           = "1977",
      pages          = "489",
      doi            = "10.1016/0370-2693(77)90852-8",
      reportNumber   = "LPTENS 77/11",
      SLACcitation   = "%%CITATION = PHLTA,B69,489;%%"
}

@article{Farrar:1978xj,
      author         = "Farrar, Glennys R. and Fayet, Pierre",
      title          = "{Phenomenology of the production, decay, and detection of
                        new hadronic states associated with supersymmetry}",
      journal        = "Phys. Lett. B",
      volume         = "76",
      year           = "1978",
      pages          = "575",
      doi            = "10.1016/0370-2693(78)90858-4",
      reportNumber   = "CALT-68-648",
      SLACcitation   = "%%CITATION = PHLTA,B76,575;%%"
}

@Article{Cacciari:2008gp,
     author    = "Cacciari, Matteo and Salam, Gavin P. and Soyez, Gregory",
     title     = "{The anti-\(k_{t}\) jet clustering algorithm}",
     journal   = "JHEP",
     volume    = "04",
     year      = "2008",
     pages     = "063",
     eprint    = "0802.1189",
     archivePrefix = "arXiv",
     primaryClass  =  "hep-ph",
     doi       = "10.1088/1126-6708/2008/04/063",
     SLACcitation  = "%%CITATION = 0802.1189;%%"
}

@Article{Fastjet,
      author         = "Cacciari, Matteo and Salam, Gavin P. and Soyez, Gregory",
      title          = "{FastJet user manual}",
      journal        = "Eur. Phys. J. C",
      volume         = "72",
      year           = "2012",
      pages          = "1896",
      doi            = "10.1140/epjc/s10052-012-1896-2",
      eprint         = "1111.6097",
      archivePrefix  = "arXiv",
      primaryClass   = "hep-ph",
      reportNumber   = "CERN-PH-TH-2011-297",
      SLACcitation   = "%%CITATION = ARXIV:1111.6097;%%"
}

@Article{Sjostrand:2007gs,
	author    = "Sj{\"o}strand, T. and Mrenna, S. and Skands, P.",
	title     = "{A brief introduction to PYTHIA 8.1}",
	journal   = "Comput. Phys. Commun.",
	volume    = "178",
	year      = "2008",
	pages     = "852-867",
	eprint    = "0710.3820",
	archivePrefix = "arXiv",
	primaryClass  =  "hep-ph",
	doi       = "10.1016/j.cpc.2008.01.036",
	SLACcitation  = "%%CITATION = 0710.3820;%%"
}

@Article{Alwall:2014hca,
      author         = "Alwall, J. and Frederix, R. and Frixione, S. and Hirschi,
                        V. and Maltoni, F. and Mattelaer, O. and Shao, H. -S. and
                        Stelzer, T. and Torrielli, P. and Zaro, M.",
      title          = "{The automated computation of tree-level and
                        next-to-leading order differential cross sections, and
                        their matching to parton shower simulations}",
      journal        = "JHEP",
      volume         = "07",
      year           = "2014",
      pages          = "079",
      doi            = "10.1007/JHEP07(2014)079",
      eprint         = "1405.0301",
      archivePrefix  = "arXiv",
      primaryClass   = "hep-ph",
      reportNumber   = "CERN-PH-TH-2014-064, CP3-14-18, LPN14-066, MCNET-14-09,
                        ZU-TH-14-14",
      SLACcitation   = "%%CITATION = ARXIV:1405.0301;%%"
}

@Article{Alioli:2010xd,
      author         = "Alioli, Simone and Nason, Paolo and Oleari, Carlo and Re,
                        Emanuele",
      title          = "{A general framework for implementing NLO calculations in
                        shower Monte Carlo programs: the POWHEG BOX}",
      journal        = "JHEP",
      volume         = "06",
      pages          = "043",
      doi            = "10.1007/JHEP06(2010)043",
      year           = "2010",
      eprint         = "1002.2581",
      archivePrefix  = "arXiv",
      primaryClass   = "hep-ph",
}

@Article{PERF-2007-01,
    author         = "{ATLAS Collaboration}",
    title          = "{The ATLAS Experiment at the CERN Large Hadron Collider}",
    journal        = "JINST",
    volume         = "3",
    year           = "2008",
    pages          = "S08003",
    doi            = "10.1088/1748-0221/3/08/S08003",
    primaryClass   = "hep-ex",
}

@Article{PERF-2012-04,
    author         = "{ATLAS Collaboration}",
    title          = "{Performance of \(b\)-jet identification in the ATLAS experiment}",
    journal        = "JINST",
    volume         = "11",
    year           = "2016",
    pages          = "P04008",
    doi            = "10.1088/1748-0221/11/04/P04008",
    reportNumber   = "CERN-PH-EP-2015-216",
    eprint         = "1512.01094",
    archivePrefix  = "arXiv",
    primaryClass   = "hep-ex",
}

@Article{PERF-2014-03,
    author         = "{ATLAS Collaboration}",
    title          = "{Performance of pile-up mitigation techniques for jets in \(pp\) collisions at \(\sqrt{s} = 8\,\text{TeV}\) using the ATLAS detector}",
    journal        = "Eur. Phys. J. C",
    volume         = "76",
    year           = "2016",
    pages          = "581",
    doi            = "10.1140/epjc/s10052-016-4395-z",
    reportNumber   = "CERN-PH-EP-2015-206",
    eprint         = "1510.03823",
    archivePrefix  = "arXiv",
    primaryClass   = "hep-ex",
}

@Article{EXOT-2015-04,
    author         = "{ATLAS Collaboration}",
    title          = "{Search for heavy particles decaying into top-quark pairs using lepton-plus-jets events in proton--proton collisions at \(\sqrt{s} = 13\,\text{TeV}\) with the ATLAS detector}",
    journal        = "Eur. Phys. J. C",
    volume         = "78",
    year           = "2018",
    pages          = "565",
    doi            = "10.1140/epjc/s10052-018-5995-6",
    reportNumber   = "CERN-EP-2018-48",
    eprint         = "1804.10823",
    archivePrefix  = "arXiv",
    primaryClass   = "hep-ex",
}

@Article{PERF-2015-09,
    author         = "{ATLAS Collaboration}",
    title          = "{Jet reconstruction and performance using particle flow with the ATLAS Detector}",
    journal        = "Eur. Phys. J. C",
    volume         = "77",
    year           = "2017",
    pages          = "466",
    doi            = "10.1140/epjc/s10052-017-5031-2",
    reportNumber   = "CERN-EP-2017-024",
    eprint         = "1703.10485",
    archivePrefix  = "arXiv",
    primaryClass   = "hep-ex",
}

@Article{EXOT-2016-20,
    author         = "{ATLAS Collaboration}",
    title          = "{Search for Low-Mass Dijet Resonances Using Trigger-Level Jets with the ATLAS Detector in \(pp\) Collisions at \(\sqrt{s} = 13\,\text{TeV}\)}",
    journal        = "Phys. Rev. Lett.",
    volume         = "121",
    year           = "2018",
    pages          = "081801",
    doi            = "10.1103/PhysRevLett.121.081801",
    reportNumber   = "CERN-EP-2018-033",
    eprint         = "1804.03496",
    archivePrefix  = "arXiv",
    primaryClass   = "hep-ex",
}

@Article{EXOT-2016-21,
    author         = "{ATLAS Collaboration}",
    title          = "{Search for new phenomena in dijet events using \(37\,\text{fb}^{-1}\) of \(pp\) collision data collected at \(\sqrt{s} = 13\,\text{TeV}\) with the ATLAS detector}",
    journal        = "Phys. Rev. D",
    volume         = "96",
    year           = "2017",
    pages          = "052004",
    doi            = "10.1103/PhysRevD.96.052004",
    reportNumber   = "CERN-EP-2017-042",
    eprint         = "1703.09127",
    archivePrefix  = "arXiv",
    primaryClass   = "hep-ex",
}

@Article{PERF-2016-01,
    author         = "{ATLAS Collaboration}",
    title          = "{Electron efficiency measurements with the ATLAS detector using 2012 LHC proton--proton collision data}",
    journal        = "Eur. Phys. J. C",
    volume         = "77",
    year           = "2017",
    pages          = "195",
    doi            = "10.1140/epjc/s10052-017-4756-2",
    reportNumber   = "CERN-EP-2016-262",
    eprint         = "1612.01456",
    archivePrefix  = "arXiv",
    primaryClass   = "hep-ex",
}

@Article{PERF-2016-07,
    author         = "{ATLAS Collaboration}",
    title          = "{Performance of missing transverse momentum reconstruction with the ATLAS detector using proton--proton collisions at \(\sqrt{s} = 13\,\text{TeV}\)}",
    journal        = "Eur. Phys. J. C",
    volume         = "78",
    year           = "2018",
    pages          = "903",
    doi            = "10.1140/epjc/s10052-018-6288-9",
    reportNumber   = "CERN-EP-2017-274",
    eprint         = "1802.08168",
    archivePrefix  = "arXiv",
    primaryClass   = "hep-ex",
}

@Article{TRIG-2016-01,
    author         = "{ATLAS Collaboration}",
    title          = "{Performance of the ATLAS trigger system in 2015}",
    journal        = "Eur. Phys. J. C",
    volume         = "77",
    year           = "2017",
    pages          = "317",
    doi            = "10.1140/epjc/s10052-017-4852-3",
    reportNumber   = "CERN-EP-2016-241",
    eprint         = "1611.09661",
    archivePrefix  = "arXiv",
    primaryClass   = "hep-ex",
}

@Article{EXOT-2017-01,
    author         = "{ATLAS Collaboration}",
    title          = "{Search for light resonances decaying to boosted quark pairs and produced in association with a photon or a jet in proton--proton collisions at \(\sqrt{s} = 13\,\text{TeV}\) with the ATLAS detector}",
    journal        = "Phys. Lett. B",
    volume         = "788",
    year           = "2019",
    pages          = "316",
    doi            = "10.1016/j.physletb.2018.09.062",
    reportNumber   = "CERN-EP-2017-280",
    eprint         = "1801.08769",
    archivePrefix  = "arXiv",
    primaryClass   = "hep-ex",
}

@Article{EXOT-2017-32,
    author         = "{ATLAS Collaboration}",
    title          = "{Constraints on mediator-based dark matter and scalar dark energy models using \(\sqrt{s} = 13\,\text{TeV}\) \(pp\) collision data collected by the ATLAS detector}",
    journal        = "JHEP",
    volume         = "05",
    year           = "2019",
    pages          = "142",
    doi            = "10.1007/JHEP05(2019)142",
    reportNumber   = "CERN-EP-2018-334",
    eprint         = "1903.01400",
    archivePrefix  = "arXiv",
    primaryClass   = "hep-ex",
}

@Article{EGAM-2018-01,
    author         = "{ATLAS Collaboration}",
    title          = "{Electron and photon performance measurements with the ATLAS detector using the 2015--2017 LHC proton--proton collision data}",
    journal        = "JINST",
    volume         = "14",
    year           = "2019",
    pages          = "P12006",
    doi            = "10.1088/1748-0221/14/12/P12006",
    reportNumber   = "CERN-EP-2019-145",
    eprint         = "1908.00005",
    archivePrefix  = "arXiv",
    primaryClass   = "hep-ex",
}

@Article{EXOT-2018-06,
    author         = "{ATLAS Collaboration}",
    title          = "{Search for new phenomena in events with an energetic jet and missing transverse momentum in \(pp\) collisions at \(\sqrt{s} = 13\,\text{TeV}\) with the ATLAS detector}",
    journal        = "Phys. Rev. D",
    volume         = "103",
    year           = "2021",
    pages          = "112006",
    doi            = "10.1103/PhysRevD.103.112006",
    reportNumber   = "CERN-EP-2020-238",
    eprint         = "2102.10874",
    archivePrefix  = "arXiv",
    primaryClass   = "hep-ex",
}

@Article{EXOT-2018-08,
    author         = "{ATLAS Collaboration}",
    title          = "{Search for high-mass dilepton resonances using \(139\,\text{fb}^{-1}\) of \(pp\) collision data collected at \(\sqrt{s} = 13\,\text{TeV}\) with the ATLAS detector}",
    journal        = "Phys. Lett. B",
    volume         = "796",
    year           = "2019",
    pages          = "68",
    doi            = "10.1016/j.physletb.2019.07.016",
    reportNumber   = "CERN-EP-2019-030",
    eprint         = "1903.06248",
    archivePrefix  = "arXiv",
    primaryClass   = "hep-ex",
}

@Article{EXOT-2018-32,
    author         = "{ATLAS Collaboration}",
    title          = "{Search for dijet resonances in events with an isolated charged lepton using \(\sqrt{s} = 13\,\text{TeV}\) proton--proton collision data collected by the ATLAS detector}",
    journal        = "JHEP",
    volume         = "06",
    year           = "2020",
    pages          = "151",
    doi            = "10.1007/JHEP06(2020)151",
    reportNumber   = "CERN-EP-2019-276",
    eprint         = "2002.11325",
    archivePrefix  = "arXiv",
    primaryClass   = "hep-ex",
}

@Article{EXOT-2018-43,
    author         = "{ATLAS Collaboration}",
    title          = "{Search for dark matter produced in association with a single top quark in \(\sqrt{s} = 13\,\text{TeV}\) \(pp\) collisions with the ATLAS detector}",
    journal        = "Eur. Phys. J. C",
    volume         = "81",
    year           = "2021",
    pages          = "860",
    doi            = "10.1140/epjc/s10052-021-09566-y",
    reportNumber   = "CERN-EP-2020-184",
    eprint         = "2011.09308",
    archivePrefix  = "arXiv",
    primaryClass   = "hep-ex",
}

@Article{EXOT-2018-48,
    author         = "{ATLAS Collaboration}",
    title          = "{Search for  \(t\bar{t}\) resonances in fully hadronic final states in \(pp\) collisions at \(\sqrt{s} = 13\,\text{TeV}\) with the ATLAS detector}",
    journal        = "JHEP",
    volume         = "10",
    year           = "2020",
    pages          = "061",
    doi            = "10.1007/JHEP10(2020)061",
    reportNumber   = "CERN-EP-2020-055",
    eprint         = "2005.05138",
    archivePrefix  = "arXiv",
    primaryClass   = "hep-ex",
}

@Article{EXOT-2018-63,
    author         = "{ATLAS Collaboration}",
    title          = "{Search for dark matter in association with an energetic photon in \(pp\) collisions at \(\sqrt{s} = 13\,\text{TeV}\) with the ATLAS detector}",
    journal        = "JHEP",
    volume         = "02",
    year           = "2021",
    pages          = "226",
    doi            = "10.1007/JHEP02(2021)226",
    reportNumber   = "CERN-EP-2020-178",
    eprint         = "2011.05259",
    archivePrefix  = "arXiv",
    primaryClass   = "hep-ex",
}

@Article{FTAG-2018-01,
    author         = "{ATLAS Collaboration}",
    title          = "{ATLAS \(b\)-jet identification performance and efficiency measurement with \(t\bar{t}\) events in \(pp\) collisions at \(\sqrt{s} = 13\,\text{TeV}\)}",
    journal        = "Eur. Phys. J. C",
    volume         = "79",
    year           = "2019",
    pages          = "970",
    doi            = "10.1140/epjc/s10052-019-7450-8",
    reportNumber   = "CERN-EP-2019-132",
    eprint         = "1907.05120",
    archivePrefix  = "arXiv",
    primaryClass   = "hep-ex",
}

@Article{HIGG-2018-26,
    author         = "{ATLAS Collaboration}",
    title          = "{Search for associated production of a \(Z\) boson with an invisibly decaying Higgs boson or dark matter candidates at \(\sqrt{s} = 13\,\text{TeV}\) with the ATLAS detector}",
    journal        = "Phys. Lett. B",
    volume         = "829",
    year           = "2022",
    pages          = "137066",
    doi            = "10.1016/j.physletb.2022.137066",
    reportNumber   = "CERN-EP-2021-204",
    eprint         = "2111.08372",
    archivePrefix  = "arXiv",
    primaryClass   = "hep-ex",
}

@Article{JETM-2018-02,
    author         = "{ATLAS Collaboration}",
    title          = "{In situ calibration of large-radius jet energy and mass in \(13\,\text{TeV}\) proton--proton collisions with the ATLAS detector}",
    journal        = "Eur. Phys. J. C",
    volume         = "79",
    year           = "2019",
    pages          = "135",
    doi            = "10.1140/epjc/s10052-019-6632-8",
    reportNumber   = "CERN-EP-2018-191",
    eprint         = "1807.09477",
    archivePrefix  = "arXiv",
    primaryClass   = "hep-ex",
}

@Article{JETM-2018-03,
    author         = "{ATLAS Collaboration}",
    title          = "{Performance of top-quark and \(W\)-boson tagging with ATLAS in Run~2 of the LHC}",
    journal        = "Eur. Phys. J. C",
    volume         = "79",
    year           = "2019",
    pages          = "375",
    doi            = "10.1140/epjc/s10052-019-6847-8",
    reportNumber   = "CERN-EP-2018-192",
    eprint         = "1808.07858",
    archivePrefix  = "arXiv",
    primaryClass   = "hep-ex",
}

@Article{JETM-2018-05,
    author         = "{ATLAS Collaboration}",
    title          = "{Jet energy scale and resolution measured in proton--proton collisions at \(\sqrt{s} = 13\,\text{TeV}\) with the ATLAS detector}",
    journal        = "Eur. Phys. J. C",
    volume         = "81",
    year           = "2021",
    pages          = "689",
    doi            = "10.1140/epjc/s10052-021-09402-3",
    reportNumber   = "CERN-EP-2020-083",
    eprint         = "2007.02645",
    archivePrefix  = "arXiv",
    primaryClass   = "hep-ex",
}

@Article{JETM-2018-06,
    author         = "{ATLAS Collaboration}",
    title          = "{Optimisation of large-radius jet reconstruction for the ATLAS detector in \(13\,\text{TeV}\) proton--proton collisions}",
    journal        = "Eur. Phys. J. C",
    volume         = "81",
    year           = "2021",
    pages          = "334",
    doi            = "10.1140/epjc/s10052-021-09054-3",
    reportNumber   = "CERN-EP-2020-134",
    eprint         = "2009.04986",
    archivePrefix  = "arXiv",
    primaryClass   = "hep-ex",
}

@Article{MUON-2018-03,
    author         = "{ATLAS Collaboration}",
    title          = "{Muon reconstruction and identification efficiency in ATLAS using the full Run~2 \(pp\) collision data set at \(\sqrt{s} = 13\,\text{TeV}\)}",
    journal        = "Eur. Phys. J. C",
    volume         = "81",
    year           = "2021",
    pages          = "578",
    doi            = "10.1140/epjc/s10052-021-09233-2",
    reportNumber   = "CERN-EP-2020-199",
    eprint         = "2012.00578",
    archivePrefix  = "arXiv",
    primaryClass   = "hep-ex",
}

@Article{PIX-2018-001,
    author         = "Abbott, B. and others",
    title          = "{Production and integration of the ATLAS Insertable B-Layer}",
    journal        = "JINST",
    volume         = "13",
    year           = "2018",
    pages          = "T05008",
    doi            = "10.1088/1748-0221/13/05/T05008",
    eprint         = "1803.00844",
    archivePrefix  = "arXiv",
    primaryClass   = "physics.ins-det",
}

@Article{SUSY-2018-08,
    author         = "{ATLAS Collaboration}",
    title          = "{Search for new phenomena in events with two opposite-charge leptons, jets and missing transverse momentum in \(pp\) collisions at \(\sqrt{s} = 13\,\text{TeV}\) with the ATLAS detector}",
    journal        = "JHEP",
    volume         = "04",
    year           = "2021",
    pages          = "165",
    doi            = "10.1007/JHEP04(2021)165",
    reportNumber   = "CERN-EP-2020-243",
    eprint         = "2102.01444",
    archivePrefix  = "arXiv",
    primaryClass   = "hep-ex",
}

@Article{SUSY-2018-12,
    author         = "{ATLAS Collaboration}",
    title          = "{Search for a scalar partner of the top quark in the all-hadronic \(t\bar{t}\) plus missing transverse momentum final state at \(\sqrt{s} = 13\,\text{TeV}\) with the ATLAS detector}",
    journal        = "Eur. Phys. J. C",
    volume         = "80",
    year           = "2020",
    pages          = "737",
    doi            = "10.1140/epjc/s10052-020-8102-8",
    reportNumber   = "CERN-EP-2020-044",
    eprint         = "2004.14060",
    archivePrefix  = "arXiv",
    primaryClass   = "hep-ex",
}

@Article{SUSY-2018-34,
    author         = "{ATLAS Collaboration}",
    title          = "{Search for new phenomena in final states with \(b\)-jets and missing transverse momentum in \(\sqrt{s} = 13\,\text{TeV}\) \(pp\) collisions with the ATLAS detector}",
    journal        = "JHEP",
    volume         = "05",
    year           = "2021",
    pages          = "093",
    doi            = "10.1007/JHEP05(2021)093",
    reportNumber   = "CERN-EP-2021-001",
    eprint         = "2101.12527",
    archivePrefix  = "arXiv",
    primaryClass   = "hep-ex",
}

@Article{EXOT-2019-03,
    author         = "{ATLAS Collaboration}",
    title          = "{Search for new resonances in mass distributions of jet pairs using \(139\,\text{fb}^{-1}\) of \(pp\) collisions at \(\sqrt{s} = 13\,\text{TeV}\) with the ATLAS detector}",
    journal        = "JHEP",
    volume         = "03",
    year           = "2020",
    pages          = "145",
    doi            = "10.1007/JHEP03(2020)145",
    reportNumber   = "CERN-EP-2019-162",
    eprint         = "1910.08447",
    archivePrefix  = "arXiv",
    primaryClass   = "hep-ex",
}

@Article{EXOT-2019-26,
    author         = "{ATLAS Collaboration}",
    title          = "{Search for \(t\bar{t}H/A \to t\bar{t} t\bar{t}\) production in the multilepton final state in proton--proton collisions at \(\sqrt{s} = 13\,\text{TeV}\) with the ATLAS detector}",
    journal        = "JHEP",
    volume         = "07",
    year           = "2023",
    pages          = "203",
    doi            = "10.1007/JHEP07(2023)203",
    reportNumber   = "CERN-EP-2022-170",
    eprint         = "2211.01136",
    archivePrefix  = "arXiv",
    primaryClass   = "hep-ex",
}

@Article{FTAG-2019-07,
    author         = "{ATLAS Collaboration}",
    title          = "{ATLAS flavour-tagging algorithms for the LHC Run~2 \(pp\) collision dataset}",
    journal        = "Eur. Phys. J. C",
    volume         = "83",
    year           = "2023",
    pages          = "681",
    doi            = "10.1140/epjc/s10052-023-11699-1",
    reportNumber   = "CERN-EP-2022-226",
    eprint         = "2211.16345",
    archivePrefix  = "arXiv",
    primaryClass   = "physics.data-an",
}

@Article{SUSY-2019-12,
    author         = "{ATLAS Collaboration}",
    title          = "{Constraints on spin-0 dark matter mediators and invisible Higgs decays using ATLAS \(13\,\text{TeV}\) \(pp\) collision data with two top quarks and missing transverse momentum in the final state}",
    journal        = "Eur. Phys. J. C",
    volume         = "83",
    year           = "2023",
    pages          = "503",
    doi            = "10.1140/epjc/s10052-023-11477-z",
    reportNumber   = "CERN-EP-2022-133",
    eprint         = "2211.05426",
    archivePrefix  = "arXiv",
    primaryClass   = "hep-ex",
}

@Article{DAPR-2021-01,
    author         = "{ATLAS Collaboration}",
    title          = "{Luminosity determination in \(pp\) collisions at \(\sqrt{s} = 13\,\text{TeV}\) using the ATLAS detector at the LHC}",
    journal        = "Eur. Phys. J. C",
    volume         = "83",
    year           = "2023",
    pages          = "982",
    doi            = "10.1140/epjc/s10052-023-11747-w",
    reportNumber   = "CERN-EP-2022-281",
    eprint         = "2212.09379",
    archivePrefix  = "arXiv",
    primaryClass   = "hep-ex",
}

@Article{EXOT-2021-01,
    author         = "{ATLAS Collaboration}",
    title          = "{Search for dark matter produced in association with a single top quark and an energetic \(W\) boson in \(\sqrt{s} = 13\,\text{TeV}\) \(pp\) collisions with the ATLAS detector}",
    journal        = "Eur. Phys. J. C",
    volume         = "83",
    year           = "2023",
    pages          = "603",
    doi            = "10.1140/epjc/s10052-023-11582-z",
    reportNumber   = "CERN-EP-2022-146",
    eprint         = "2211.13138",
    archivePrefix  = "arXiv",
    primaryClass   = "hep-ex",
}

@Booklet{ATL-SOFT-PUB-2021-001,
    author         = "{ATLAS Collaboration}",
    title          = "{The ATLAS Collaboration Software and Firmware}",
    howpublished   = "{ATL-SOFT-PUB-2021-001}",
    url            = "https://cds.cern.ch/record/2767187",
    year           = "2021",
}

@Booklet{ATL-SOFT-PUB-2023-001,
    author         = "{ATLAS Collaboration}",
    title          = "{ATLAS Computing Acknowledgements}",
    howpublished   = "{ATL-SOFT-PUB-2023-001}",
    url            = "https://cds.cern.ch/record/2869272",
    year           = "2023",
}

@Report{ATLAS-TDR-19,
    author         = "{ATLAS Collaboration}",
    title          = "{ATLAS Insertable B-Layer: Technical Design Report}",
    type           = "ATLAS-TDR-19; CERN-LHCC-2010-013",
    year           = "2010",
    url            = "https://cds.cern.ch/record/1291633",
    related        = "ATLAS-TDR-19-addm",
    relatedstring  = "Addendum:",
}

@Booklet{ATLAS-CONF-2018-038,
    author         = "{ATLAS Collaboration}",
    title          = "{Object-based missing transverse momentum significance in the ATLAS Detector}",
    howpublished   = "{ATLAS-CONF-2018-038}",
    url            = "https://cds.cern.ch/record/2630948",
    year           = "2018",
}

@Booklet{ATLAS-CONF-2020-022,
    author         = "{ATLAS Collaboration}",
    title          = "{Measurement of the ATLAS Detector Jet Mass Response using Forward Folding with \(80\,\text{fb}^{-1}\) of \(\sqrt{s} = 13\,\text{TeV}\) \(pp\) data}",
    howpublished   = "{ATLAS-CONF-2020-022}",
    url            = "https://cds.cern.ch/record/2724442",
    year           = "2020",
}

@Booklet{ATL-PHYS-PUB-2016-012,
    author         = "{ATLAS Collaboration}",
    title          = "{Optimisation of the ATLAS \(b\)-tagging performance for the 2016 LHC Run}",
    howpublished   = "{ATL-PHYS-PUB-2016-012}",
    url            = "https://cds.cern.ch/record/2160731",
    year           = "2016",
}

@Booklet{ATL-PHYS-PUB-2019-026,
    author         = "{ATLAS Collaboration}",
    title          = "{Forward jet vertex tagging using the particle flow algorithm}",
    howpublished   = "{ATL-PHYS-PUB-2019-026}",
    url            = "https://cds.cern.ch/record/2683100",
    year           = "2019",
}
\clearpage
 
\begin{flushleft}
\hypersetup{urlcolor=black}
{\Large The ATLAS Collaboration}

\bigskip

\AtlasOrcid[0000-0002-6665-4934]{G.~Aad}$^\textrm{\scriptsize 104}$,
\AtlasOrcid[0000-0001-7616-1554]{E.~Aakvaag}$^\textrm{\scriptsize 17}$,
\AtlasOrcid[0000-0002-5888-2734]{B.~Abbott}$^\textrm{\scriptsize 123}$,
\AtlasOrcid[0000-0002-0287-5869]{S.~Abdelhameed}$^\textrm{\scriptsize 119a}$,
\AtlasOrcid[0000-0002-1002-1652]{K.~Abeling}$^\textrm{\scriptsize 56}$,
\AtlasOrcid[0000-0001-5763-2760]{N.J.~Abicht}$^\textrm{\scriptsize 50}$,
\AtlasOrcid[0000-0002-8496-9294]{S.H.~Abidi}$^\textrm{\scriptsize 30}$,
\AtlasOrcid[0009-0003-6578-220X]{M.~Aboelela}$^\textrm{\scriptsize 45}$,
\AtlasOrcid[0000-0002-9987-2292]{A.~Aboulhorma}$^\textrm{\scriptsize 36e}$,
\AtlasOrcid[0000-0001-5329-6640]{H.~Abramowicz}$^\textrm{\scriptsize 154}$
% ...
\bigskip
\\

% ...
$^{17}$Department for Physics and Technology, University of Bergen, Bergen; Norway.\\
% ...
$^{30}$Physics Department, Brookhaven National Laboratory, Upton NY; United States of America.\\
% ...
$^{36}$$^{(a)}$Facult\'e des Sciences Ain Chock, Universit\'e Hassan II de Casablanca;$^{(b)}$Facult\'{e} des Sciences, Universit\'{e} Ibn-Tofail, K\'{e}nitra;$^{(c)}$Facult\'e des Sciences Semlalia, Universit\'e Cadi Ayyad, LPHEA-Marrakech;$^{(d)}$LPMR, Facult\'e des Sciences, Universit\'e Mohamed Premier, Oujda;$^{(e)}$Facult\'e des sciences, Universit\'e Mohammed V, Rabat;$^{(f)}$Institute of Applied Physics, Mohammed VI Polytechnic University, Ben Guerir; Morocco.\\
% ...
$^{45}$Physics Department, Southern Methodist University, Dallas TX; United States of America.\\
% ...
$^{50}$Fakult\"{a}t Physik , Technische Universit{\"a}t Dortmund, Dortmund; Germany.\\
% ...
$^{56}$II. Physikalisches Institut, Georg-August-Universit\"{a}t G\"ottingen, G\"ottingen; Germany.\\
% ...
$^{104}$CPPM, Aix-Marseille Universit\'e, CNRS/IN2P3, Marseille; France.\\
% ...
$^{119}$$^{(a)}$New York University Abu Dhabi, Abu Dhabi;$^{(b)}$United Arab Emirates University, Al Ain; United Arab Emirates.\\
% ...
$^{123}$Homer L. Dodge Department of Physics and Astronomy, University of Oklahoma, Norman OK; United States of America.\\
% ...
$^{154}$Raymond and Beverly Sackler School of Physics and Astronomy, Tel Aviv University, Tel Aviv; Israel.\\
% ...
$^{a}$ Also Affiliated with an institute covered by a cooperation agreement with CERN.\\
$^{b}$ Also at An-Najah National University, Nablus; Palestine.\\
$^{c}$ Also at Borough of Manhattan Community College, City University of New York, New York NY; United States of America.\\
$^{d}$ Also at Center for Interdisciplinary Research and Innovation (CIRI-AUTH), Thessaloniki; Greece.\\
$^{e}$ Also at Centro Studi e Ricerche Enrico Fermi; Italy.\\
$^{f}$ Also at CERN, Geneva; Switzerland.\\
% ...

\end{flushleft}


%

%
%

%

%
%
%
%

\end{document}